\documentclass[reprint,aps,prd,longbibliography,floatfix,superscriptaddress,showkeys]{revtex4-2} 
\usepackage{amsmath}
\usepackage{amsfonts}
\usepackage{amssymb}
\usepackage{hyperref}
\usepackage{tabulary}
\usepackage[dvipsnames]{xcolor}
\usepackage{soul}
\usepackage{siunitx} %
\usepackage{selinput}%
\usepackage{todonotes}
\usepackage{cleveref} %
\usepackage{upgreek} %

\usepackage{graphicx}
\usepackage[export]{adjustbox} %
\usepackage{blindtext}
\usepackage{tabulary}
\usepackage{rotating}
\usepackage{afterpage}

\usepackage{multirow}
\usepackage{makecell}

\usepackage{booktabs}

\newcommand{\cpp}{\hbox{C\hspace{-0.5ex}
			    \protect\raisebox{0.5ex}
			    {\protect\scalebox{0.67}{++}}}}

\usepackage[toc,xindy,nonumberlist,shortcuts,acronym,automake]{glossaries}
\usepackage{csquotes}
\glsdisablehyper

\newacronym{AOM}{AOM}{acousto-optic modulator}
\newacronym{ASD}{ASD}{amplitude spectral density}

\newacronym{BH}{BH}{black hole}
\newacronym[firstplural={black hole binaries (BHBs)}]{BHB}{BHB}{black hole binary}

\newacronym{CAS}{CAS}{constellation acquisition sensor}
\newacronym{CBE}{CBE}{current best estimate}
\newacronym{CMM}{CMM}{coordinate measuring machine}
\newacronym{CMS}{CMS}{charge management system}
\newacronym{CMNT}{CMNT}{colloid micro-newton thruster}
\newacronym{CMB}{CMB}{cosmic microwave background}
\newacronym{CoM}{CoM}{centre of mass}
\newacronym{CQP}{CQP}{calibrated quadrant photodiode pair}
\newacronym{CTP}{CTP}{core technology program}

\newacronym{D/A}{D/A}{digital-to-analogue converter}
\newacronym{DAL}{DAL}{days after launch}
\newacronym{DF}{DF}{drag-free}
\newacronym{DFACS}{DFACS}{Drag-Free Attitude Control System}
\newacronym{DMU}{DMU}{Data Management Unit}
\newacronym{DoF}{DoF}{degree of freedom}
\newacronym{DOY}{DOY}{day of year}
\newacronym{DP}{DP}{diagnostic package}
\newacronym{DPC}{DPC}{Data Processing Centre}
\newacronym{DPLL}{DPLL}{digital phase locked loop}
\newacronym{DRS}{DRS}{disturbance reduction system}
\newacronym{DSC}{Daughter-S/C}{\enquote{Daughter} spacecraft}
\newacronym{DS}{DS}{Diagnostics Subsystem}
\newacronym{DTM}{DTM}{deterministic transfer manoeuvre}
\newacronym{DWS}{DWS}{differential wavefront sensing}

\newacronym{ED}{ED}{engineering days}
\newacronym{EGSE}{EGSE}{Electrical Ground Support Equipment}
\newacronym{EH}{EH}{Electrode housing}
\newacronym{EM}{EM}{electro-magnetic}
\newacronym{EOL}{EOL}{end-of-life}
\newacronym{EOM}{EOM}{electro-optical modulator}
\newacronym{ESA}{ESA}{European Space Agency}

\newacronym[first={finite-element}]{FE}{FE}{finite-element (methods)}
\newacronym{FIOS}{FIOS}{Fibre Injector Optical Subassembly}
\newacronym{FM}{FM}{Flight Model}
\newacronym{FSU}{FSU}{fibre switching unit}
\newacronym{FR}{FR}{laser frequency reference}
\newacronym{FS}{FS}{frequency separated}

\newacronym{GRACE-FO}{GRACE-FO}{Gravity Recovery and Climate Explorer Follow On}%
\newacronym{GPRM}{GPRM}{Grabbing Positioning Release Mechanism}
\newacronym{GR}{GR}{General Theory of Relativity}
\newacronym{GRS}{GRS}{Gravitational Reference Sensor}
\newacronym{GS}{GS}{ground station}
\newacronym{GW}{GW}{gravitational wave}

\newacronym{HR}{HR}{High Resolution}
                    
\newacronym{IFO}{IFO}{Interferometer}
\newacronym{IMS}{IMS}{interferometric measurement system}

\newacronym[firstplural={light-emitting diodes
  (LEDs)}]{LED}{LED}{light-emitting diode}
\newacronym{LIGO}{LIGO}{Laser Interferemeter Gravitational Wave Observatory}%
\newacronym{LISA}{LISA}{Laser Interferometer Space Antenna}%
\newacronym{LIST}{LIST}{LISA International Science Team}%
\newacronym{LPF}{LPF}{LISA Pathfinder}%
\newacronym{LPS}{LPS}{Laser Pre-stabilization System}
\newacronym{LTPDA}{LTPDA}{LPF Data Analysis package}
\newacronym{LH}{LH}{laser head}
\newacronym{LRI}{LRI}{Laser Ranging Instrument (on GRACE-FO)}
\newacronym{LS}{LS}{laser system}
\newacronym{LSO}{LSO}{last stable orbit}
\newacronym{LTP}{LTP}{LISA Pathfinder technology package}
\newacronym{LXE}{LXE}{long cross-talk experiment}

\newacronym{MCMC}{MCMC}{Markov-chain Monte Carlo}
\newacronym{MOC}{MOC}{Mission Operations Centre}
\newacronym{MSC}{Mother-S/C}{\enquote{Mother} spacecraft}

\newacronym{NASA}{NASA}{National Aeronautic and Space Administration}%
\newacronym{NPRO}{NPRO}{non-planar ring oscillator}
\newacronym{NR}{NR}{numerical relativity}

\newacronym[firstplural={Optical Benches
  (OBs)}]{OB}{OB}{Optical Bench}
\newacronym[firstplural={Optical Bench Assemblies
  (OBAs)}]{OBA}{OBA}{Optical Bench Assembly}
\newacronym{OBC}{OBC}{on-board computer}
\newacronym{OGSE}{OGSE}{Optical Ground Support Equipment}
\newacronym{OMS}{OMS}{Optical Metrology System}
\newacronym{ORO}{ORO}{optical read-out}

\newacronym{PA}{PA}{power amplifier}
\newacronym{PAA}{PAA}{point-ahead angle}
\newacronym{PCU}{PCU}{power conditioning unit}
\newacronym{PCDU}{PCDU}{power control and distribution unit}
\newacronym{PCP}{PCP}{Payload Commanding and Processing}
\newacronym{PDF}{PDF}{probability density function}
\newacronym{PDH}{PDH}{Pound Drever Hall}
\newacronym{PD}{PD}{photo diode}
\newacronym{PDS}{PDS}{photo detection system}
\newacronym{PLL}{PLL}{phase-locked loop}
\newacronym{P/L}{P/L}{payload}
\newacronym{P/M}{P/M}{propulsion module}
\newacronym{PM}{PM}{Phasemeter}
\newacronym{PMF}{PMF}{Polarization-Maintaining Fibre}
\newacronym{PMS}{PMS}{Phase Measurement Subsystem}
\newacronym{PRDS}{PRDS}{Phase Reference Distribution System}
\newacronym{PSD}{PSD}{power spectral density}

\newacronym{QPD}{QPD}{quadrant photodetector}

\newacronym{REF}{REF}{reference}
\newacronym{RF}{RF}{radio frequency}
\newacronym{RIN}{RIN}{relative intensity noise}
\newacronym{RM}{RM}{Radiation Monitor}
\newacronym{RX}{RX}{received signal}

\newacronym[firstplural={spacecraft
  (S/C)},plural=S/C]{S/C}{S/C}{spacecraft}
\newacronym[firstplural={subsystems
  (S/S)},plural=S/Ss]{S/S}{S/S}{subsystem}
\newacronym{S/C-P/M}{S/C-P/M}{spacecraft/propulsion-module}
\newacronym{SAU}{SAU}{sensing and actuation unit}
\newacronym{SEPD}{SEPD}{single-element photo diode}
\newacronym{SIM}{SIM}{Space Interferometry Mission}
\newacronym{SNR}{SNR}{Signal-to-Noise Ratio}
\newacronym{SOC}{SOC}{Science Operation Centre}
\newacronym{SXE}{SXE}{short cross-talk experiment}

\newacronym{TBC}{TBC}{to be confirmed}
\newacronym{TBD}{TBD}{to be determined}
\newacronym{TC/TM}{TC/TM}{telecommand/telemetry}
\newacronym{TDI}{TDI}{time-delay interferometry}
\newacronym[firstplural={test masses (TMs)}]{TM}{TM}{test mass}
\newacronym{TRP}{TRP}{temperature reference points}
\newacronym{TS}{TS}{Telescope}
\newacronym{TTC}{TT\&C}{telemetry, tracking, and command}
\newacronym{TTL}{TTL}{tilt to length}

\newacronym{USO}{USO}{ultra-stable oscillator}
\newacronym{UTC}{UTC}{Coordinated Universal Time}
\newacronym{UV}{UV}{ultra-violet}

\newacronym{WR}{WR}{Wide Range}

\makeglossaries

\usepackage{ulem}
\usepackage{todonotes}

\begin{document}
\title{Tilt-to-length coupling in LISA Pathfinder: a data analysis}

\def\addressb{Max Planck Institute for Gravitational Physics (Albert-Einstein-Institut), 
30167 Hannover, Germany}
\def\addressluh{Leibniz Universit\"at Hannover, 30167 Hannover, Germany}
\def\addressbp{University of Potsdam, Institute of Physics and Astronomy, 14476 Potsdam, Germany}
\def\addressbn{Department of Physics, Aristotle University of Thessaloniki, Thessaloniki 54124, Greece}
\def\addressa{European Space Astronomy Centre, European Space Agency, Villanueva de la
    Ca\~{n}ada, 28692 Madrid, Spain}
\def\addressc{APC, Universit\'e de Paris, CNRS, Astroparticule et Cosmologie, F-75006 
    Paris, France}
\def\addresscy{IRFU, CEA, Universit\'e Paris-Saclay, F-91191 Gif-sur-Yvette, France}
\def\addressd{High Energy Physics Group, Physics Department, Imperial College London, 
    Blackett Laboratory, Prince Consort Road, London, SW7 2BW, UK }
\def\addresse{Dipartimento di Fisica, Universit\`a di Roma ``Tor Vergata'',  and INFN, 
    sezione Roma Tor Vergata, I-00133 Roma, Italy}
\def\addressf{Department of Industrial Engineering, University of Trento, via Sommarive 
    9, 38123 Trento, 
    and Trento Institute for Fundamental Physics and Application / INFN}
\def\addressh{European Space Technology Centre, European Space Agency, 
    Keplerlaan 1, 2200 AG Noordwijk, The Netherlands}
\def\addressi{Dipartimento di Fisica, Universit\`a di Trento and Trento Institute for 
    Fundamental Physics and Application / INFN, 38123 Povo, Trento, Italy}
\def\addressk{Istituto di Fotonica e Nanotecnologie, CNR-Fondazione Bruno Kessler, 
    I-38123 Povo, Trento, Italy}
\def\addressj{The School of Physics and Astronomy, University of
    Birmingham, Birmingham, UK}
\def\addressl{Institut f\"ur Geophysik, ETH Z\"urich, Sonneggstrasse 5, CH-8092, 
    Z\"urich, Switzerland}
\def\addressm{The UK Astronomy Technology Centre, Royal Observatory, Edinburgh, Blackford 
    Hill, Edinburgh, EH9 3HJ, UK}
\def\addressn{Institut de Ci\`encies de l'Espai (ICE, CSIC), Campus UAB, Carrer de Can 
    Magrans s/n, 08193 Cerdanyola del Vall\`es, Spain}
\def\addresso{DISPEA, Universit\`a di Urbino Carlo Bo, Via S. Chiara, 27 61029 
    Urbino/INFN, Italy}
\def\addressp{European Space Operations Centre, European Space Agency, 64293 Darmstadt, 
    Germany}
\def\addressq{Physik Institut, 
    Universit\"at Z\"urich, Winterthurerstrasse 190, CH-8057 Z\"urich, Switzerland}
\def\addressr{SUPA, Institute for Gravitational Research, School of Physics and 
    Astronomy, University of Glasgow, Glasgow, G12 8QQ, UK}
\def\addresss{Department d'Enginyeria Electr\`onica, Universitat Polit\`ecnica de 
    Catalunya,  08034 Barcelona, Spain}
\def\addresst{Institut d'Estudis Espacials de Catalunya (IEEC), C/ Gran Capit\`a 2-4, 
    08034 Barcelona, Spain}
\def\addressu{Gravitational Astrophysics Lab, NASA Goddard Space Flight Center, 8800 
    Greenbelt Road, Greenbelt, MD 20771 USA}
\def\addressbb{Department of Mechanical and Aerospace Engineering, MAE-A, P.O. Box 
    116250, University of Florida, Gainesville, Florida 32611, USA}
\def\addressbbb{Department of Physics,
    2001 Museum Road, University of Florida, Gainesville, Florida 32611, USA}
\def\addressbh{Institut für Theoretische Physik, Universität Heidelberg, Philosophenweg 16, 69120 Heidelberg, Germany}
\def\addresscc{Istituto di Fotonica e Nanotecnologie, CNR-Fondazione Bruno Kessler, 
    I-38123 Povo, Trento, Italy}
\def\addressdd{isardSAT SL, Marie Curie 8-14, 08042 Barcelona, Catalonia, Spain}
\def\addressee{Escuela Superior de Ingenier\'ia, Universidad de C\'adiz, 11519 C\'adiz, 
    Spain}
\def\addressnn{Observatoire de la C\^{o}te d'Azur, Boulevard de l'Observatoire CS 34229 - 
    F 06304 NICE, France}
\def\addressff{City University of Applied Sciences, Flughafenallee 10, 28199 Bremen, 
    Germany}
\def\addressgg{Texas A\&M University, 701 H.R. Bright Bldg,
    College Station, TX 77843-3141 USA}
\def\addresshh{OHB System AG, Universit\"atsallee 27-29, 28359 Bremen, Germany}
\def\addressii{Airbus Defence and Space, Claude-Dornier-Strasse, 88090 Immenstaad, 
    Germany}
\def\addressrr{Department of Quantitative Methods, Universidad Loyola Andalucia, Avenida 
de las Universidades s/n, 41704, Dos Hermanas, Sevilla, Spain}

\author{M~Armano}\affiliation{\addressh}
\author{H~Audley}\affiliation{\addressb}\affiliation{\addressluh}
\author{J~Baird}\affiliation{\addressc}
\author{P~Binetruy}\thanks{Deceased 30 March 2017}\affiliation{\addresscy}%
\author{M~Born}\affiliation{\addressb}\affiliation{\addressluh}
\author{D~Bortoluzzi}\affiliation{\addressf}
\author{E~Castelli}\affiliation{\addressi}
\author{A~Cavalleri}\affiliation{\addresscc}
\author{A~Cesarini}\affiliation{\addresso}
\author{A\,M~Cruise}\affiliation{\addressj}
\author{K~Danzmann}\affiliation{\addressb}\affiliation{\addressluh}
\author{M~de Deus Silva}\affiliation{\addressa}
\author{I~Diepholz}\affiliation{\addressb}\affiliation{\addressluh}
\author{G~Dixon}\affiliation{\addressj}
\author{R~Dolesi}\affiliation{\addressi}
\author{L~Ferraioli}\affiliation{\addressl}
\author{V~Ferroni}\affiliation{\addressi}
\author{E\,D~Fitzsimons}\affiliation{\addressm}
\author{M~Freschi}\affiliation{\addressa}
\author{L~Gesa}\thanks{Deceased 29 May 2020}\affiliation{\addressn}\affiliation{\addresst}
\author{D~Giardini}\affiliation{\addressl}
\author{F~Gibert}\affiliation{\addressi}%
\author{R~Giusteri}\affiliation{\addressb}\affiliation{\addressluh}
\author{C~Grimani}\affiliation{\addresso}
\author{J~Grzymisch}\affiliation{\addressh}
\author{I~Harrison}\affiliation{\addressp}
\author{M-S~Hartig}\email{marie-sophie.hartig@aei.mpg.de}\affiliation{\addressb}\affiliation{\addressluh}
\author{G~Heinzel}\affiliation{\addressb}\affiliation{\addressluh}
\author{M~Hewitson}\affiliation{\addressb}\affiliation{\addressluh}
\author{D~Hollington}\affiliation{\addressd}
\author{D~Hoyland}\affiliation{\addressj}
\author{M~Hueller}\affiliation{\addressi}
\author{H~Inchausp\'e}\affiliation{\addressbh}%
\author{O~Jennrich}\affiliation{\addressh}
\author{P~Jetzer}\affiliation{\addressq}
\author{U~Johann}\affiliation{\addressii}
\author{B~Johlander}\affiliation{\addressh}
\author{N~Karnesis}\email{karnesis@auth.gr}\affiliation{\addressbn}
\author{B~Kaune}\affiliation{\addressb}\affiliation{\addressluh}
\author{C\,J~Killow}\affiliation{\addressr}
\author{N~Korsakova}\affiliation{\addressc}%
\author{J\,A~Lobo}\thanks{Deceased 30 September 2012}\affiliation{\addressn}\affiliation{\addresst}
\author{J\,P~L\'opez-Zaragoza}\affiliation{\addressn}
\author{R~Maarschalkerweerd}\affiliation{\addressp}
\author{D~Mance}\affiliation{\addressl}
\author{V~Mart\'{i}n}\affiliation{\addressn}\affiliation{\addresst}
\author{L~Martin-Polo}\affiliation{\addressa}
\author{F~Martin-Porqueras}\affiliation{\addressa}
\author{J~Martino}\affiliation{\addressc}
\author{P\,W~McNamara}\affiliation{\addressh}
\author{J~Mendes}\affiliation{\addressp}
\author{L~Mendes}\affiliation{\addressa}
\author{N~Meshksar}\affiliation{\addressl}
\author{M~Nofrarias}\affiliation{\addressn}\affiliation{\addresst}
\author{S~Paczkowski}\affiliation{\addressb}\affiliation{\addressluh}
\author{M~Perreur-Lloyd}\affiliation{\addressr}
\author{A~Petiteau}\affiliation{\addressc}\affiliation{\addresscy}
\author{E~Plagnol}\affiliation{\addressc}
\author{J~Ramos-Castro}\affiliation{\addresss}%
\author{J~Reiche}\affiliation{\addressb}\affiliation{\addressluh}
\author{F~Rivas}\affiliation{\addressrr}
\author{D\,I~Robertson}\affiliation{\addressr}
\author{G~Russano}\affiliation{\addressi}
\author{J~Sanjuan}\affiliation{\addressbbb}
\author{J~Slutsky}\affiliation{\addressu}
\author{C\,F~Sopuerta}\affiliation{\addressn}\affiliation{\addresst}
\author{T~Sumner}\affiliation{\addressd}\affiliation{\addressbbb}
\author{L~Tevlin}\affiliation{\addressb}\affiliation{\addressluh}\affiliation{\addressbp}
\author{D~Texier}\affiliation{\addressa}
\author{J\,I~Thorpe}\affiliation{\addressu}
\author{D~Vetrugno}\affiliation{\addressi}
\author{S~Vitale}\affiliation{\addressi}
\author{G~Wanner}\email{gudrun.wanner@aei.mpg.de}\affiliation{\addressluh}\affiliation{\addressb}
\author{H~Ward}\affiliation{\addressr}
\author{P\,J~Wass}\affiliation{\addressd}\affiliation{\addressbb}
\author{W\,J~Weber}\affiliation{\addressi}
\author{L~Wissel}\affiliation{\addressb}\affiliation{\addressluh}
\author{A~Wittchen}\affiliation{\addressb}\affiliation{\addressluh}
\author{P~Zweifel}\affiliation{\addressl}

\date{\today}

\begin{abstract}
We present a study of the tilt-to-length coupling noise during the LISA Pathfinder mission and how it depended on the system's alignment. 
Tilt-to-length coupling noise is the unwanted coupling of angular and lateral spacecraft or test mass motion into the primary interferometric displacement readout. 
It was one of the major noise sources in the LISA Pathfinder mission and is likewise expected to be a primary noise source in LISA. 
We demonstrate here that a recently derived and published analytical model describes the dependency of the LISA Pathfinder tilt-to-length coupling noise on the alignment of the two freely falling test masses. 
This was verified with the data taken before and after the realignments performed in March (engineering days) and June 2016, and during a two-day experiment in February 2017 (long cross-talk experiment). The latter was performed with the explicit goal of testing the tilt-to-length coupling noise dependency on the test mass alignment.
Using the analytical model, we show that all realignments performed during the mission were only partially successful and explain the reasons why.
In addition to the analytical model, we computed another physical tilt-to-length coupling model via a minimising routine making use of the long cross-talk experiment data. A similar approach could prove useful for the LISA mission.
\end{abstract}
\maketitle

\section{Introduction}
\label{sec:Intro}

In December 2015, \gls{LPF} \cite{McNamara2008,Armano2018,Armano2016} was launched as technology demonstrator mission for the \gls{LISA} \cite{Danzmann2011,elisa13ARXIV,LISAMission}, the first space-based gravitational-wave detector.
\gls{LPF} not only successfully proved the technology planned to be implemented in \gls{LISA} but also exceeded its requirements, already reaching the \gls{LISA} performance requirements in the entire frequency band \cite{Armano2018,Armano2016}.

\begin{figure}
\centering
  \includegraphics[width=0.9\columnwidth]{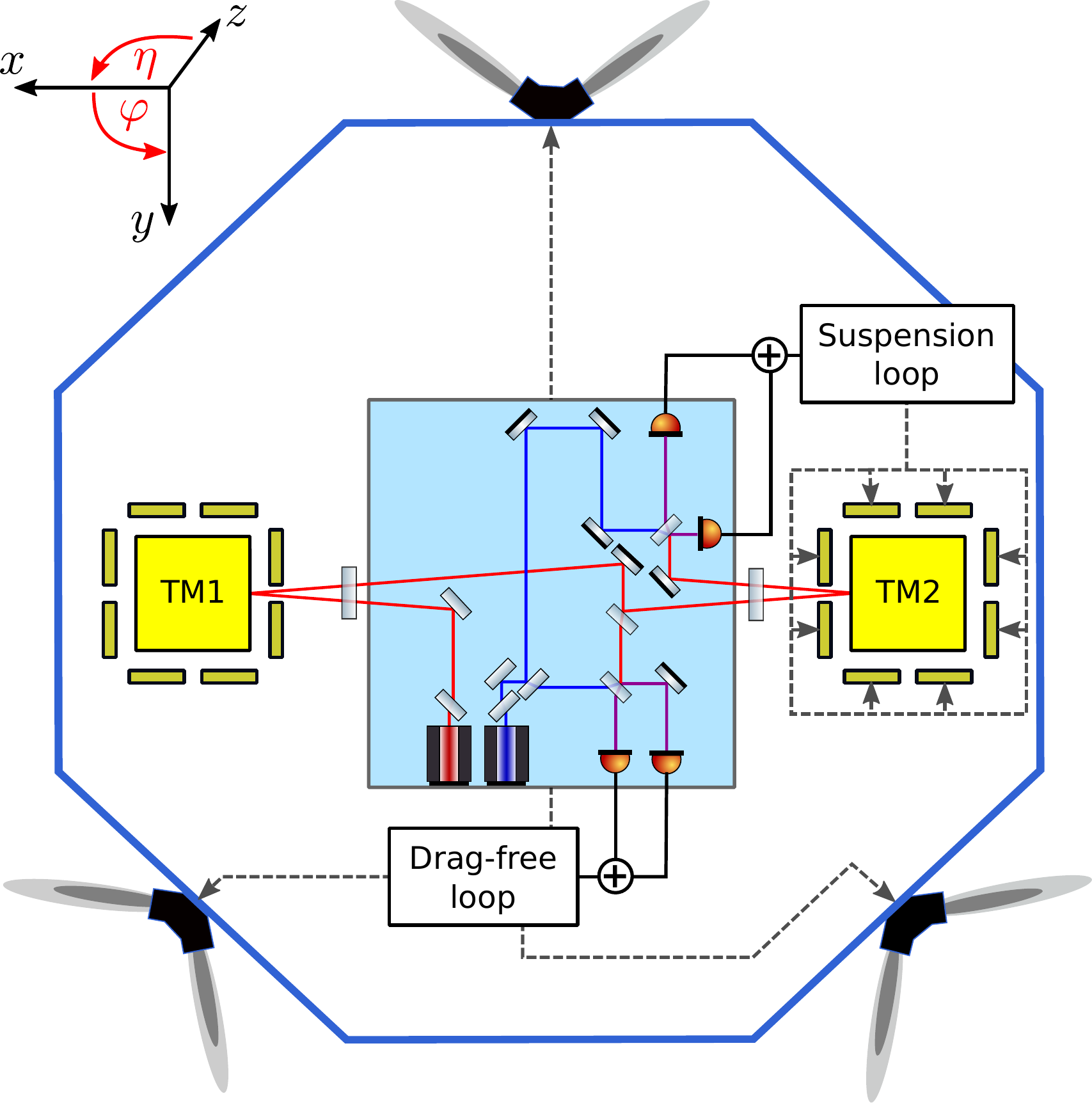}
  \caption{Simplified sketch of the LPF setup. The yellow squares represent the two TMs which are surrounded by the electrodes from the electrode housing. 
  The light-blue square in the centre is the optical bench. The optical setup is reduced to the components relevant for the TM distance changes and control loop measurements. The same holds for the shown beam paths.
  Further, the principle of the nominal control mechanisms is shown.
  The drag-free loop utilised the measurement of the S/C relative to TM1 to define S/C accelerations via the thrusters.
  The suspension loop is fed in by the readout of the S/C-TM2 alignment used to control the TM2 alignment via electrostatic forces.
  Figure reproduced from \cite{Armano2019_stability}.}
\label{fig:LPFsketch}
\end{figure}

A residual noise originated from the lateral and angular jitter of the \gls{S/C} (and subsequently the optical bench) with respect to its two hosted \glspl{TM} in free fall. 
The jitter affected the path of the beam being reflected at the \glspl{TM} (red beam in Fig.~\ref{fig:LPFsketch}) and induced alterations of the measured length signal.
This coupling noise is called \gls{TTL} coupling \cite{G21,NG21}.
It limited the mission performance between 20 and 200\,mHz before its subtraction in post-processing \cite{Wanner2017}.
\gls{TTL} coupling will also be a major noise source in \gls{LISA} \cite{Paczkowski2022,George2022}. 

For the suppression of the \gls{TTL} coupling in \gls{LPF}, realignments of the two \glspl{TM} were conducted three times during the mission (in March and twice in June 2016) \cite{Wanner2017,dlr2020}. 
These realignments significantly reduced the overall \gls{TTL} coupling. 
However, the observed TTL noise was not fully understood then, and the existent model described the data insufficiently.
Consequently, the realignments did not fully mitigate the noise, making a further subtraction strategy necessary.

In \citep{LPFana22}, a new analytical model for the \gls{TTL} coupling in \gls{LPF} was presented.
Within this work, we show that this model, for the first time, successfully explains the dependency of the \gls{TTL} noise in \gls{LPF} on the \gls{TM} alignments.
With this model, we were able to reproduce the \gls{TTL} noise changes due to the \gls{TM} realignments for \gls{TTL} suppression, e.g., during the \gls{ED} in March 2016.
Furthermore, we derived how the \glspl{TM} should have been aligned during the mission for the \gls{TTL} noise mitigation and compare these with the angles that were commanded instead.

We start with a brief summary of the functionalities of \gls{LPF} relevant for our analysis in Sec.~\ref{sec:Basics_Instrument} and a characterisation of the \gls{TTL} coupling noise in \gls{LPF} in Sec.~\ref{sec:Basics_TTLinLPF}.
In Sec.~\ref{sec:TTLmodel}, we then introduce the two \gls{TTL} models, which are the core of our analysis: the fit model that was used for \gls{TTL} subtraction during the mission and the recently published analytical \gls{TTL} model.
Both models are applied to mission data of experiments, that we explain in Sec.~\ref{sec:Basics_Realignments}.
The results for the data of the \gls{LXE}, that was designed to test \gls{TTL} coupling, are presented in Sec.~\ref{sec:LXE}, and for the data around the \gls{TM} realignments carried out for \gls{TTL} suppression in Sec.~\ref{sec:TMalignments}.
Finally, we summarise our findings in Sec.~\ref{sec:summary}.

\section{The LISA Pathfinder Instrument}
\label{sec:Basics_Instrument}

\gls{LPF} hosted two \glspl{TM} in free fall, which were each placed inside an electrode housing.
In the following, we briefly introduce the four interferometers in \gls{LPF} and explain how their measurements were used to control the stability of the optical setup.

\subsection{The interferometers}
\label{sec:Basics_Instrument_ifos}

\gls{LPF} had a total of four interferometers, measuring the heterodyne interference signal of two local Gaussian beams.
Out of these interferometers, two were maintaining the reliability of the readouts of the other two:
the reference and the frequency interferometer.
The signal of the \textit{frequency interferometer} was used to stabilise the laser frequency, thus minimising the noise due to frequency fluctuations.
The \textit{reference interferometer} provided a reference phase signal, which was subtracted from the readouts of the other interferometers.  

The \textit{x1-interferometer} measured the interference signal of the local reference beam and a beam that was reflected at the first \gls{TM}. From this, the longitudinal displacement (i.e.\ along the $x$-axis in Fig.~\ref{fig:LPFsketch}) and angular alignment of the \gls{S/C} with respect to the first \gls{TM} were obtained.
Further, the \textit{x12-interferometer} detected the interference signal of the local reference beam and a beam that was reflected at both \glspl{TM}. The angular alignment changes of the second \gls{TM}, as well as the distance changes between both \glspl{TM}, were obtained from this signal.
In the two latter interferometers, the angular alignments were measured utilising \gls{DWS} \cite{Wanner2012,Morrison1994}.
In general, the measurements of the x1- and x12-interferometers were used to control the stability of the whole system (Sec.~\ref{sec:Basics_Instrument_dfacs}) and contained the main science signal: the distance variations between the two \glspl{TM}.

More detailed information about the interferometers in \gls{LPF} is given in \cite{Heinzel2003,Heinzel2004,Armano2015}.

\subsection{The control mechanisms}
\label{sec:Basics_Instrument_dfacs}

The stability of the whole system was maintained by the \gls{DFACS} \cite{Schleicher2018,Armano2019_stability}.
Its control loops used the readouts of the x1- and the x12-interferometers as well as the displacement readout by the \gls{GRS} \cite{Dolesi2003,Armano2017,Armano2020_actua-electr} as input.
The latter determined the relative positions of the \glspl{TM} with respect to their housings via electrostatic measurements. Electrodes inside the electrode housings of the \glspl{TM} were used partly for these measurements and partly to apply forces for alignment correction. 

In nominal operation, the \gls{DFACS} ensured that the \gls{S/C} followed the first \gls{TM} (\textit{drag-free control loop}), which was not controlled along the sensitive $x$-axis. 
Then, the second \gls{TM} followed the \gls{S/C} (\textit{suspension control loop}) to sustain the alignment of the full optical system.
For the \gls{S/C} and \gls{TM} alignment corrections, either actuations via the \gls{S/C} thrusters or the electrodes inside the housings were applied, see Fig.~\ref{fig:LPFsketch}.
To prevent the electrostatic forces and torques from perturbing the science signal, they were applied at frequencies below the \gls{LPF} measurement band. 
Additionally, they were subtracted in post-processing.

To preserve the angular stability of the system, its angular alignment was continuously tracked (mostly via the \gls{DWS} measurements of the x1- and x12-interferometer) and kept at predefined nominal offsets by the \gls{DFACS}.
These nominal alignments were redefined three times during the \gls{LPF} mission for \gls{TTL} suppression, see Sec.~\ref{sec:Basics_Realignments}.

\section{TTL coupling in LISA Pathfinder}
\label{sec:Basics_TTLinLPF}
The main science signal in \gls{LPF} was the residual acceleration between the two \glspl{TM} along the $x$-axis. 
Angular and lateral jitter of both the \gls{S/C} and \glspl{TM} coupled into that readout and added \gls{TTL} noise.
The \glspl{TM} were very quiet
since they were nominally in free fall and the forces acting on them were kept small.
Hence, most of the jitter leading to the \gls{TTL} coupling noise in \gls{LPF} originated from the \gls{S/C} itself. 
However, for an easier modelling of this coupling, we can equally interpret the \gls{S/C} jitter as a simultaneous jitter of both \glspl{TM} with respect to the optical bench in Fig.~\ref{fig:LPFsketch} around the \gls{S/C}'s centre of rotation.

We consider two types of jitter: \textit{angular} and \textit{lateral} jitter.
The angular jitter of the \gls{S/C} changed the orientation of the beam that was reflected at the first \gls{TM}. Consequently, the beam's path length and detection point at the photodiodes changed. Both yielded \gls{TTL} coupling noise on the final signal \cite{G21,NG21}.
Lateral jitter of the \gls{S/C} added \gls{TTL} noise if the \gls{TM} surfaces were tilted with respect to the optical bench. In this case, the \gls{TM} surfaces shifted into and out of the beam path due to the lateral jitter \cite{G21}.

In general, angular realignments of the \glspl{TM} affect the beam's orientation after reflection and the degree of lateral jitter coupling.
Therefore, \gls{TTL} coupling noise can be reduced by optimal rotations of the \glspl{TM}.
Such a realignment also was part of the \gls{TTL} suppression strategy of the \gls{LPF} mission. In total, the \glspl{TM} were realigned three times. This is discussed in Secs.~\ref{sec:Basics_Realignments} and~\ref{sec:TMalignments}.
We present a corresponding analytical model in the following section.

\section{TTL coupling models}
\label{sec:TTLmodel}

Modelling \gls{TTL} coupling is necessary for the successful suppression of this noise.
First, simple \gls{TTL} models are used for the subtraction of the coupling noise in the measurement band. 
In addition, more complex models that describe the dependency of the coupling on the setup parameters can be used for a suppression via design or realignment.

In this work, we investigate the \gls{TTL} coupling in \gls{LPF} with two models: a fit model that was used during the \gls{LPF} mission for \gls{TTL} subtraction (Sec.~\ref{sec:TTLmodel_fit}) and the analytical model derived in \cite{LPFana22}, which additionally describes how the \gls{TTL} coupling depends on the \gls{TM} alignments (Sec.~\ref{sec:TTLmodel_ana}).
We show in Sec.~\ref{sec:LXE_minimiser} that a third model can be derived from the mission data. This model describes the dependency of the \gls{TTL} coupling on the alignment like the analytical model but was computed using only the fit results of different alignment configurations.

\subsection{The linear fit model}
\label{sec:TTLmodel_fit}

The \gls{TM} realignments performed at the beginning of the \gls{LPF} mission did not fully mitigate the \gls{TTL} coupling, see Sec.~\ref{sec:realingments_description}.
Therefore, a linear \gls{TTL} model was fitted and then subtracted from the data.
In detail, the dependency of the measured acceleration $\Delta g$ on the mean \gls{TM} accelerations in the orthogonal degrees of freedom ($\ddot{\bar{\varphi}}$,\,$\ddot{\bar{\eta}}$,\,$\ddot{\bar{y}}$,\,$\ddot{\bar{z}}$) and on the displacement of the \glspl{TM} via stiffnesses ($\bar{y}$,\,$\bar{z}$) was fitted (see Fig.~\ref{fig:LPFsketch} for the coordinate system definition).
These relative accelerations and displacements describe, in a good approximation, the inverse \gls{S/C} jitter.
Additionally, the coupling of \gls{S/C} jitter along the sensitive axis between both \glspl{TM} ($\ddot{o}_1$) was considered in the fit. These movements did not alter the distance between both \glspl{TM} but nonetheless coupled into the signal due to imperfections in the symmetry of the setup. 
These contributors added up to the following linear fit model \cite{Armano2016,Armano2018}:
\begin{align}
\begin{split}
\Delta g^\text{fit}_\text{xtalk} &=
C_\varphi^\text{fit}\,\ddot{\overline{\varphi}} 
+ C_\eta^\text{fit}\,\ddot{\overline{\eta}} 
+ C_y^\text{fit}\,\ddot{\overline{y}} 
+ C_z^\text{fit}\,\ddot{\overline{z}} \\
&+ C_{y,s}^\text{fit}\,\overline{y} + C_{z,s}^\text{fit}\, \overline{z} + C_{o_1}^\text{fit}\,\ddot{o}_1 \,.
\end{split}
\label{eq:fitmodel}
\end{align}

We make use of this fit model in our \gls{TTL} data analysis presented in this paper. 
Analogously to the data analysis and the \gls{TTL} coupling subtraction during the \gls{LPF} mission, we performed our computations in MATLAB using the \gls{LTPDA} toolbox \cite{LTPDA}.
For the fit of the model Eq.~\eqref{eq:fitmodel} to the $\Delta g$ measurements, we applied a \gls{LTPDA} function,
which performs a least-square fit in the frequency domain.
We chose to consider all frequencies between 2\,mHz and 70\,mHz.
At the upper frequency bound, noise originating from the thrusters coupled into the readout \cite{Armano2019_thrusters}, which affected the cross-coupling results. 
The lower frequency limit was set to 2\,mHz %
to cover the differential frequencies in the \gls{LXE} (Sec.~\ref{sec:LXE}). 
Further, the fit procedure used the 4-term Blackman-Harris window (BH92).
The algorithm stopped when the previous and the current relative residuals differed by less than 10$^{-10}$ (tolerance criterion).

The lateral accelerations and displacements ($\ddot{\bar{y}}$,\,$\ddot{\bar{z}}$,\,$\bar{y}$,\,$\bar{z}$) considered in the fit were measured electrostatically by the \gls{GRS} within the \glspl{TM} housings \cite{Armano2017}. 
These readouts were processed via the \gls{DFACS} \cite{Schleicher2018}. 
Therefore, we also used the \gls{DFACS} data for the angular (\gls{DWS}) readouts and the $o_1$ and $\Delta g$ measurements in our data analysis. 
Note that the \gls{DWS} readouts were also available as optical metrology system \cite{Armano2021_OMS,Armano2022_OMS} data.
However, there were small shifts between the time sets that would have to be corrected.

Due to cross-sensing in general and the similar pattern of the lateral accelerations ($\ddot{\bar{y}},\, \ddot{\bar{z}}$) and displacements ($\bar{y},\, \bar{z}$) in particular, the jitter readouts were partially correlated.
The coefficient errors provided by the fit algorithm are caused by these correlations and the measurement noise.

\subsection{The linear analytical model}
\label{sec:TTLmodel_ana}

In \gls{LPF}, \gls{TTL} coupling originated from the lateral and angular jitter of its two hosted \glspl{TM} or the hosting \gls{S/C} itself, which can be interpreted as a simultaneous jitter of the \glspl{TM} relative to the \gls{S/C}.
This jitter coupling can be described via the lever arm and the piston effect. 
Additionally, the windows between the optical bench and the vacuum housing of the \glspl{TM} added a small amount of \gls{TTL} coupling.
All these coupling effects can be computed analytically, as shown in \cite{G21,NG21}.

A full analytical \gls{TTL} coupling model was derived for the \gls{LPF} setup in \cite{LPFana22} considering these effects and further wavefront and detector geometry-dependent coupling terms.
The parameters used to describe the \gls{LPF} optical setup relied on the measurement of the \gls{LPF} in-flight model \cite{TNoptocad}.  
The result was then verified against the numerical computations by the \cpp\ library IfoCAD \cite{Wanner2012,IfoCAD,Kochkina2013}.
We transformed the model to the form of the fit model (Eq.~\eqref{eq:fitmodel}) -- with the exception of the displacements terms $\bar{y}$ and $\bar{z}$, and the longitudinal term $\ddot{o}_1$, which were not modelled analytically -- to make both models comparable. 
This yielded
\begin{align}
\Delta g_\text{xacc}^\text{ana} = 
C_\varphi^\text{ana}\,\ddot{\overline{\varphi}} 
+ C_\eta^\text{ana}\,\ddot{\overline{\eta}} 
+ C_y^\text{ana}\,\ddot{\overline{y}} 
+ C_z^\text{ana}\,\ddot{\overline{z}} \,
\label{eq:anamodel}
\end{align}
with 
\begin{subequations}
\begin{align}
C_\varphi^\text{ana} =&\ C_{\varphi,0} 
+0.210^{+0.017}_{-0.016}\,\frac{\mathrm{m}}{\mathrm{rad}^2} \,\varphi_{1}
+0.182^{+0.018}_{-0.020}\,\frac{\mathrm{m}}{\mathrm{rad}^2} \,\varphi_{2} 
\label{eq:anamodel_Cphi}\\
\begin{split}
C_\eta^\text{ana} =&\ C_{\eta,0} 
+0.209^{+0.017}_{-0.015}\,\frac{\mathrm{m}}{\mathrm{rad}^2} \,\eta_{1}
 +0.177^{+0.018}_{-0.019}\,\frac{\mathrm{m}}{\mathrm{rad}^2}\,\eta_{2}  \\
&-0.005^{+0}_{-0}\,\frac{\mathrm{m}}{\mathrm{rad}^2} \,\varphi_{1}
 +0.005^{+0}_{-0}\,\frac{\mathrm{m}}{\mathrm{rad}^2}\,\varphi_{2} 
\end{split}
 \label{eq:anamodel_Ceta}\\
C_y^\text{ana} =&\ C_{y,0}
+\, 1.000^{+0}_{-0} \ \frac{1}{\text{rad}}\,(-\varphi_{1}+\varphi_{2}) 
\label{eq:anamodel_Cy}\\
C_z^\text{ana} =&\ C_{z,0}
+\, 1.000^{+0}_{-0} \ \frac{1}{\text{rad}}\,(\eta_{1}-\eta_{2}) \,.
\label{eq:anamodel_Cz}
\end{align}
\label{eq:anamodel_C}
\end{subequations}

The model shown here provides the dependency of the coefficients on the \gls{TM} alignments.
These alignments are not the \gls{TM} tilts relative to the optical bench ($\hat{\varphi}_i,\ \hat{\eta}_i$, but are defined relative to a nominal offset ($\varphi_i = \hat{\varphi}_i - \varphi_{0i},\ \eta_i = \hat{\eta}_i - \eta_{0i}$). 
As nominal offsets, we understand here the unknown mean alignment of the \glspl{TM} relative to the optical bench at the beginning of the investigated timespan.

The errors shown in Eq.~\eqref{eq:anamodel_C} for the coefficient estimates were derived by implementing measurement uncertainties of the beam parameters \cite{Killow2016} and the location of the \gls{S/C}'s centre of mass \cite{TNdfacs} into the analytical model. Furthermore, we considered a drift of the \gls{S/C}'s centre of mass due to non-balanced fuel consumption. %
These errors were derived along with the coefficients in \cite{LPFana22}.
Note that the errors given for the lateral coupling coefficients are not precisely zero but seven orders of magnitude smaller than the coefficients themselves.

The constant offsets $C_{j,0}$, $j\in\{\varphi,\eta,y,z\}$, of the coefficients depend on various setup parameters, which are not all precisely known.
Additionally, the unknown nominal rotations of the \glspl{TM} add into these coefficients.
In the following analysis, we therefore substitute these constants by their fitted correspondences. 
These are either the fitted coefficients during a noise run prior to realignments or the mean of the coefficients computed for different timespans with non-altered \gls{TTL} coupling within an experiment, e.g.\ the \gls{LXE}.

\begin{figure*}
\centering
\includegraphics[scale=0.22]{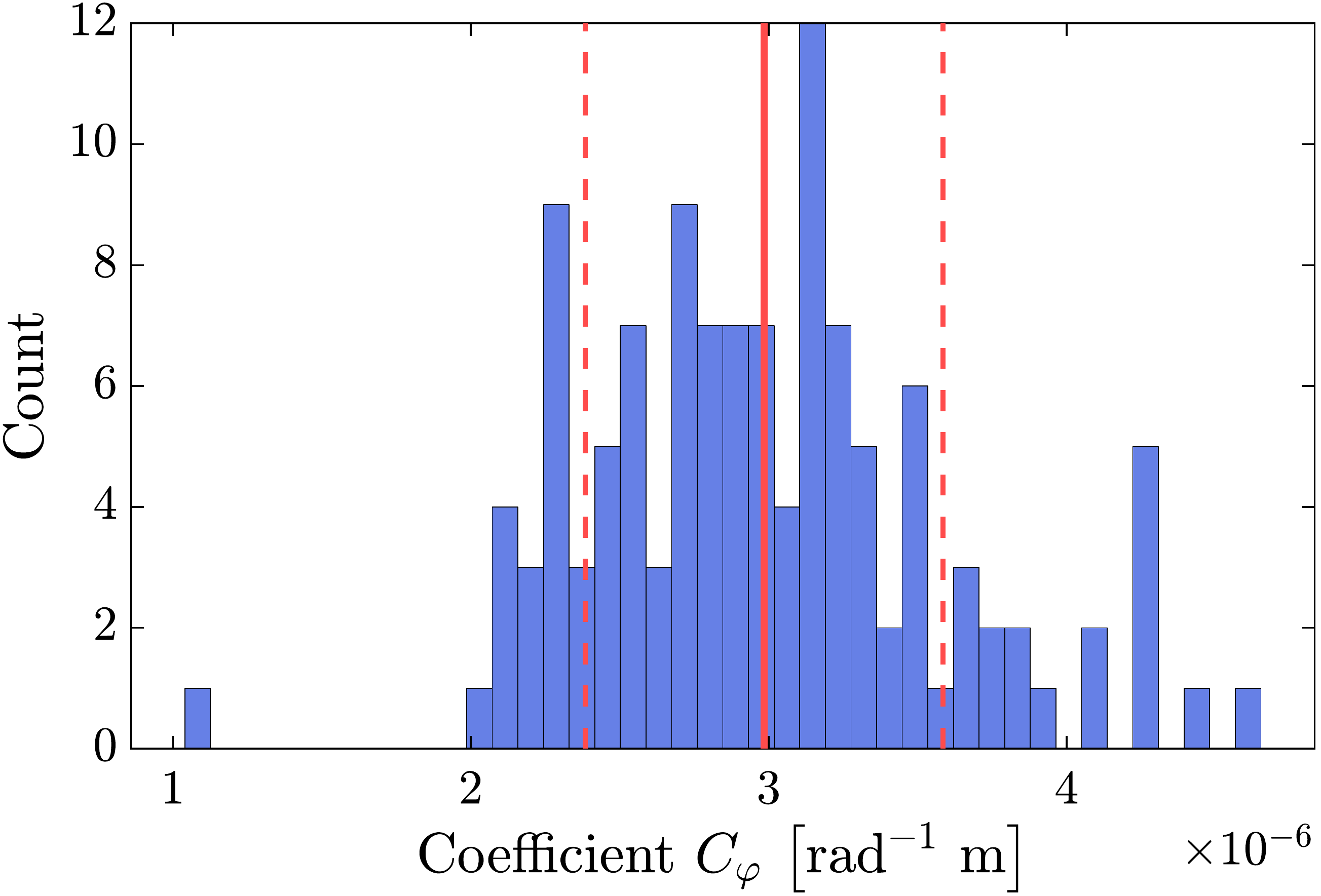}\quad\ \
\includegraphics[scale=0.22]{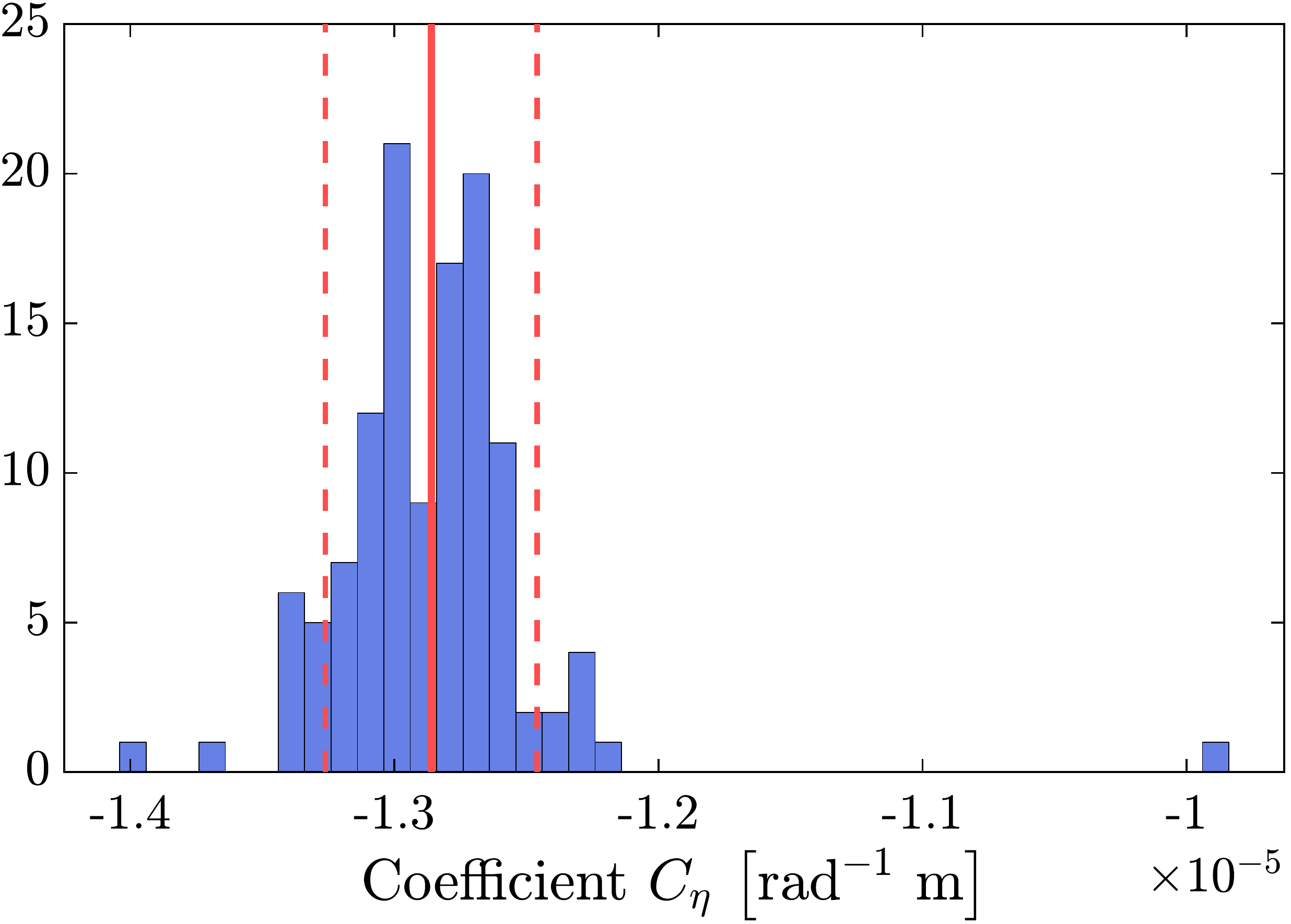}\hfill
\includegraphics[scale=0.22]{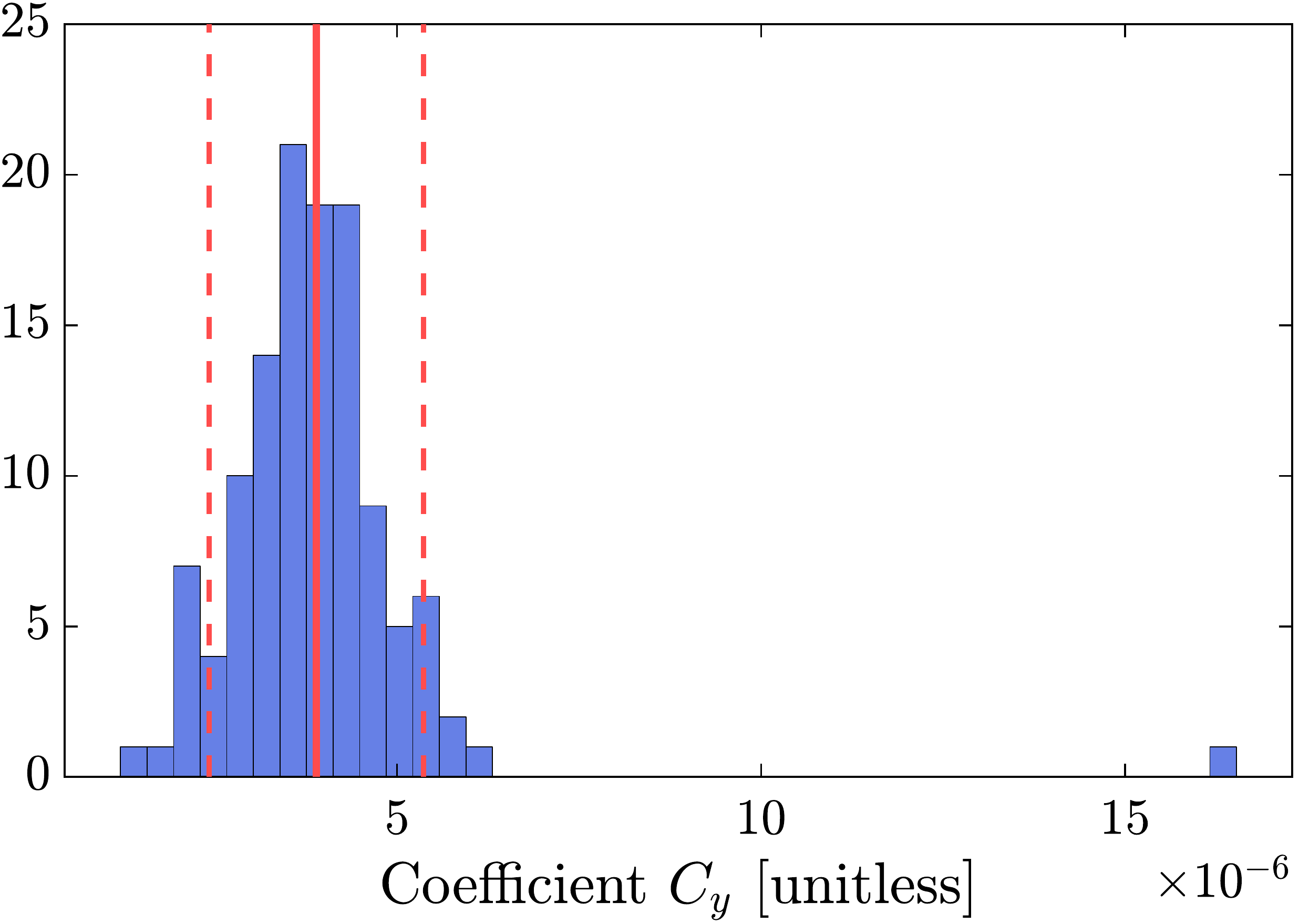} \\[1ex]
\includegraphics[scale=0.22]{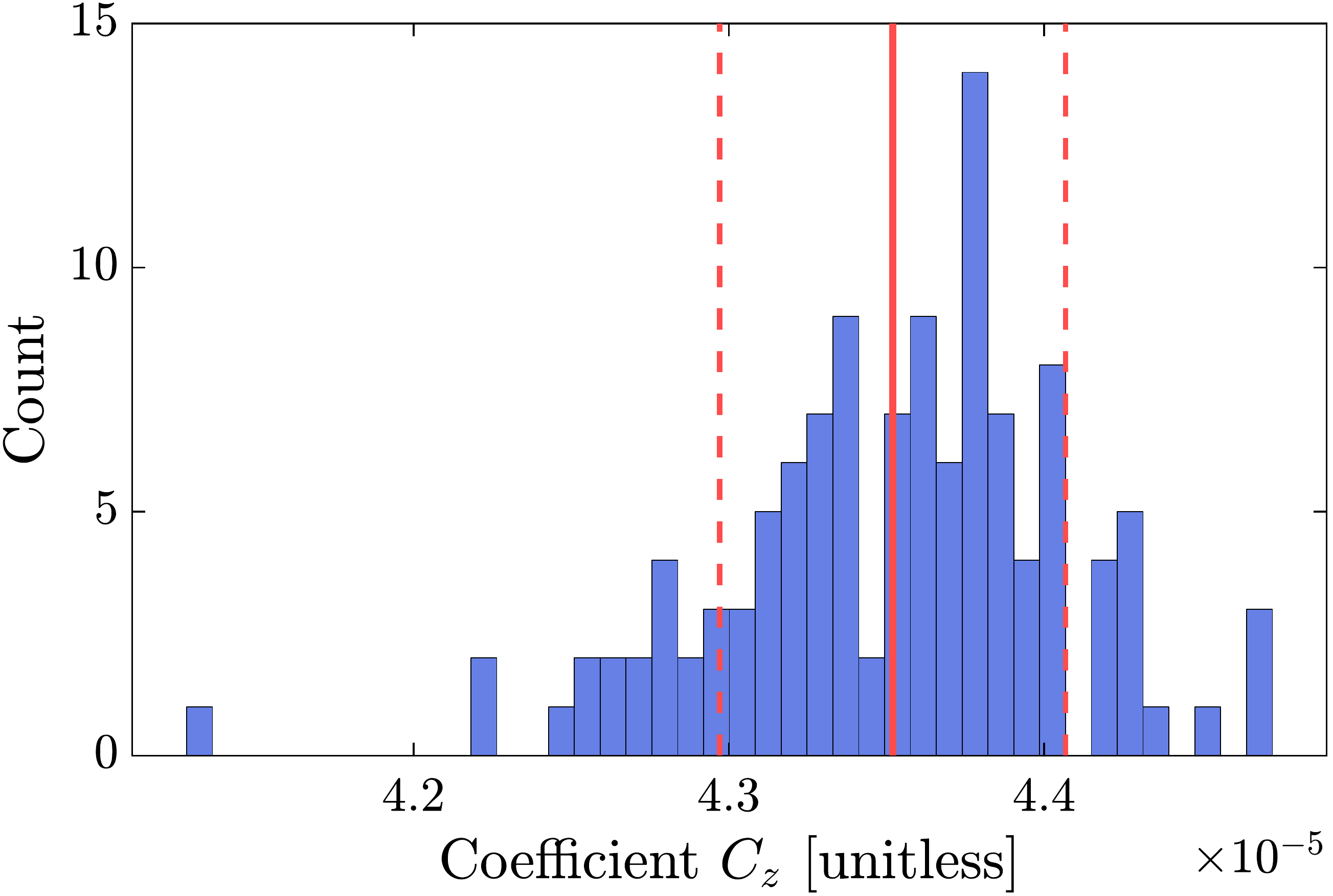}\quad\ \
\includegraphics[scale=0.22]{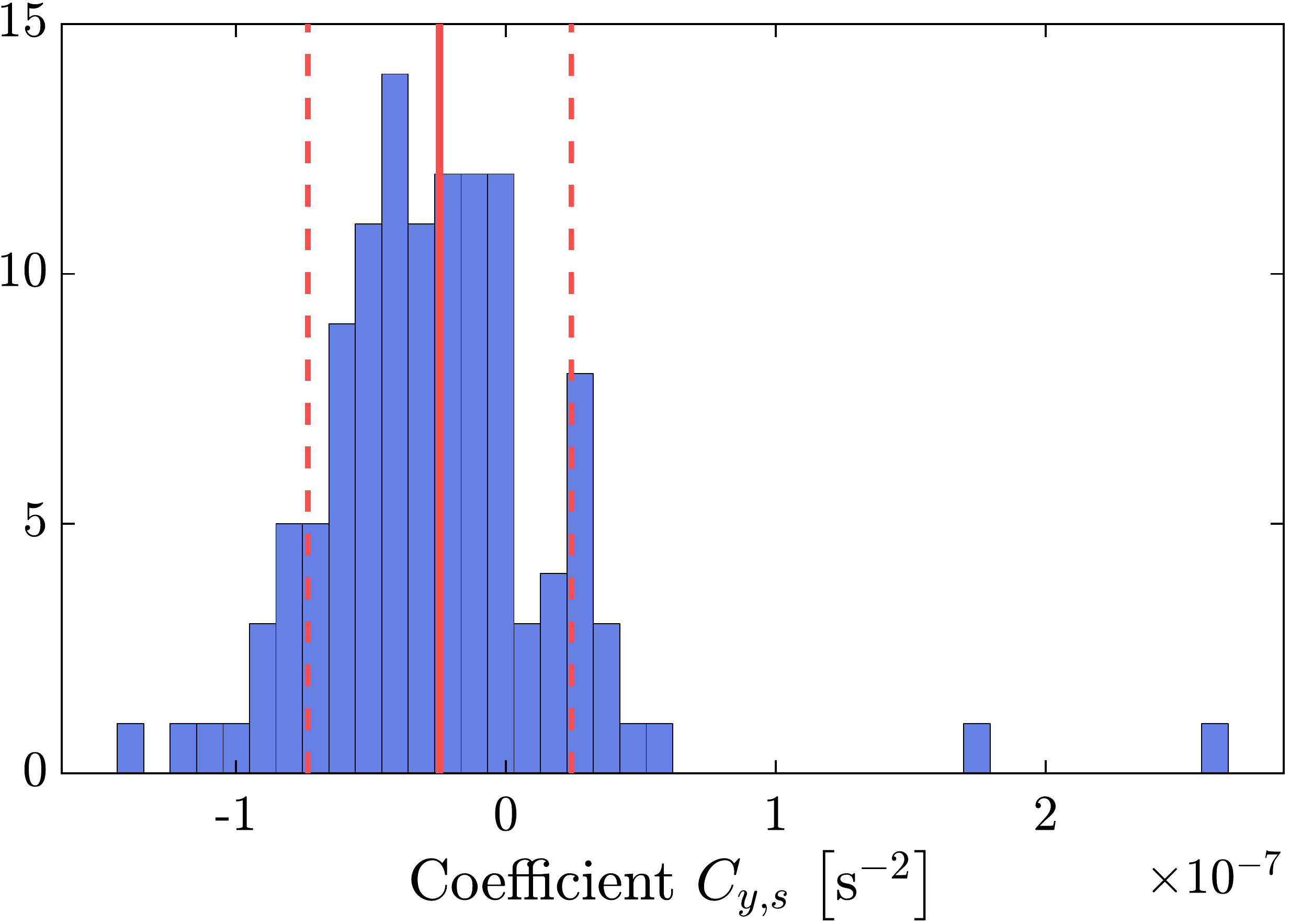}\hfill
\includegraphics[scale=0.22]{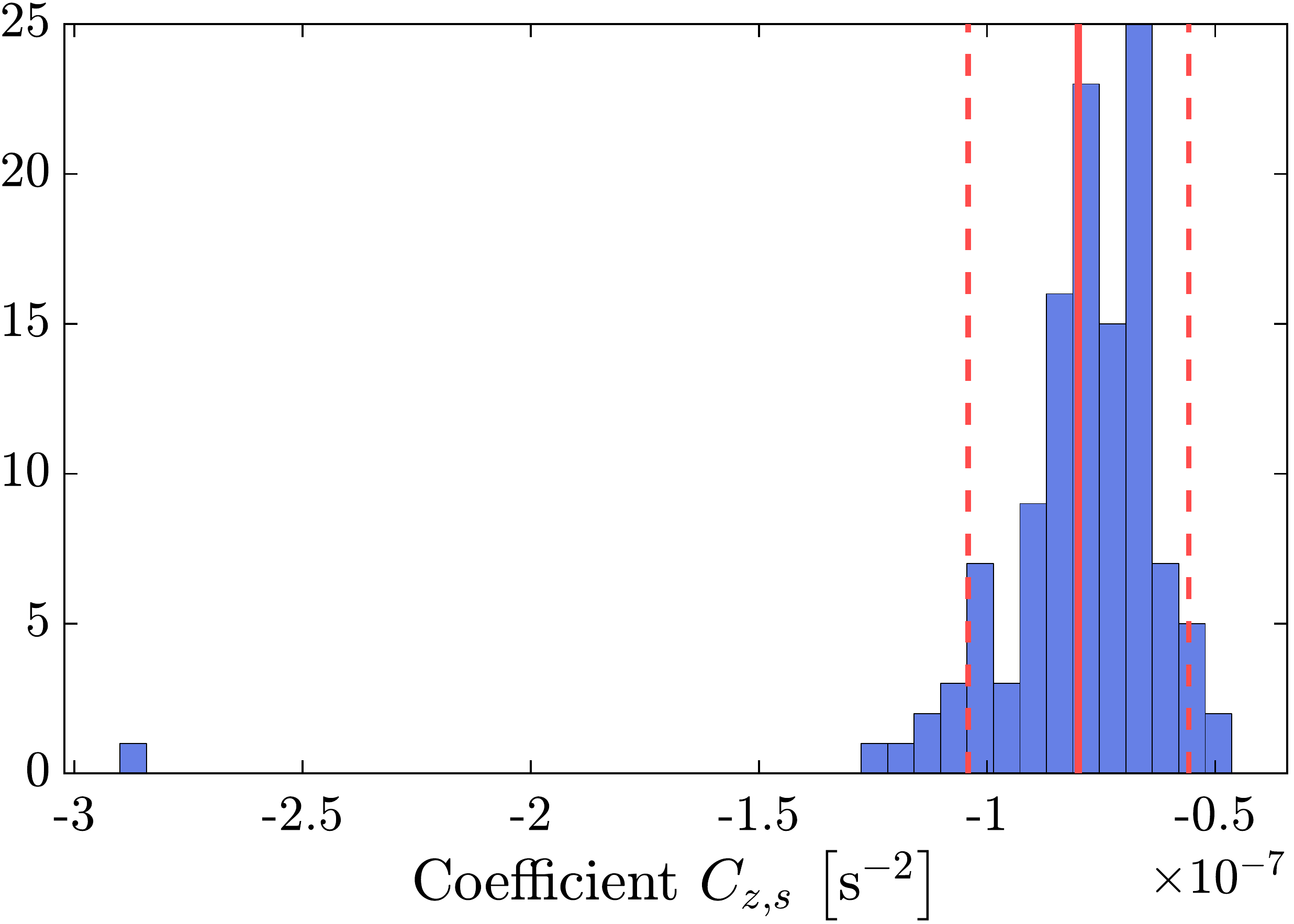} \\[1ex]
\includegraphics[scale=0.22]{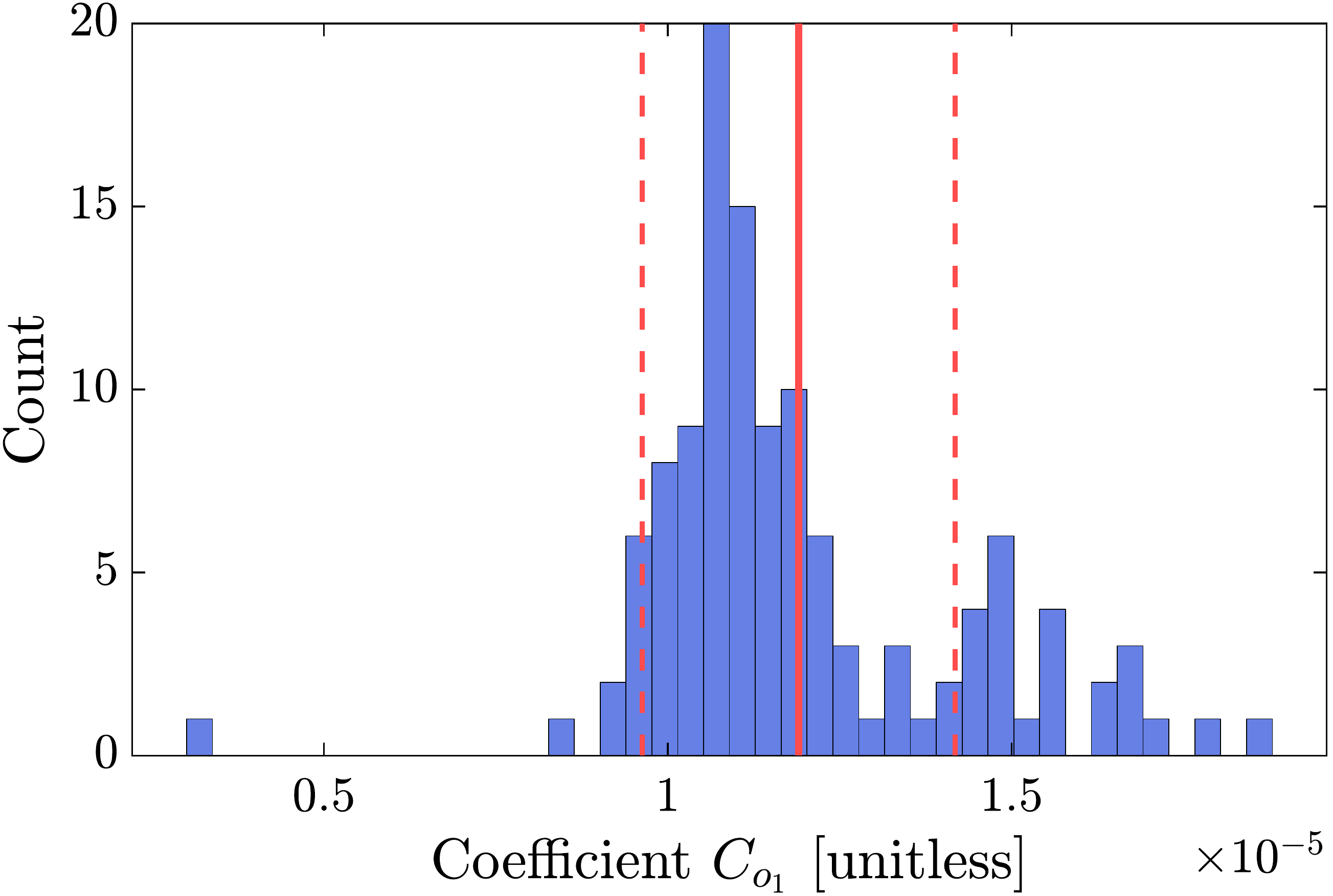}
\caption{Histogram of the coupling coefficients fitted for 120 2-hour timespans during the noise run which started on 13th Feb.\ 2017. The coefficients are not constant but spread about a mean. 
The dashed red lines are offset from the mean by the root-mean-square. With twice this deviation, we would cover all coefficient estimates but the outliers.}
\label{fig:LXE_histogram}
\end{figure*}

Note that the effect of tiny instabilities in the system (e.g.\ temperature-related) and noise variations on the fit almost canceled out for longer noise runs, but altered the results for shorter segments. 
The corresponding variations of the fitted coefficients were neither covered by their error bars nor modelled analytically.
For a better comparison of the fitted coefficients with the analytical computation, we additionally computed the variations of the fit results during a noise run. 
These could be seen as uncertainties of the offsets $C_{j,0}$, $j\in\{\varphi,\eta,y,z\}$, in the analytical model due to the unmodelled instabilities.
We split ten days of a noise run in mid-February 2017 into two-hour segments and investigated the variance of the fitted coefficients (Fig.~\ref{fig:LXE_histogram}). This yielded the deviations (root-mean-squares) of the coefficient estimates presented in Tab.~\ref{tab:LXE_errors}.
In the following, we will plot the analytical coefficients always with twice these root-mean-squares, which cover all fitted coupling coefficients but the outliers. Since we cannot certainly say whether these deviations are uncorrelated to analytically computed errors, we add both linearly (worst-case estimate).
Note that, while the variance of the coefficients can differ over the mission time, these estimates are considered to be useful for the analysis of experiments with small temporal distance and no lasting changes in the setup, e.g., the \gls{LXE} analysed in Sec.~\ref{sec:LXE}. 
However, mind also that these variations could even be bigger during the \gls{LXE} due to the experiments being performed.

\begin{table}
\begin{tabular}{lllccc}
\toprule
\multicolumn{2}{l}{coefficient} & & mean  & absolute   & root-mean-  \\
name & [unit]                   & & value & deviations & square \\
\midrule
$C_\varphi$ &[$\upmu$m\,rad$^{-1}$] &&   2.98 & $\pm$1.76 & 0.60 \\
$C_\eta$    &[$\upmu$m\,rad$^{-1}$] && -12.86 & $\pm$2.05 & 0.40 \\
$C_y$ &[$10^{-6}$]      &&  3.89 & $\pm$7.55 & 1.47 \\
$C_z$ &[$10^{-6}$]      && 43.52 & $\pm$1.70 & 0.55 \\
$C_{y,s}$ &[$10^{-6}$\,s$^{-2}$]    && -0.02 & $\pm$0.20 & 0.05 \\
$C_{z,s}$ &[$10^{-6}$\,s$^{-2}$]    && -0.08 & $\pm$0.12 & 0.02 \\
$C_{o_1}$ &[$10^{-6}$]  && 11.90 & $\pm$7.76 & 2.28 \\
\bottomrule
\end{tabular}
\caption{Fitted coefficients and their variations during the noise measurement starting on the 13th Feb.\ 2017, 8\,AM \gls{UTC}. The coefficients were computed for 120 2-hour segments of this noise run. Their distribution is shown in Fig.~\ref{fig:LXE_histogram}.
Twice the root-mean-squares of the coefficient deviations cover most of these fit results except for the extreme outliers.}
\label{tab:LXE_errors}
\end{table}

When comparing the presented analytical model to the \gls{TTL} models available during the \gls{LPF} mission, we find some significant differences. 
First, the model Eq.~\eqref{eq:anamodel} only depends on the angular \gls{TM} alignments, i.e.\ any lateral displacements of the \glspl{TM} would not affect the coupling. %
This finding depends on the centre of rotation we take in our evaluation into account: since the \gls{S/C} jitter is dominant, we apply the rotations in our analysis around its centre of mass. 
In the early models used during the mission, the relevance of the location of the centre of rotation was underestimated.
For simplicity, the jitter of the \glspl{TM} was analysed individually with the centre of rotation lying in the \gls{TM}'s centre of mass.
In this case, lateral \gls{TM} displacements would have shifted the centre of rotation relative to the beam axis.
Since such offsets significantly affect the \gls{TTL} coupling \cite{G21}, the old models included lateral \gls{TM} alignment terms not present in the new model presented here.
Also, the longitudinal location of the centre of rotation was neglected before. This contributed non-negligibly to $C_\varphi^\text{ana}$ and $C_\eta^\text{ana}$.
Furthermore, non-geometric coupling effects have been shown to contribute significantly to the overall coupling \cite{LPFana22,NG21}.

\section{The TTL coupling experiments during the mission}
\label{sec:Basics_Realignments}
As we explained in Sec.~\ref{sec:Basics_TTLinLPF} and  will further show in Secs.~\ref{sec:LXE} and \ref{sec:TMalignments}, the \gls{TTL} coupling during the \gls{LPF} mission depended on the alignment of the \glspl{TM}.
By realigning either of the \glspl{TM}, we modified the coupling of the different \gls{TTL} mechanisms acting together in the case of the \gls{LPF} setup \cite{G21,NG21}. Therefore, suitable \gls{TM} tilts can suppress the full \gls{TTL} coupling if the single effects counteracted each other.
During the mission, the \glspl{TM} were realigned three times for a \gls{TTL} noise suppression: on the 16th March, 19th June and 26th June 2016.

Furthermore, we performed an experiment during the \gls{LPF} mission extension phase with the aim of studying the dependency of the \gls{TTL} coupling on the \gls{TM} positions, the \gls{LXE}. 

In the following, we introduce these two \gls{TTL} coupling investigations.

\subsection{Redefinition of the TM set-points}
\label{sec:realingments_description}
The \gls{TM} alignment set-points were redefined three times during the \gls{LPF} mission (Tab.~\ref{tab:realignments}).
The new set-points were derived from simplified geometrical \gls{TTL} models.
The analytical model (Eq.~\eqref{eq:anamodel}) used for the analyses in this work was not yet available by then.

\begin{figure}
\centering
\includegraphics[scale=0.31]{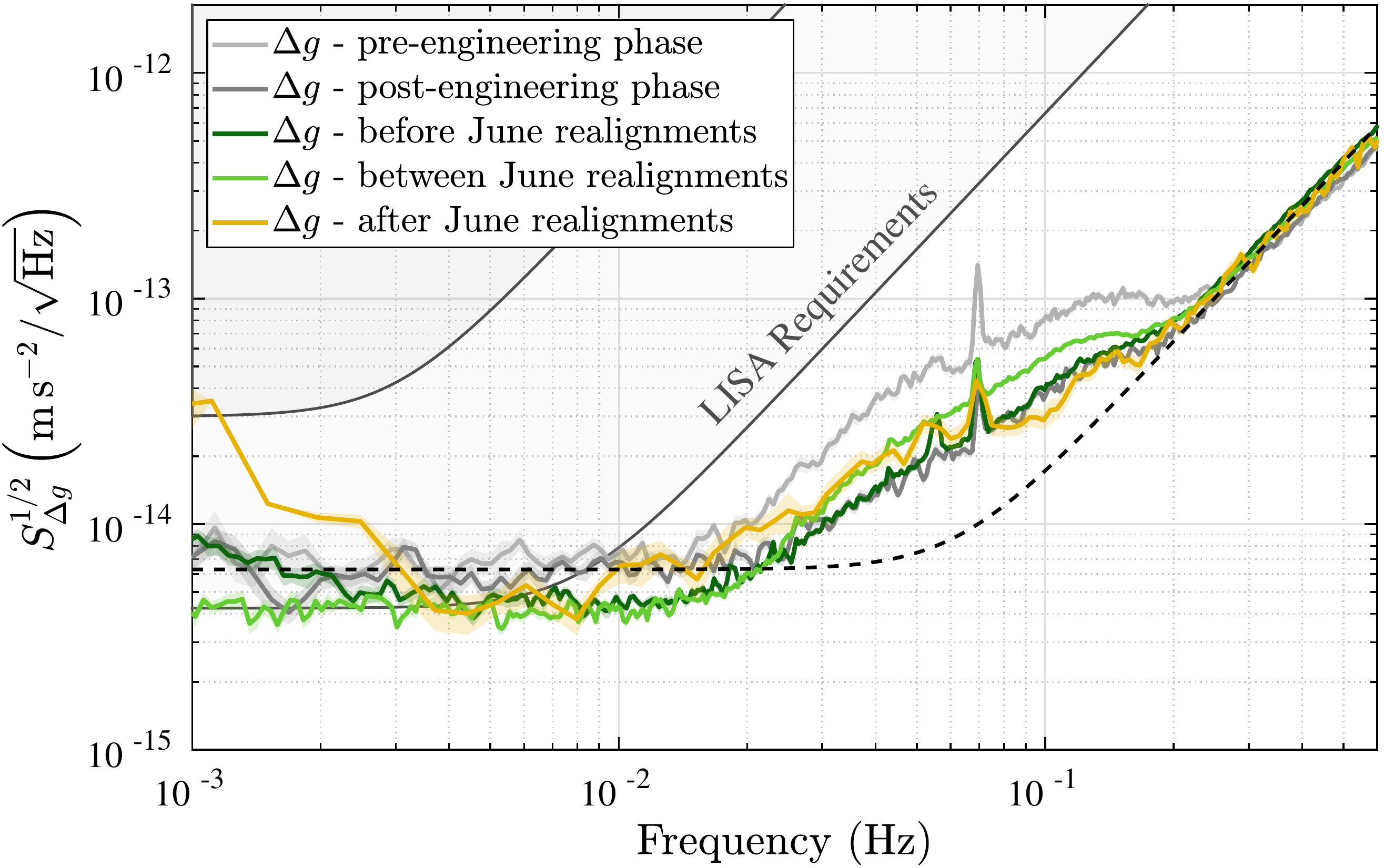}
\caption{Changes of the $\Delta g$ observable due to realignments of the \glspl{TM}.
  Dark green: $\Delta g$ before the realignments in June (three day noise run ending on the 18th June 2016).
  Light green: $\Delta g$ in between the realignments in June (noise run from the 20th to 24th June 2016).
  Yellow: $\Delta g$ after the realignments in June (one hour data with low noises at the 26th June 2016).
  These noise runs are defined in further detail in Sec.~\ref{sec:TMalignments_June_timespans}.
  The grey curves show the $\Delta g$ observable before and after the realignment during the \gls{ED} for comparison.
  Black dashed curve: LPF performance model for the time of the ED.}
\label{fig:Realignment_deltag}
\end{figure}

The first realignment, in March 2016, reduced the \gls{TTL} coupling by a factor of two. 
The residual TTL noise was then subtracted as shown in \cite{Wanner2017,Armano2018_calibration}. Yet, the question remained whether better alignment could be achieved suppressing the noise prior to subtraction even further.
Therefore, the \gls{TM} realignment was repeated in June 2016 after continuing theoretical investigations of the \gls{TTL} noise.
The first attempt on 19th June was less successful than expected and even increased the noise level. This was caused by a sign error and corrected for by a second realignment performed on 25th June. These alignment settings were then left untouched until the mission ended.
A comparison of the different levels of \gls{TTL} noise in the $\Delta g$ observable is plotted in Fig.~\ref{fig:Realignment_deltag}.
When we compare the `bump' before (dark green) and after (yellow) the two \gls{TM} realignments in June, we see that the noise was only reduced above 70\,mHz. Below these frequencies, the new alignments yielded an increased noise.
The shape of the bump depends on the dominant \gls{TTL} noise contributors, which we further discuss for these cases in Sec.~\ref{sec:TMalignments_June}.

Since neither of the realignments resulted in a full  \gls{TTL} noise suppression, it was at all times of the mission necessary to subtract the residual TTL noise from the data.

We show in Sec.~\ref{sec:TMalignments} that the noise level changes caused by the three realignments can be reconstructed and explained by the analytical model Eq.~\eqref{eq:anamodel}. Also, we present \gls{TM} alignments which presumingly would have made a full \gls{TTL} suppression via realignment possible.

\begin{table}
\begin{tabular}{cc|ccc}
\toprule
DoF & [unit] & 16.03.\ 14:28 & 19.06.\ 8:20 & 25.06.\ 8:00 \\ 
\midrule
$\varphi_1$ & [$\upmu$rad] & -59.25 & -57.32 & -61.2 \\
$\varphi_2$ & [$\upmu$rad] & -21.35 & -33.01 &  -9.7 \\
$\eta_1$    & [$\upmu$rad] &  -3.5  &  -2.14 &  -4.9 \\
$\eta_2$    & [$\upmu$rad] &   3.5  &  10.3  &  -3.3 \\
\bottomrule
\end{tabular}
\caption{Adaptions of the \gls{TM} set-points in their angular degrees of freedom (DoF) for \gls{TTL} suppression relative to the \gls{TM} alignment before
the 16th March 2016. 
Times are given in UTC and correspond to the year 2016. The angles were partially corrected compared to \cite{dlr2020}.}
\label{tab:realignments}
\end{table}

\subsection{Description of the cross-talk experiments}
\label{sec:LXE_description}
For a detailed study of the \gls{TTL} coupling in \gls{LPF} and the insufficiency of the former analytical models,
two cross-talk experiments were performed during the mission extension phase of \gls{LPF} (starting in Dec.\ 2016).
First, a short experiment was executed on the 20th Jan.\ 2017, followed by the \gls{LXE} about two weeks later \cite{dlr2020}.

The \gls{SXE} consisted of a series of sinusoidal injections applied to the \glspl{TM} about their nominal alignment, causing a sinusoidal \gls{S/C} motion due to the attitude control loops. %
At first, parallel lateral displacements of the \glspl{TM} along $y$ were commanded, inducing a sinusoidal lateral translations of the \gls{S/C}. 
After a short break, anti-parallel lateral displacements of the \glspl{TM} in the same plane were injected. These were compensated by a sinusoidal rotation of the \gls{S/C}.
Finally, a sinusoidal injection with one frequency was applied to the first \gls{TM}, while at the same time a sinusoidal injection with another frequency was applied to the second \gls{TM}. This corresponded to a more complex superposition of translations and rotations of the \gls{S/C}. 
This series of injections was then repeated along $z$, resulting in the total of six different injection types summarised in Tab.~\ref{tab:shortxtalk}. 

The \gls{LXE} was performed between the 4th Feb.\ 2017, 21:00 \gls{UTC}, and 6th Feb.\ 2017, 23:00 \gls{UTC}, in order to provide data for an exhaustive test and correction of the physical \gls{TTL} coupling model containing the \glspl{TM} alignment information. 
The data obtained from the \gls{LXE} allowed testing the analytical model Eq.~\eqref{eq:anamodel_C}.
The experiment consisted of twelve \gls{SXE}-type sub-experiments with different set-points. 
This means that first, one or both \glspl{TM} were shifted or rotated with respect to their nominal alignment. 
An overview of the chosen set-points is given in Tab.~\ref{tab:longxtalk}.
Note, that also lateral set-points were chosen since these were assumed to change the \gls{TTL} noise according to the old coupling models (see Sec.~\ref{sec:TTLmodel_ana}).
Depending on the set-point, \gls{SXE}-type injection were applied:
If the \gls{TM} position or orientation was changed only in the $xy$-plane ($y_1,\ y_2,\ \varphi_1,\ \varphi_2$ alignments), only the $y$-injections were applied.
Analogously, only $z$-injections were commanded if the respective set-point included only changes in the $xz$-plane ($z_1,\ z_2,\ \eta_1,\ \eta_2$ alignments).
For \glspl{TM} set-points in both planes, the full set of \gls{SXE} injections was chosen.
In the last step, after the injections, the \glspl{TM} were brought back to their nominal position.

\begin{table*}
    \centering
    \begin{tabular}{l cccc r} 
       	\toprule
       	injection type & amplitude & frequency & duration & ramp duration & wait time \\
       	\toprule
	 $y_1 = y_2$ & \SI{0.5}{\upmu m}  & \SI{17}{mHz} & \SI{30}{\minute} & \SI{8}{\minute} & \SI{5}{\minute}	\\
	 $y_1 = - y_2$ & \SI{0.5}{\upmu m}  & \SI{12}{mHz} & \SI{30}{\minute} & \SI{8}{\minute} & \SI{5}{\minute}	\\
	 $y_1$, $y_2$ & \SI{0.3}{\upmu m}  & \SI{10}{mHz}, \SI{17}{mHz} & \SI{50}{\minute} & \SI{8}{min} & \SI{10}{\minute}	\\
	 $z_1 = z_2$ & \SI{0.5}{\upmu m}  & \SI{5}{mHz} & \SI{50}{\minute} & \SI{8}{\minute} & \SI{5}{\minute}	\\
	 $z_1 = -z_2$ & \SI{0.5}{\upmu m}  & \SI{8}{mHz} & \SI{30}{\minute} & \SI{8}{\minute} & \SI{5}{\minute}	\\
	 $z_1$, $z_2$ & \SI{0.3}{\upmu m}  & \SI{5}{mHz}, \SI{8}{mHz} & \SI{50}{\minute} & \SI{5}{min} & -	\\
         \bottomrule
    \end{tabular}
    \caption[The short crosstalk experiment]{Sequence of performed injections during the \gls{SXE}. Table reprinted from the authors' publication \cite{dlr2020}.}
    \label{tab:shortxtalk}
\end{table*}

\begin{table*}
    \centering
    \small
    \begin{tabular}{c|lc|l} 
        \toprule
       \# &	Start time & End time  &	Set-point 	\\
       	  &	\multicolumn{2}{c|}{(month.day hour:minutes)}  & (Relative to initial)	\\
       \toprule
1 	&	02.04 22:02 	&	02.05 00:08	&	$y_1$: -30$\upmu\mathrm{m}$	\\ \hline
2 	&	02.05 00:36 	&	02.05 03:34	&	$z_2$: 10$\upmu\mathrm{m}$ 	\\ \hline
3 	&	02.05 04:00 	&	02.05 07:00	&	$z_1$: 21.6$\upmu\mathrm{m}$ 	\\ \hline
4 	&	02.05 07:27 	&	02.05 09:34	&	$y_2$: -22$\upmu\mathrm{m}$	\\ \hline
5 	&	02.05 10:00 	&	02.05 13:00	&	$\eta_1$: 12.1$\upmu\mathrm{rad}$	\\ \hline
6 	&	02.05 13:25 	&	02.05 15:25	&	$\varphi_2$: 20$\upmu\mathrm{rad}$	\\ \hline
7 	&	02.05 16:00 	&	02.05 18:11	&	$\varphi_1$: 30$\upmu\mathrm{rad}$	\\ \hline 
8 	&	02.05 18:35 	&	02.05 21:34	&	$\eta_2$: -20.3$\upmu\mathrm{rad}$	\\ \hline
9 	&	02.05 22:05 	&	02.06 03:27	&	$y_1$: -10$\upmu\mathrm{m}$, $z_1$: 5$\upmu\mathrm{m}$	\\ \hline
10 	&	02.06 04:32 	&	02.06 09:35	&	$y_1$: -20$\upmu\mathrm{m}$, $y_2$: -10$\upmu\mathrm{m}$, $z_1$: 15$\upmu\mathrm{m}$, $z_2$: 5$\upmu\mathrm{m}$	\\ \hline
11 	&	02.06 10:03 	&	02.06 15:34	&	$y_1$: -25$\upmu\mathrm{m}$, $y_2$: -15$\upmu\mathrm{m}$, $z_1$: 20$\upmu\mathrm{m}$, $z_2$: 15$\upmu\mathrm{m}$, $\varphi_1$: 10$\upmu\mathrm{rad}$, $\eta_1$: 5$\upmu\mathrm{rad}$	\\ \hline
12 	&	02.06 16:00 	&	02.06 21:35	&	$y_1$: -35$\upmu\mathrm{m}$, $y_2$: -25$\upmu\mathrm{m}$, $z_1$: 25$\upmu\mathrm{m}$, $z_2$: 25$\upmu\mathrm{m}$, $\varphi_1$: 20$\upmu\mathrm{rad}$, $\eta_1$: 10$\upmu\mathrm{rad}$, $\varphi_2$: 10$\upmu\mathrm{rad}$, $\eta_2$: -5$\upmu\mathrm{rad}$	\\
        \bottomrule
    \end{tabular}
    \caption[The long cross-talk experiment]{Timeline of the \gls{LXE} that was performed between the 4th and 6th of February 2017. Times are given in \gls{UTC}. The shown set-points were commanded relative to the initial TM positions. Table reprinted from the authors' publication \cite{dlr2020}.}
    \label{tab:longxtalk}
\end{table*}

\section{Long Cross Talk Experiment}
\label{sec:LXE}

In this section, we use the data of the \gls{LXE} introduced in the previous section for the analysis of the performance of both introduced \gls{TTL} models.
We first analyse
the performance of both models on subtracting the induced \gls{TTL} noise (Sec.~\ref{sec:LXE_performance}), and directly compare the corresponding \gls{TTL} coefficients (Sec.~\ref{sec:LXE_comparison}).
The residual differences between both models are discussed in Sec.~\ref{sec:LXE_noise-contributors} based on the noise contributors in the single sub-experiments of the \gls{LXE}.
In Sec.~\ref{sec:LXE_2nd-order}, we extend the former linear \gls{TTL} models to quadratic models.
Finally, we present in Sec.~\ref{sec:LXE_minimiser} a simple method how the realignment information could be derived without an analytical model. We show that this method provides comparable results for the \gls{LXE} data as the analytical model.

\subsection{Performance of the TTL models}
\label{sec:LXE_performance}

To assess the performance of the fit and the analytical \gls{TTL} model, we applied both to the data of the \gls{LXE} and investigated the residual after subtraction from the observable $\Delta g$ (see \cite{Armano2016} for the preparation of $\Delta g$).
Since the fit model Eq.~\eqref{eq:fitmodel} also considers stiffness terms and the residual acceleration of the \gls{S/C} along the optical axis ($\ddot{o}_1$-term), these needed to be added to the analytical model for the comparison of both models. We fitted these coefficients ($C_{y,s},\,C_{z,s},\,C_{o_1}$) to the difference of the $\Delta g$ measurement and the analytical model  Eq.~\eqref{eq:anamodel}.
The offsets $C_{i,0}$ of the analytical coupling coefficients (Eq.~\eqref{eq:anamodel_C}) were replaced by the mean of the fitted coefficients in sub-experiments in which they should not (or only negligibly) change according to the analytical model (i.e., the experiments 1-5 and 8-10 in the case of $C_{\varphi,0}$ and $C_{y,0}$ and experiments 1-4, 6-7 and 9-10 in the case of $C_{\eta,0}$ and $C_{z,0}$). 
While these coefficients changed during the mission due to stresses and relaxations of the optical system \cite{LPFstabi23}, they can be assumed constant in short timespans, e.g., for the time of the \gls{LXE}.

\begin{figure*}
\includegraphics[scale=0.23,valign=t]{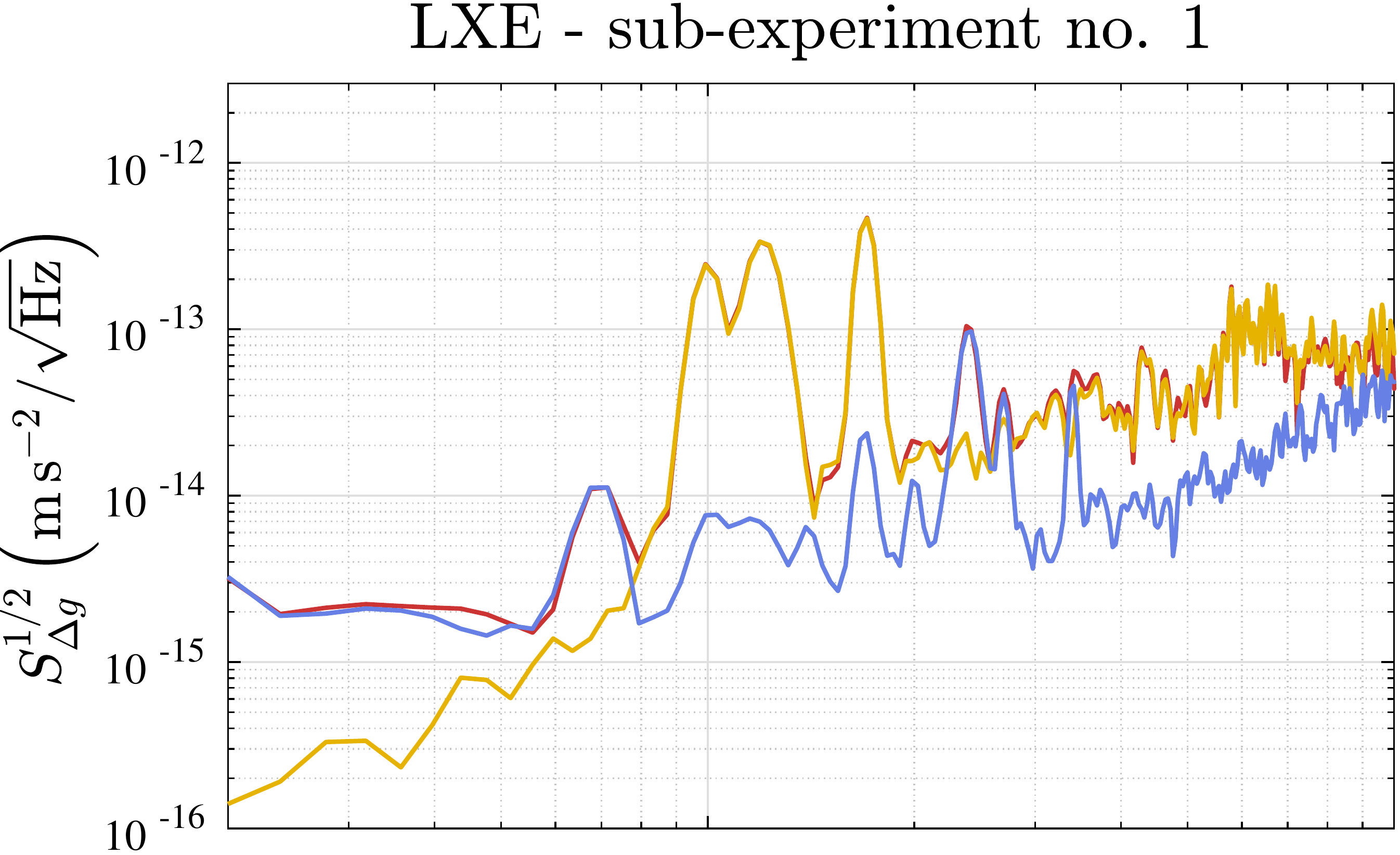}\,\quad
\includegraphics[scale=0.23,valign=t]{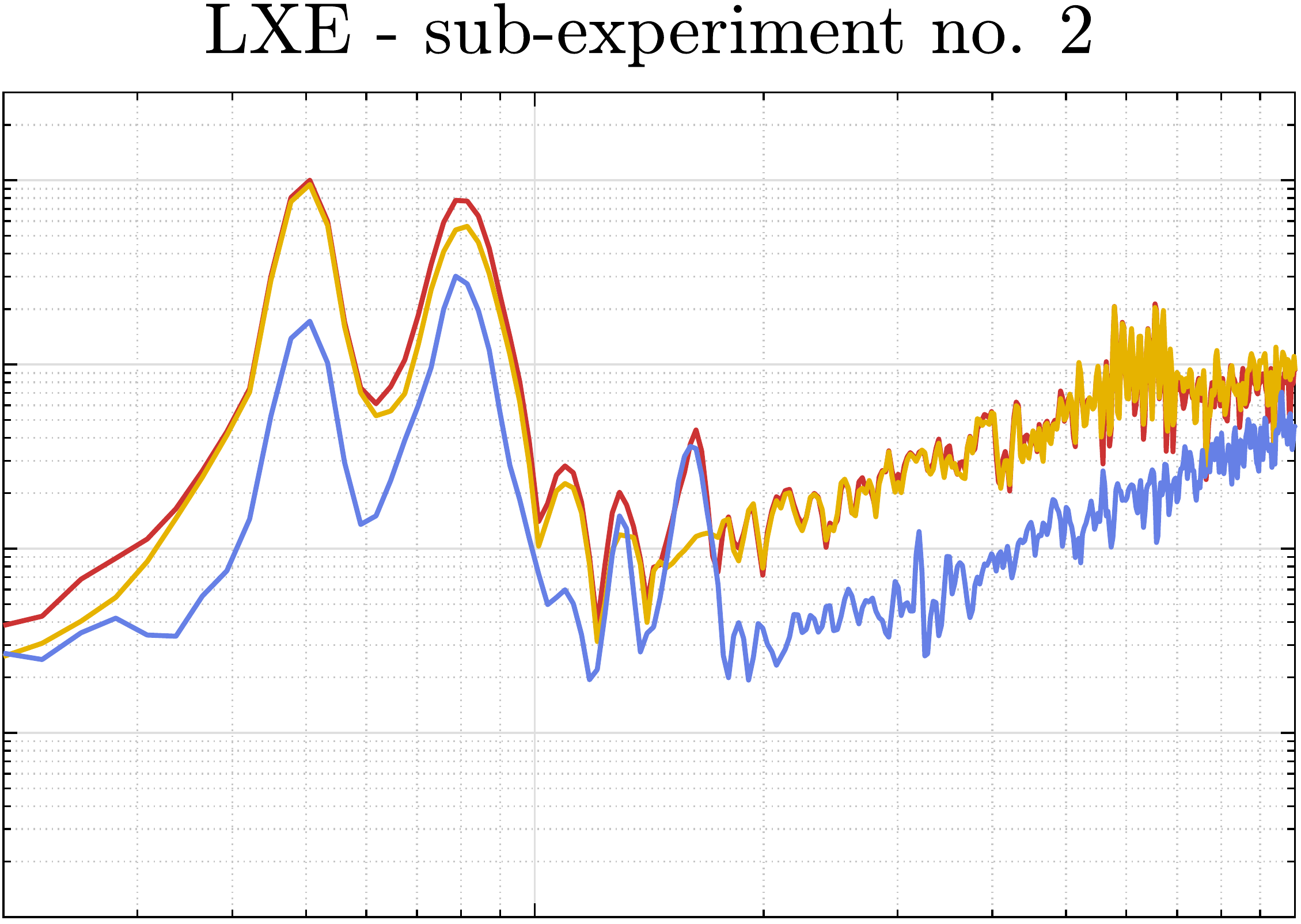}\quad
\includegraphics[scale=0.23,valign=t]{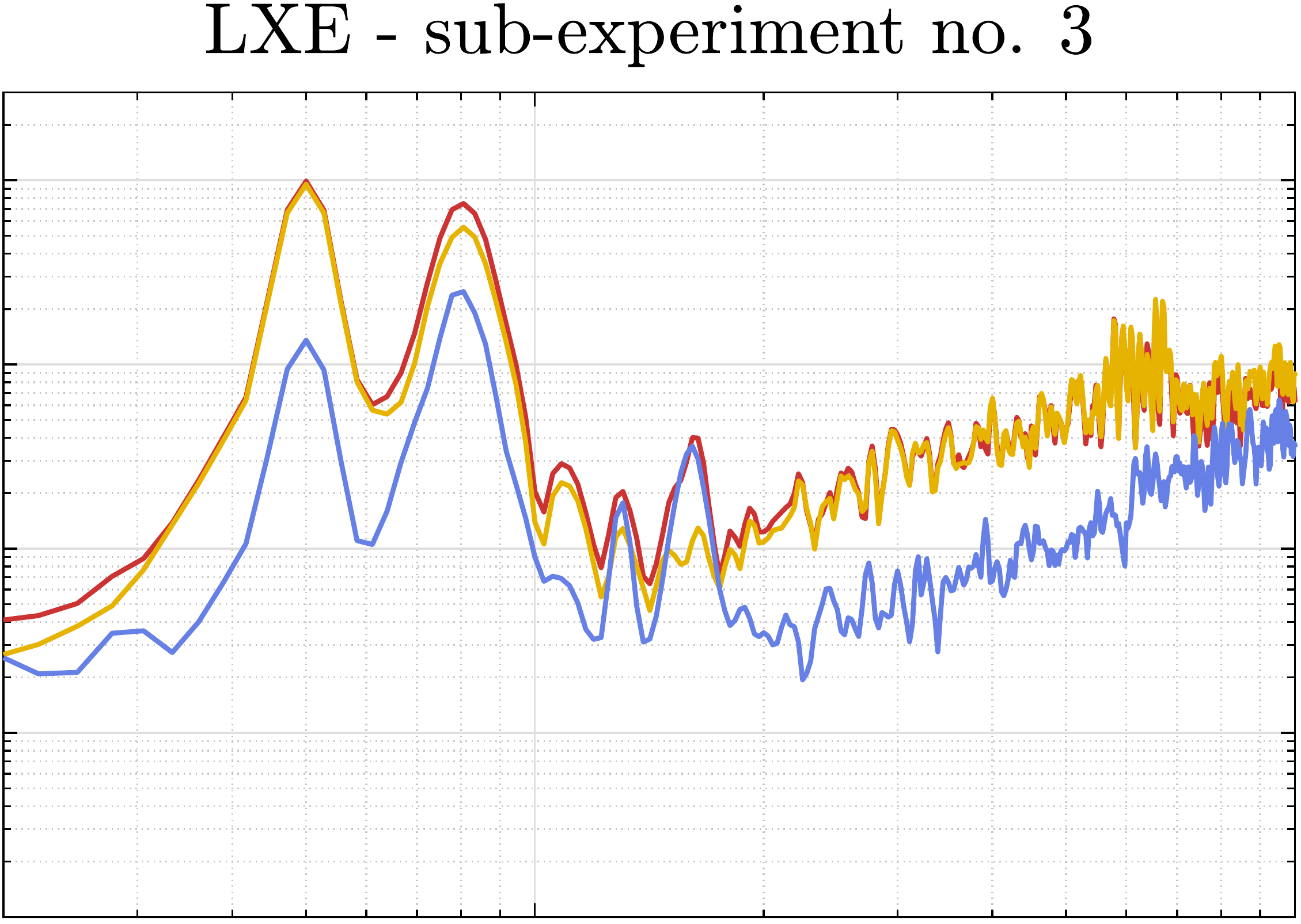} \\[1ex]
\includegraphics[scale=0.23,valign=t]{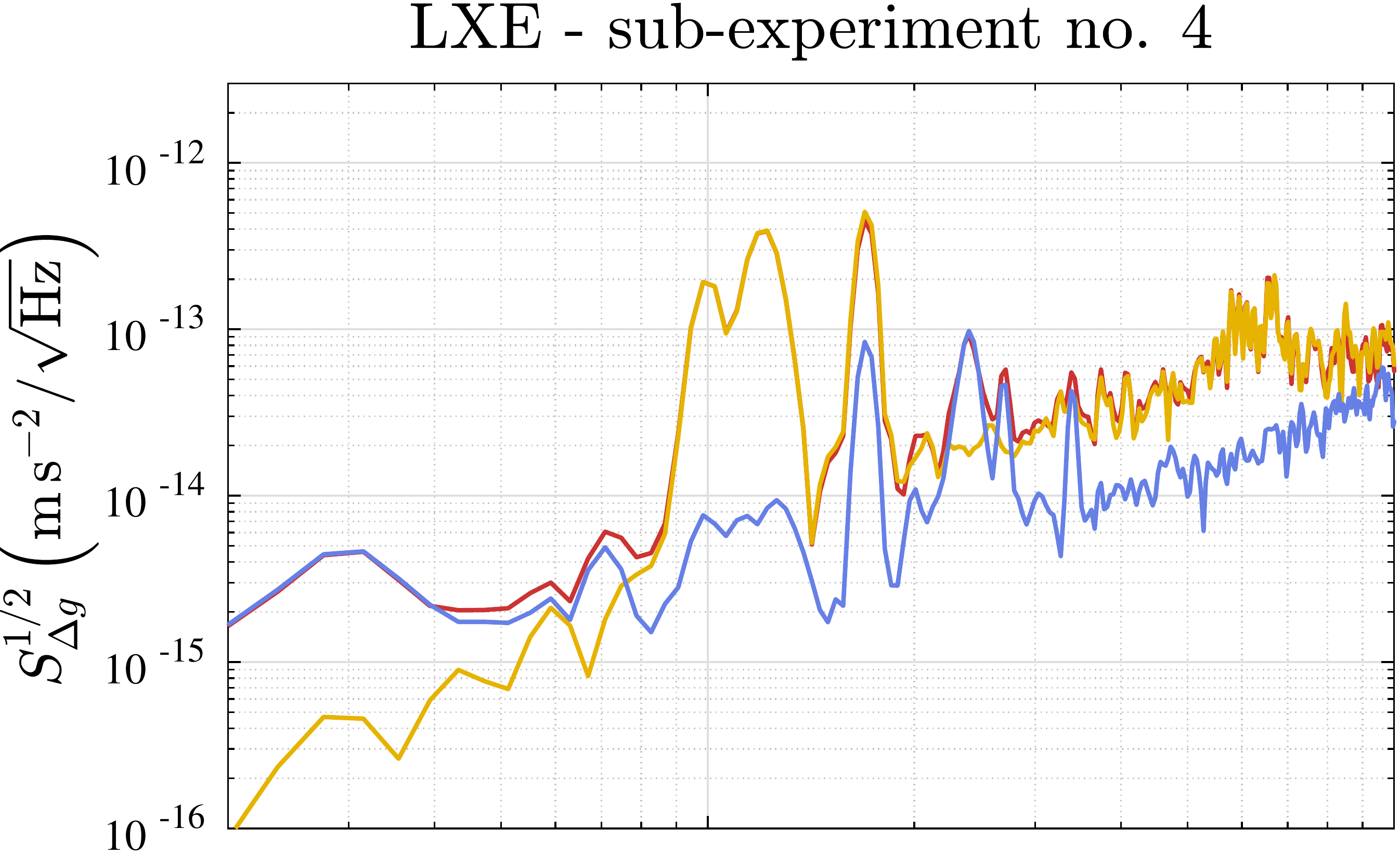}\,\quad
\includegraphics[scale=0.23,valign=t]{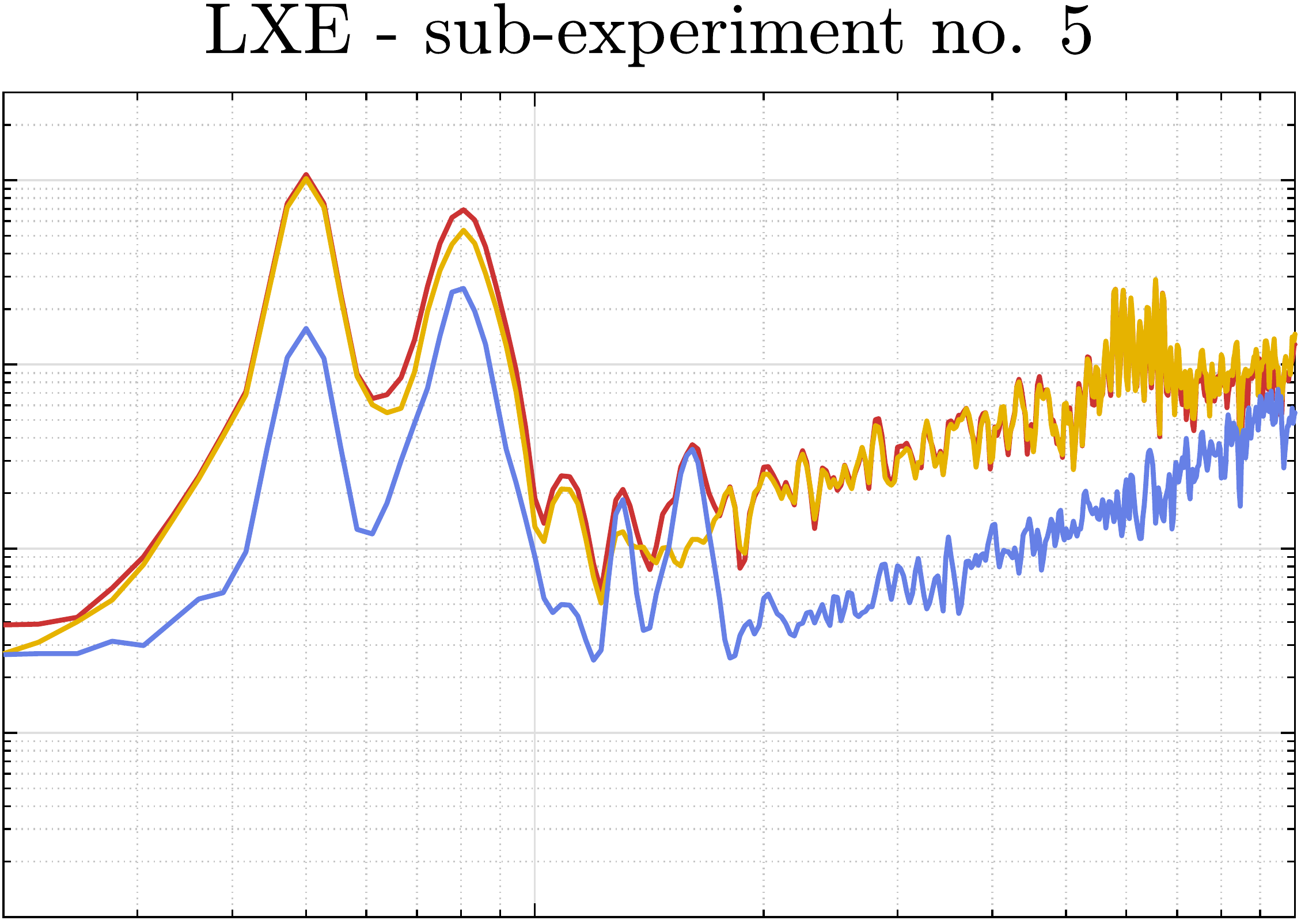}\quad
\includegraphics[scale=0.23,valign=t]{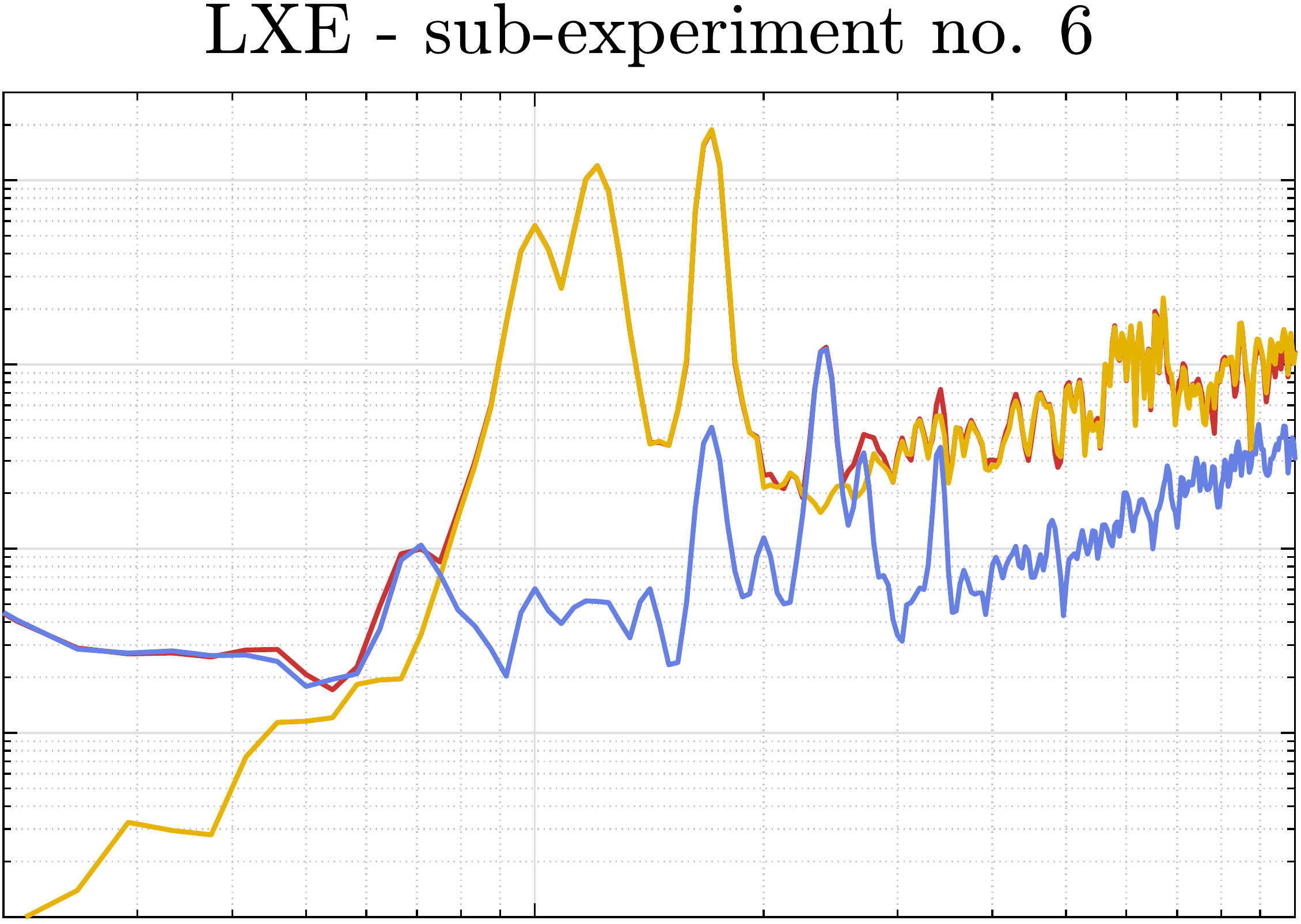} \\[1ex]
\includegraphics[scale=0.23,valign=t]{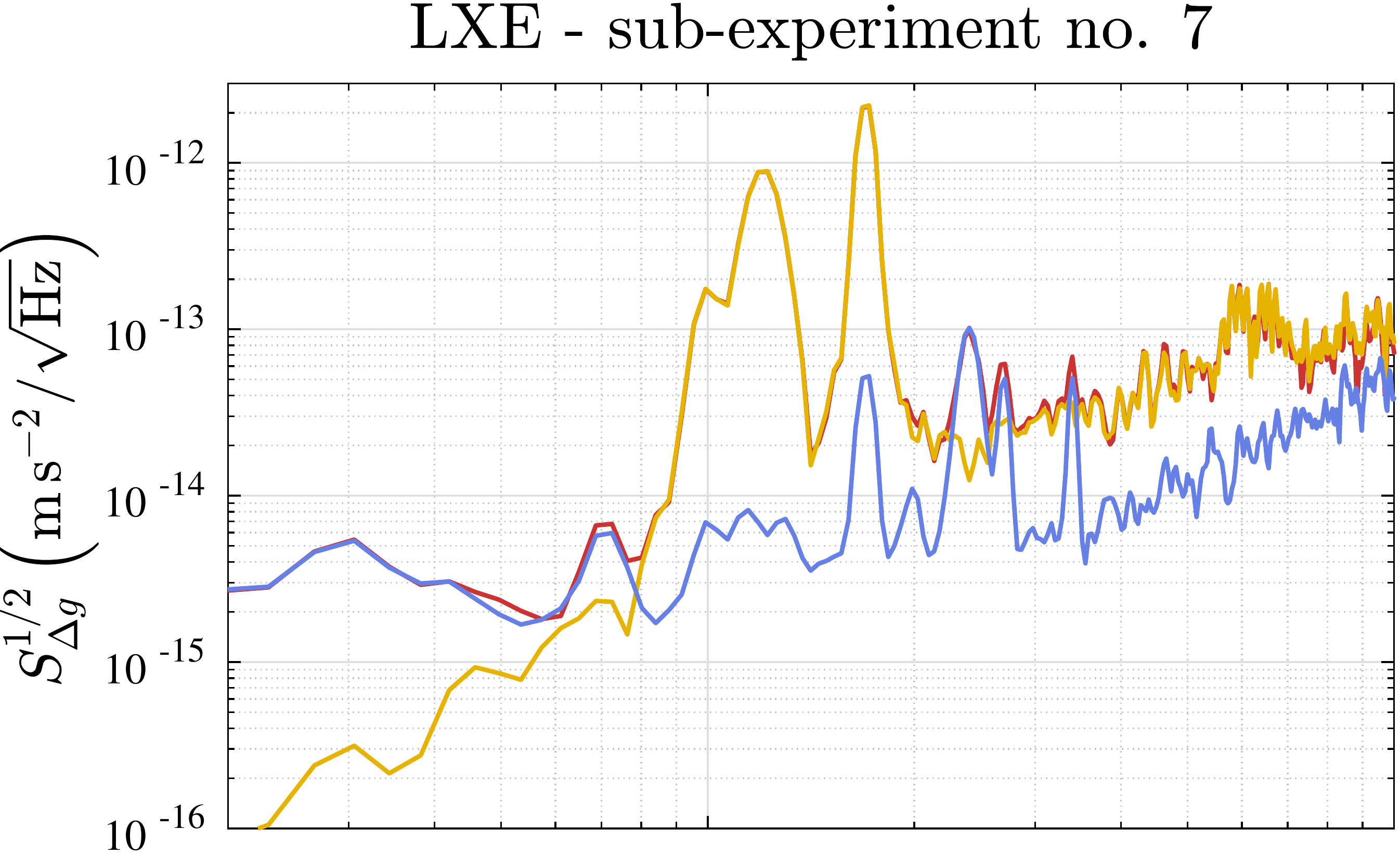}\,\quad
\includegraphics[scale=0.23,valign=t]{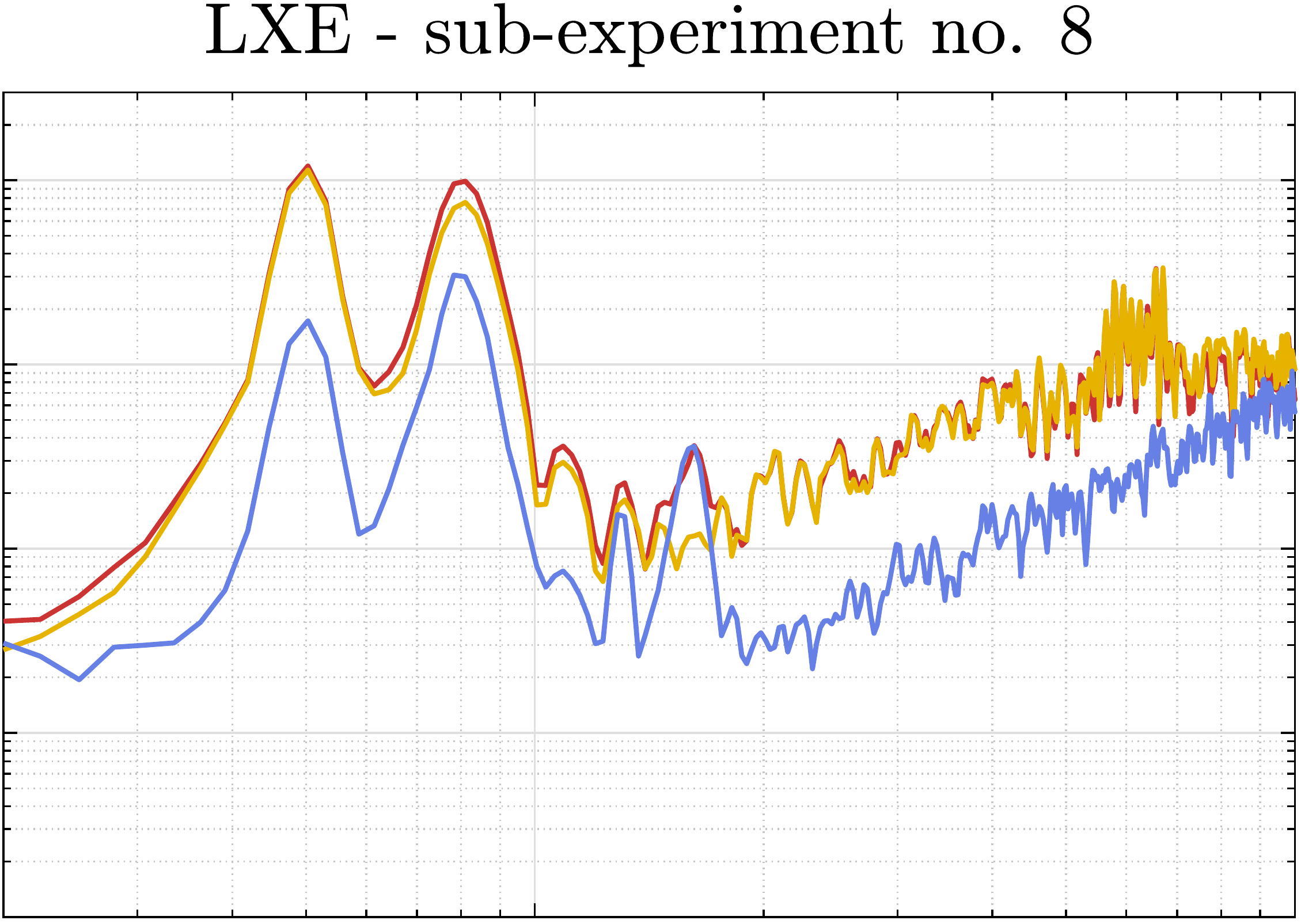}\quad
\includegraphics[scale=0.23,valign=t]{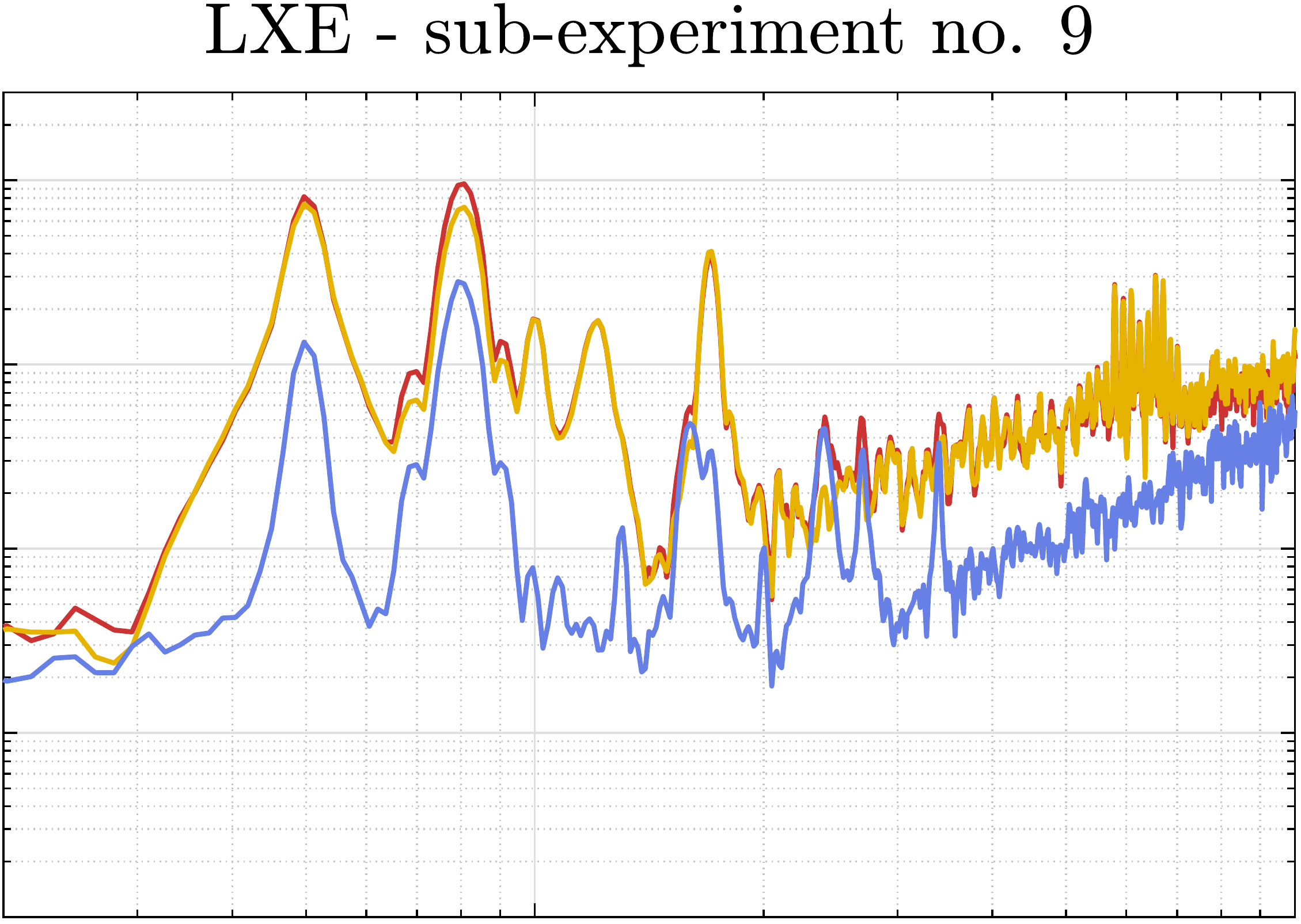} \\[1ex]
\,\ \includegraphics[scale=0.23]{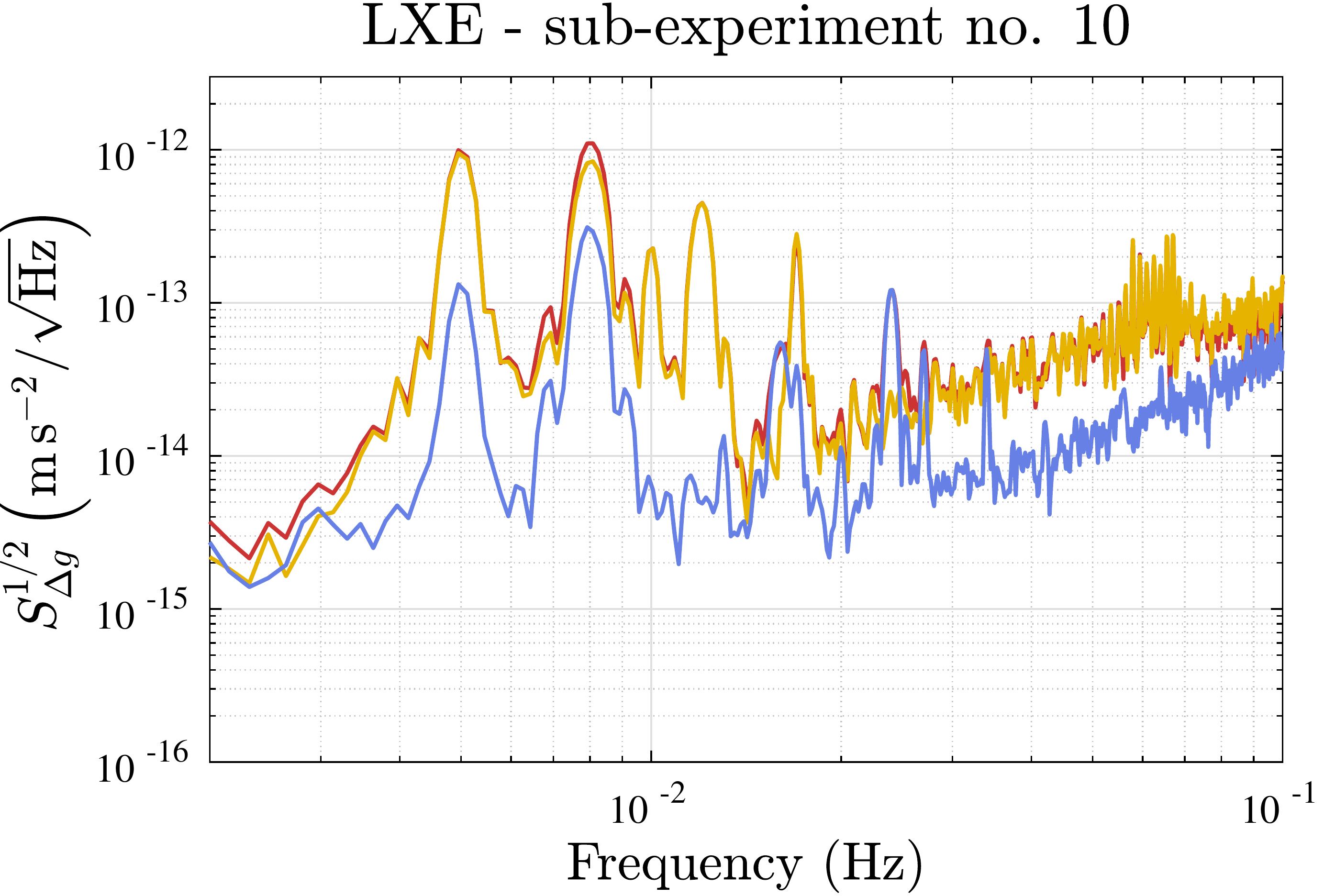}\,\,\
\includegraphics[scale=0.23]{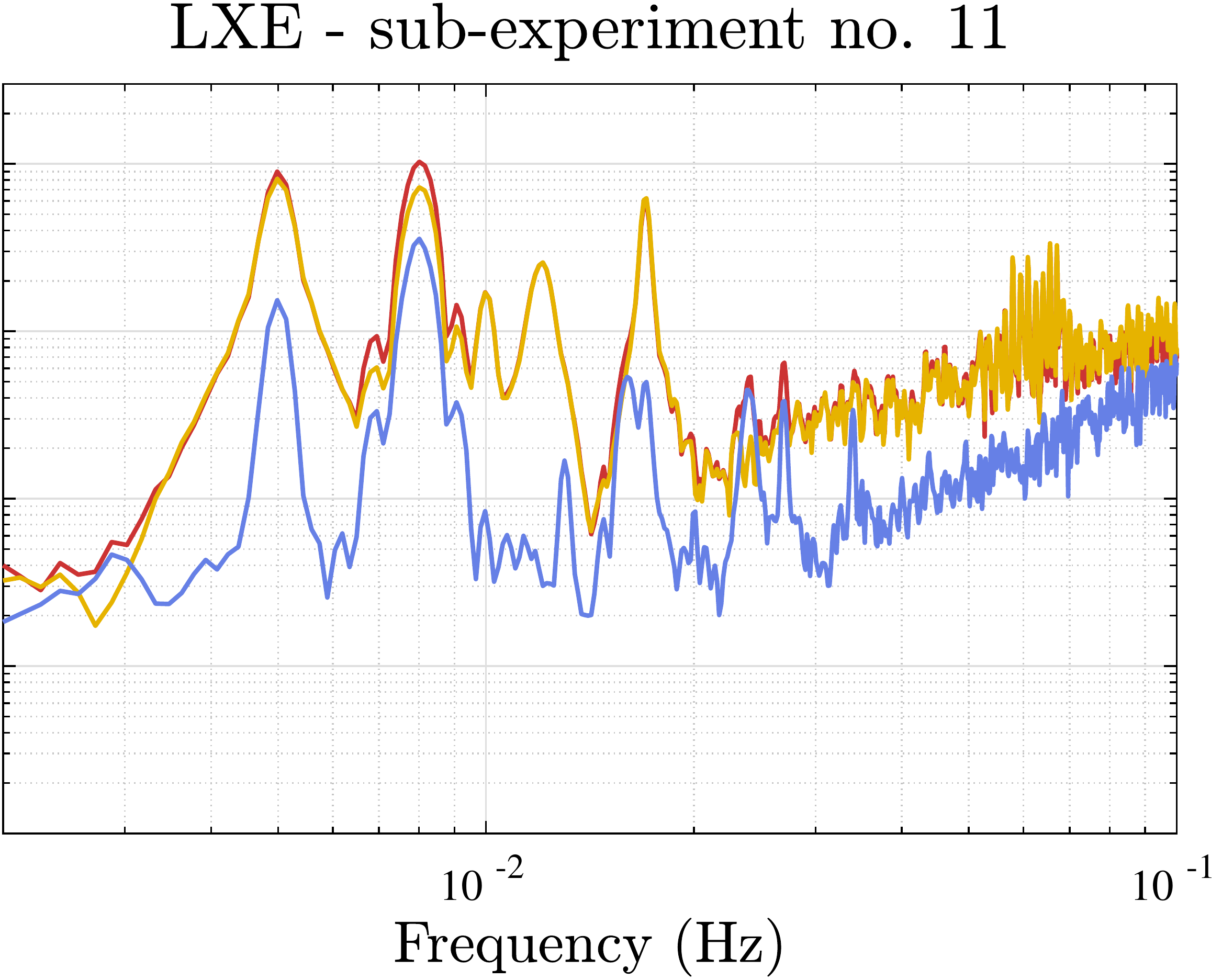}\hfill
\includegraphics[scale=0.23]{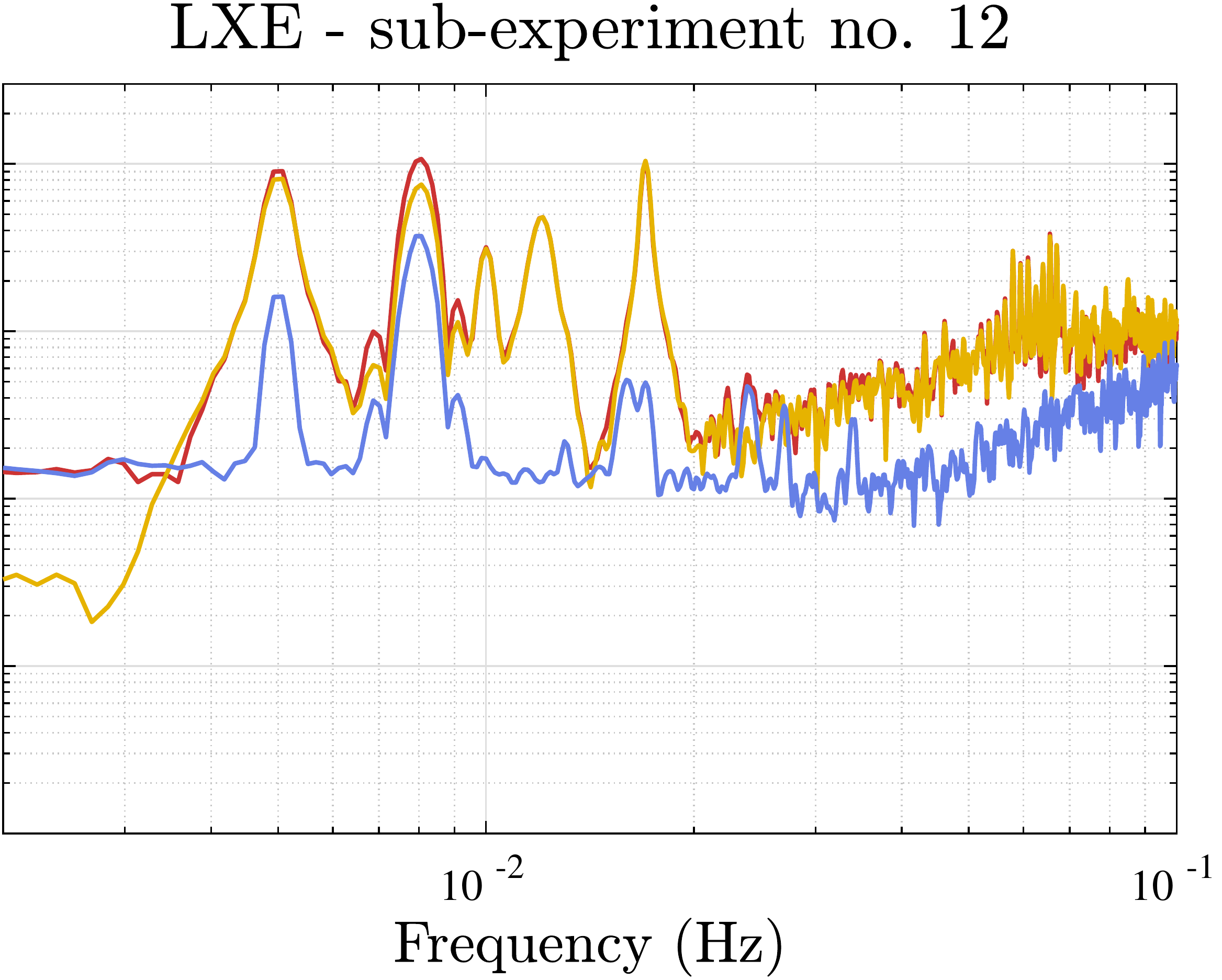} \\[2ex]
\includegraphics[scale=0.8]{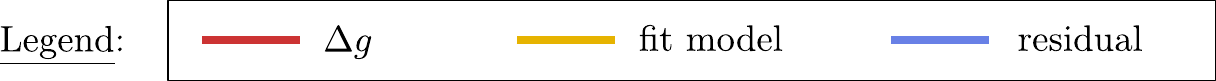}
\caption{Performance of the fit model for the twelve sub-experiments of the LXE.
The red curve shows the \gls{ASD} of the observable $\Delta g$. The yellow curve shows the respective density of the linear model fitted to the data. The blue curve shows the residual remaining after the subtraction of the fit from $\Delta g$. 
The noise peaks at the injection frequencies of the $y$-injections (10, 12 and 17\,mHz) are well subtracted, and the noise at the injection frequencies of the $z$-injections (5 and 10\,mHz) is significantly reduced. The noise above 20\,mHz is suppressed except for the multiples of the injections frequencies not covered by the linear fit model.}
\label{fig:LXE_performance_fit}
\end{figure*}

Let us first investigate the performance of the fit model shown in Fig.~\ref{fig:LXE_performance_fit}. 
Note here that the coupling coefficients of Eq.~\eqref{eq:fitmodel} were fitted for each sub-experiment individually, resulting in one set of coefficients per sub-experiment.
The subtraction of these fits from the $\Delta g$ observable removed different levels of noise in the different experiments.
The subtraction worked better at frequencies belonging to the injections along $y$ than for the $z$-injections (compare e.g.\ the blue residual in sub-experiment no.\ 1 ($y$-injections at 10,\,12 and 17\,mHz) with the blue residual in sub-experiment no.\ 2 ($z$-injections at 5 and 8\,mHz)). 
We further discuss this is Sec.~\ref{sec:LXE_noise-contributors}.
Also, the linear fit model does not well subtract the \gls{TTL} noise at sums and multiples of or the differential injection frequencies.
Most prominent are here the residual peaks at 7,\,24,\,27 and 34\,mHz for $y$-injections and at 13 and 16\,mHz for $z$-injections.
Since the noise at these frequencies originated from higher-order effects, it is naturally not covered by linear models. 
However, we show in Sec.~\ref{sec:LXE_2nd-order} that this noise is subtracted by second-order models.
In general, the \gls{TTL} coupling induced noise at frequencies above 20\,mHz (the `bump' \cite{Armano2016,Wanner2017}), where it was dominating during the mission, was well reduced in all sub-experiments. 

\begin{figure*}
\centering
\includegraphics[scale=0.23,valign=t]{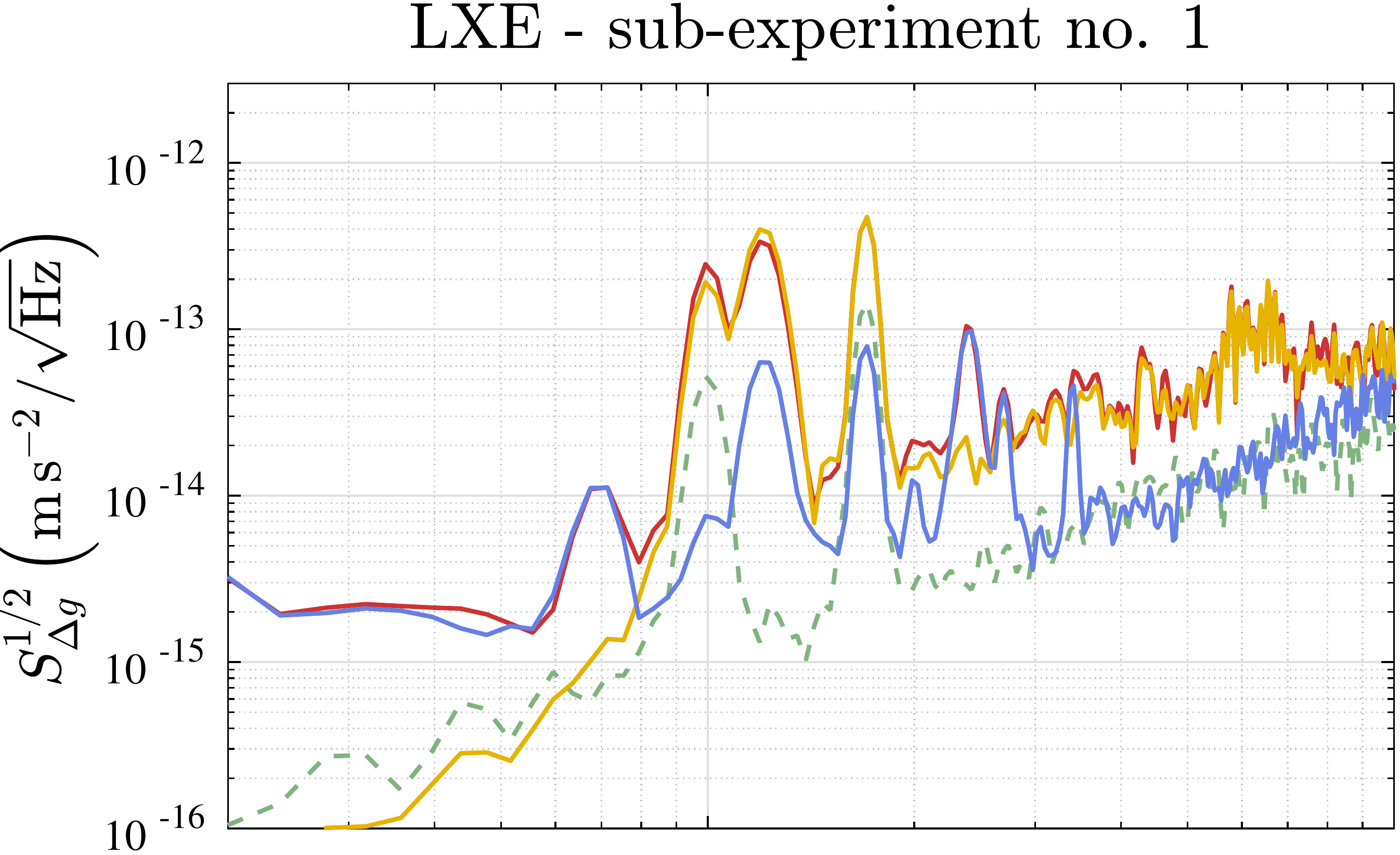}\,\quad
\includegraphics[scale=0.23,valign=t]{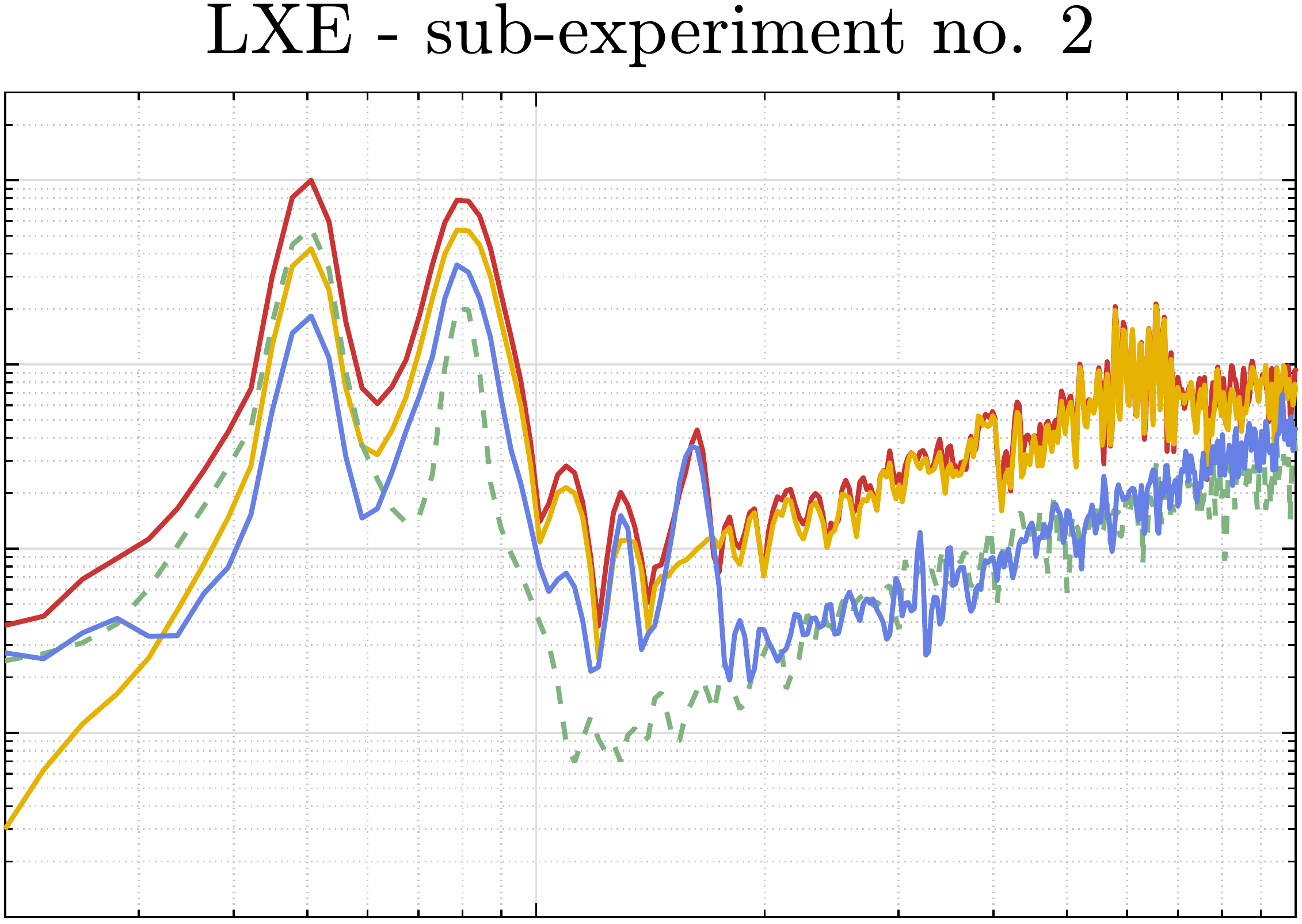}\quad
\includegraphics[scale=0.23,valign=t]{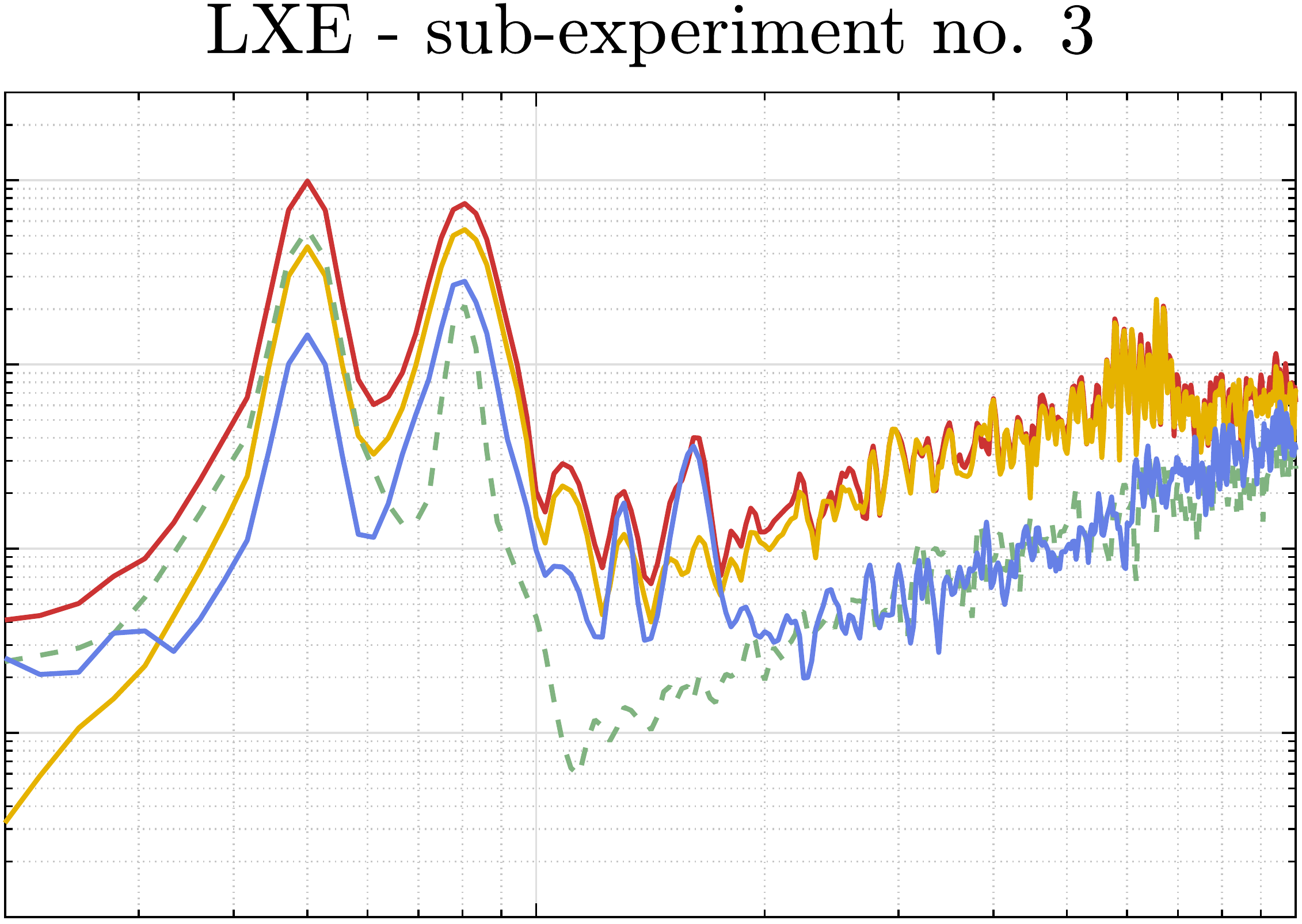} \\[1ex]
\includegraphics[scale=0.23,valign=t]{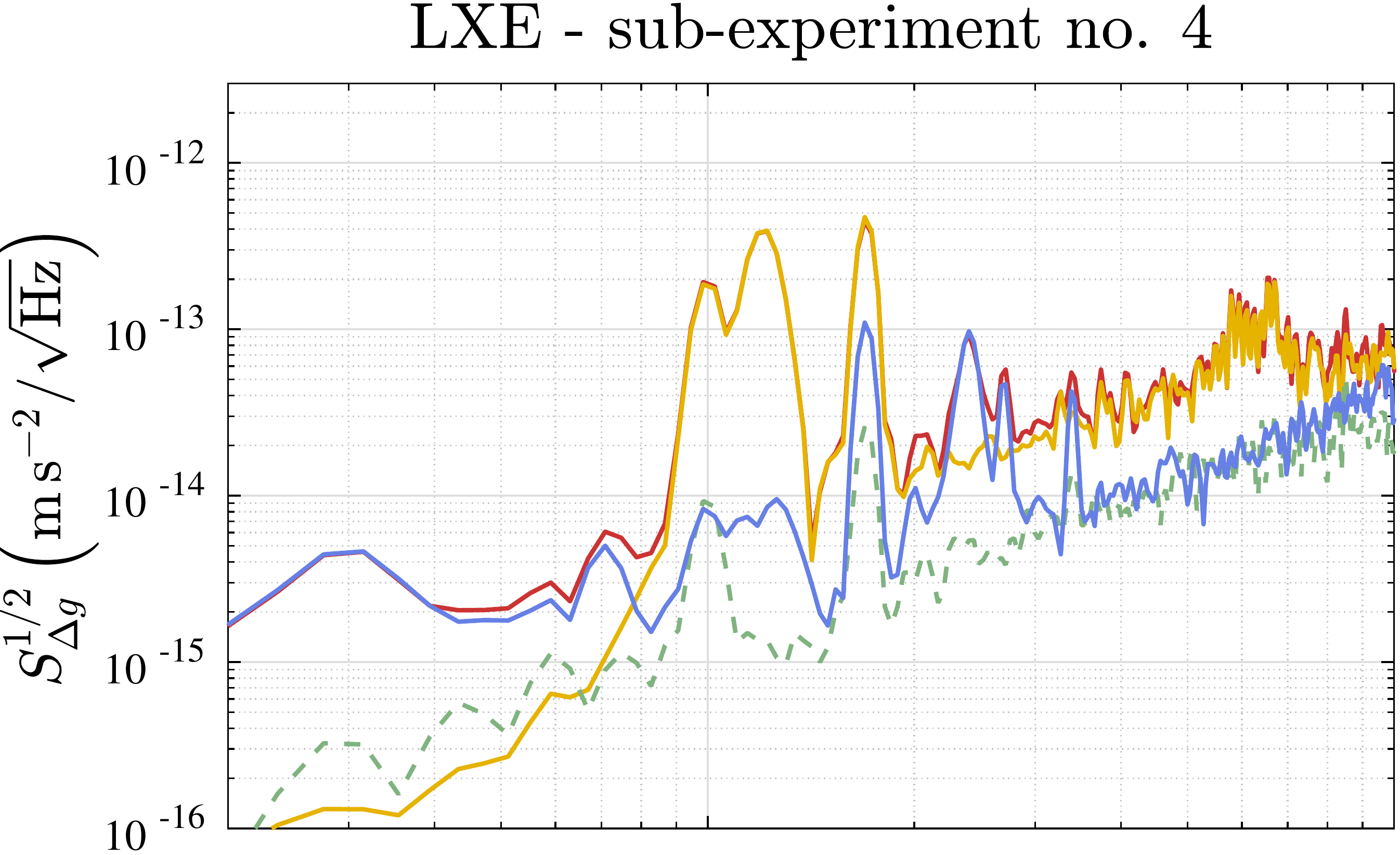}\,\quad
\includegraphics[scale=0.23,valign=t]{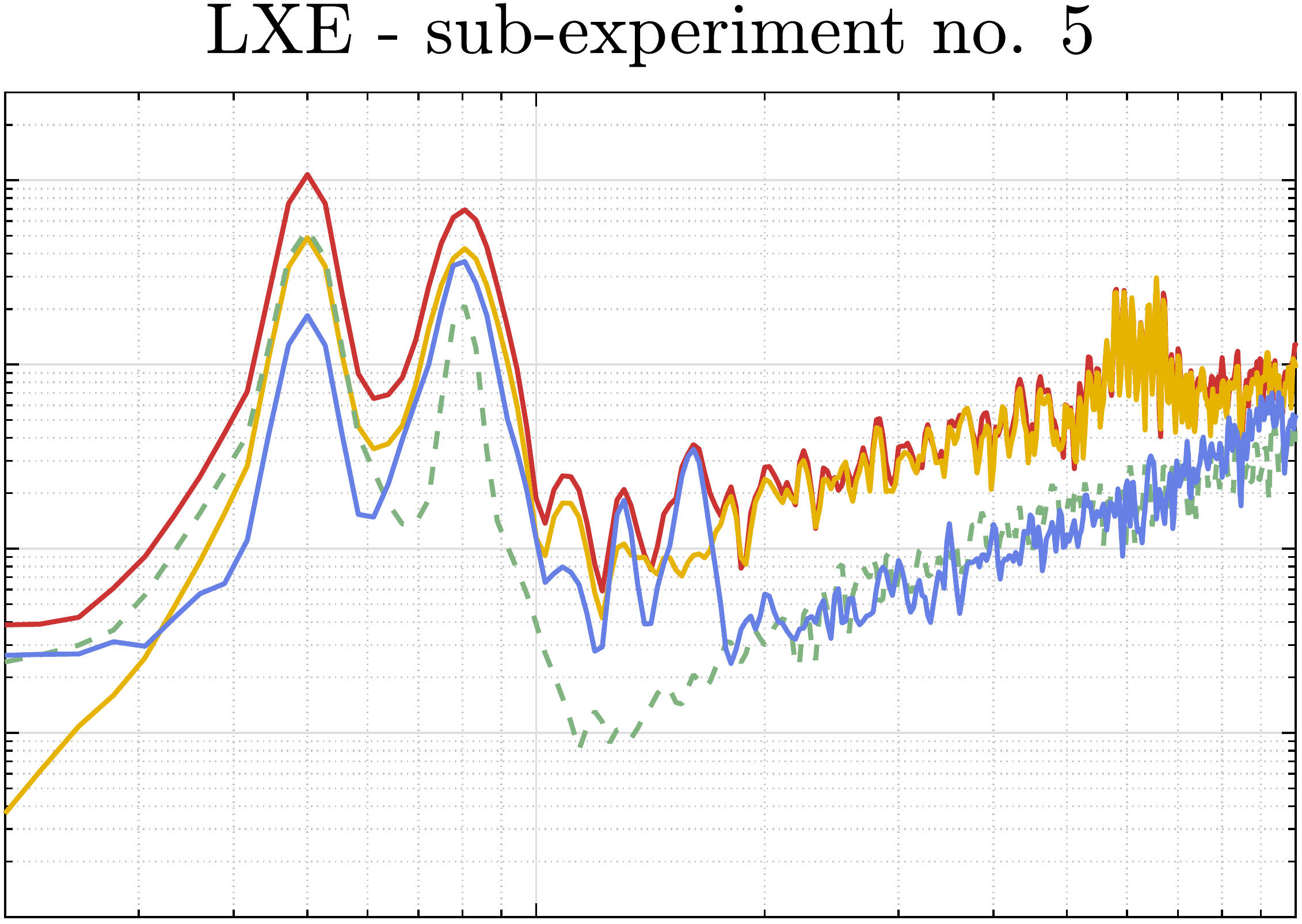}\quad
\includegraphics[scale=0.23,valign=t]{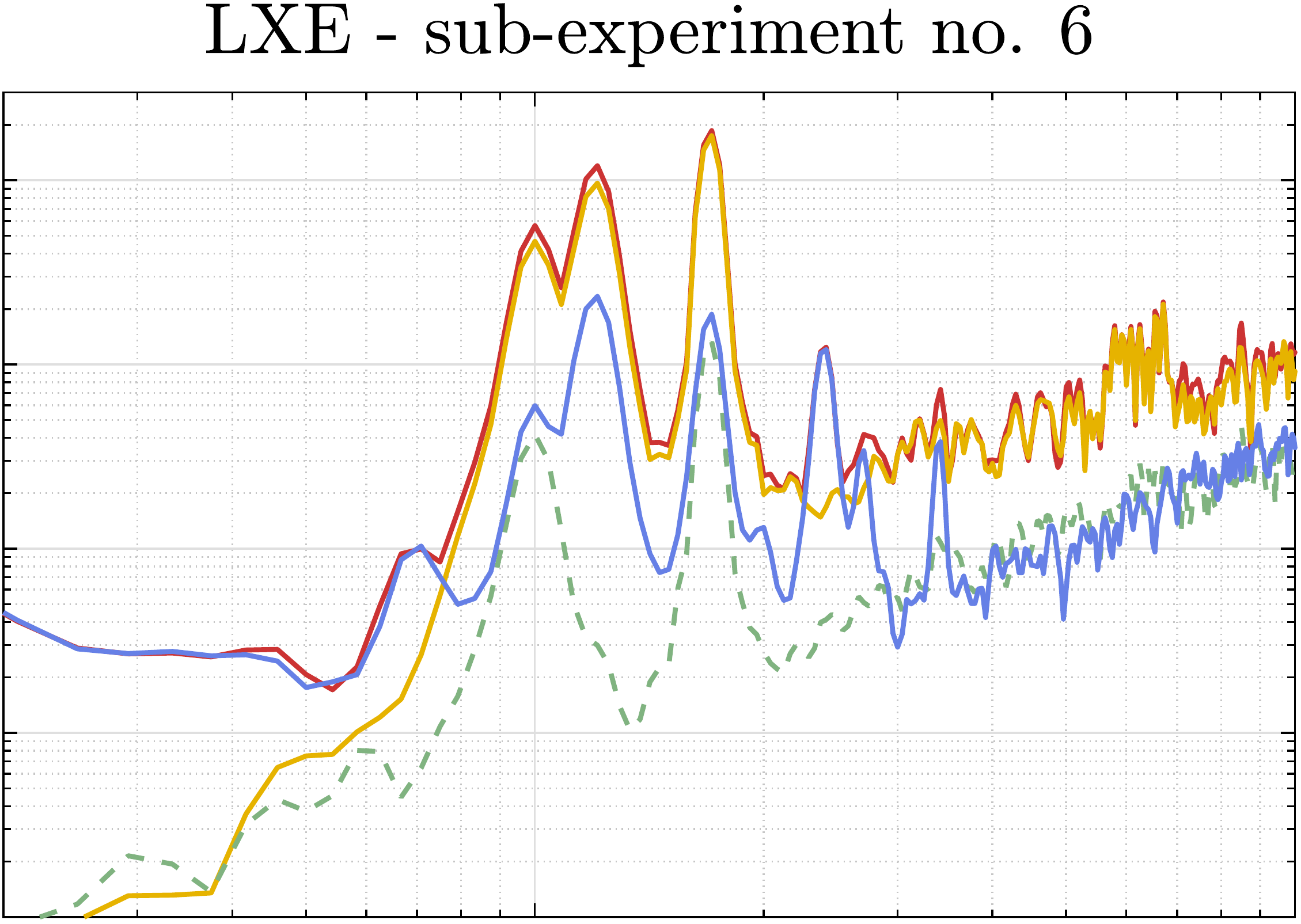} \\[1ex]
\includegraphics[scale=0.23,valign=t]{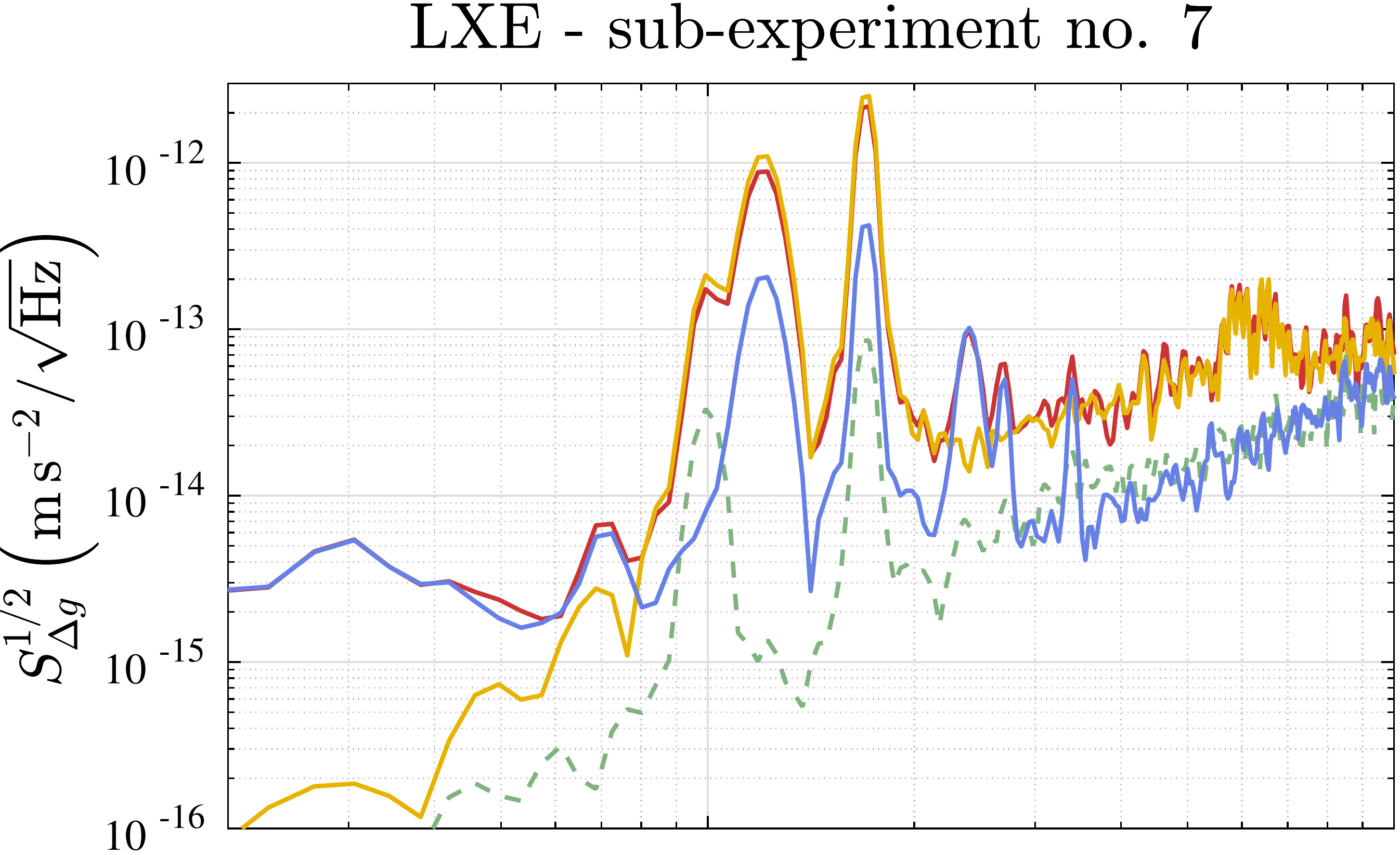}\,\quad
\includegraphics[scale=0.23,valign=t]{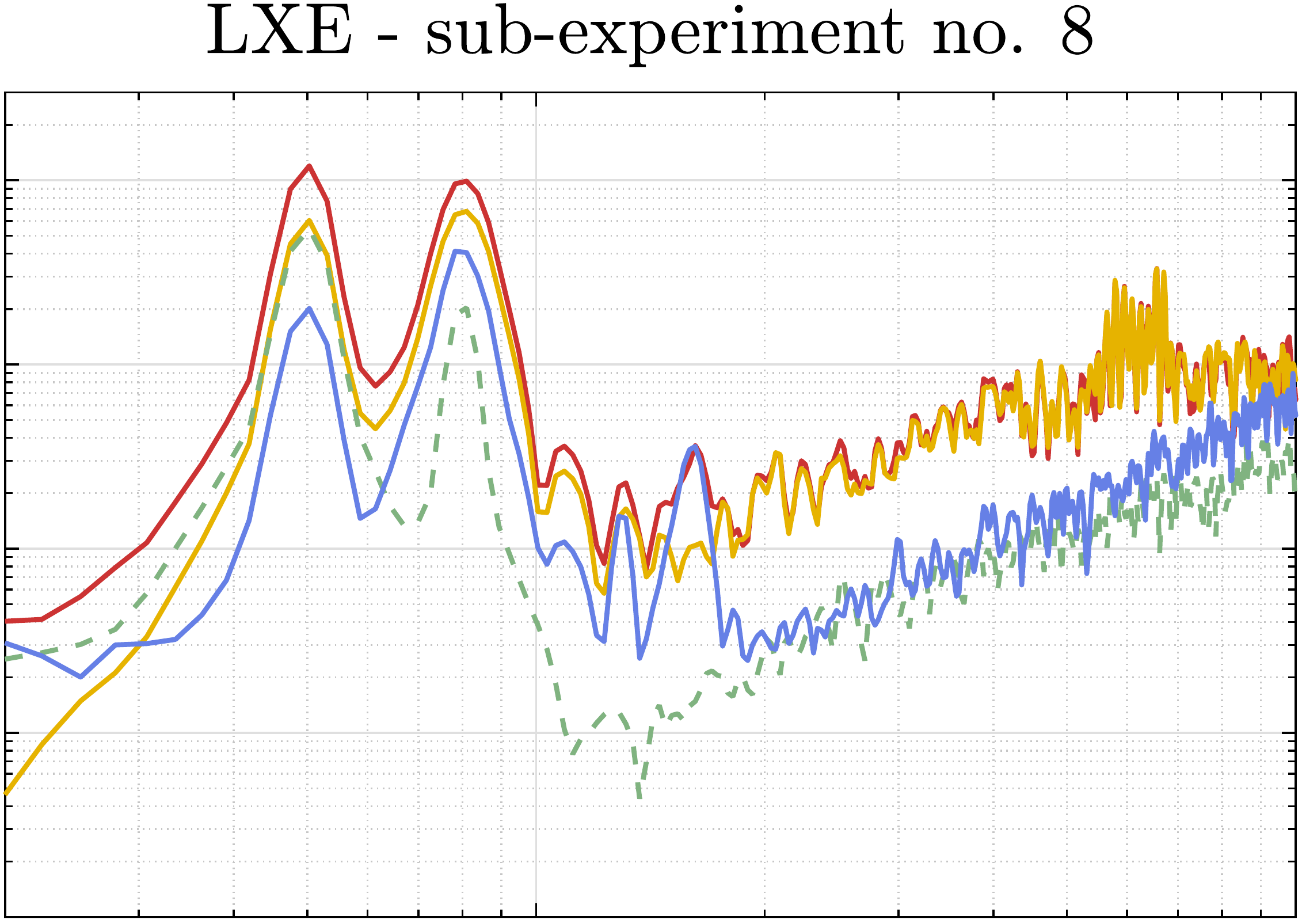}\quad
\includegraphics[scale=0.23,valign=t]{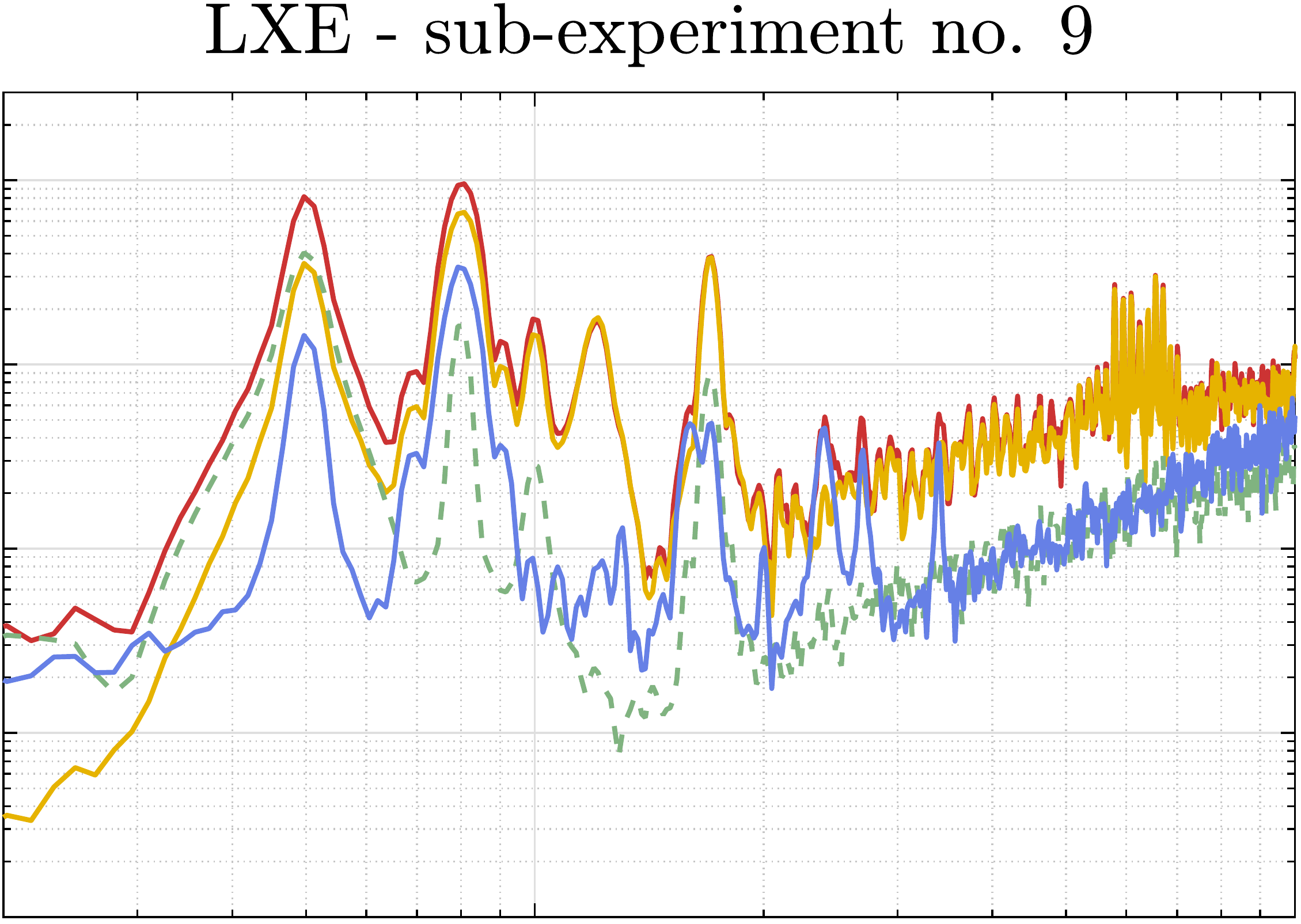} \\[1ex]
\,\ \includegraphics[scale=0.23]{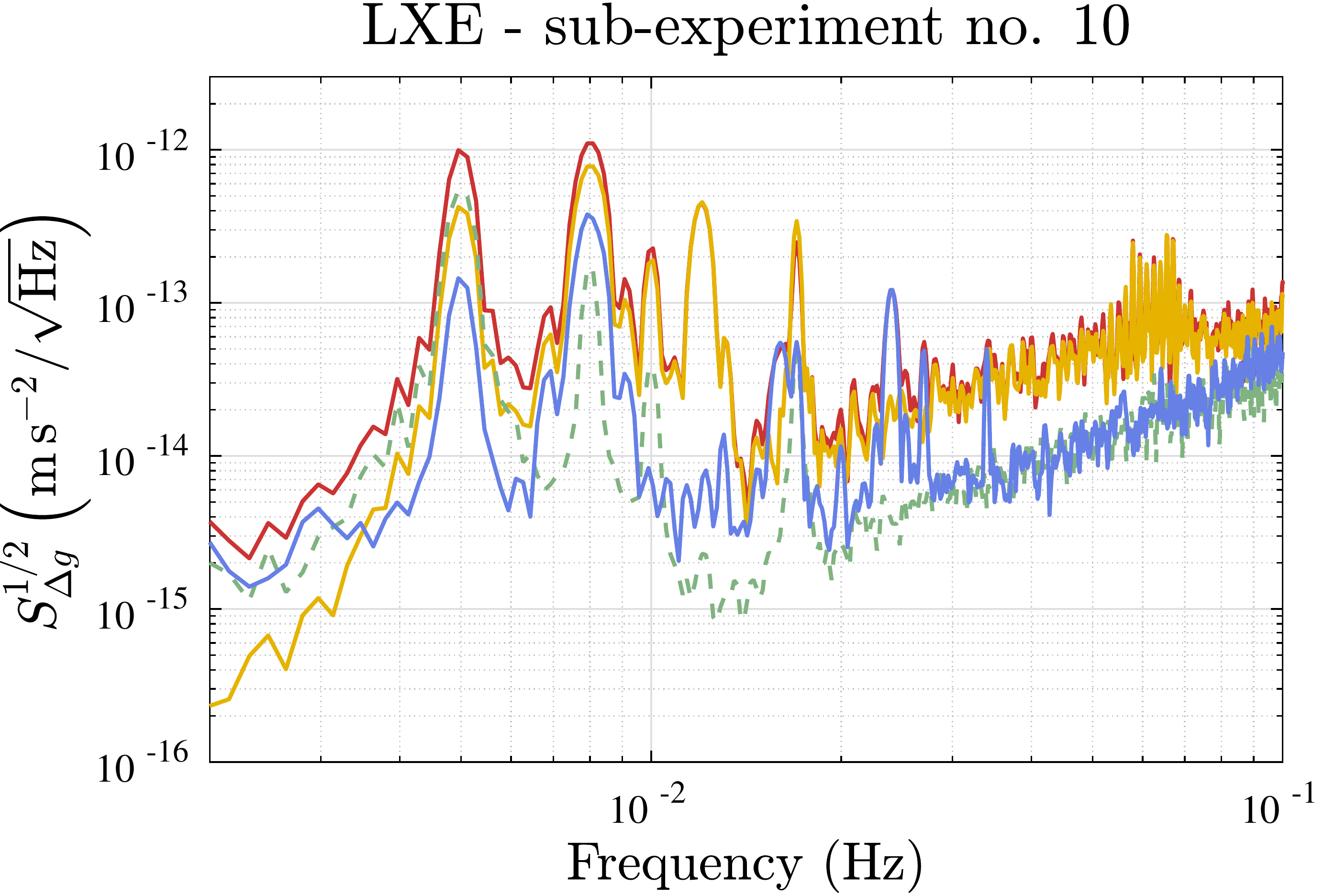}\,\,\
\includegraphics[scale=0.23]{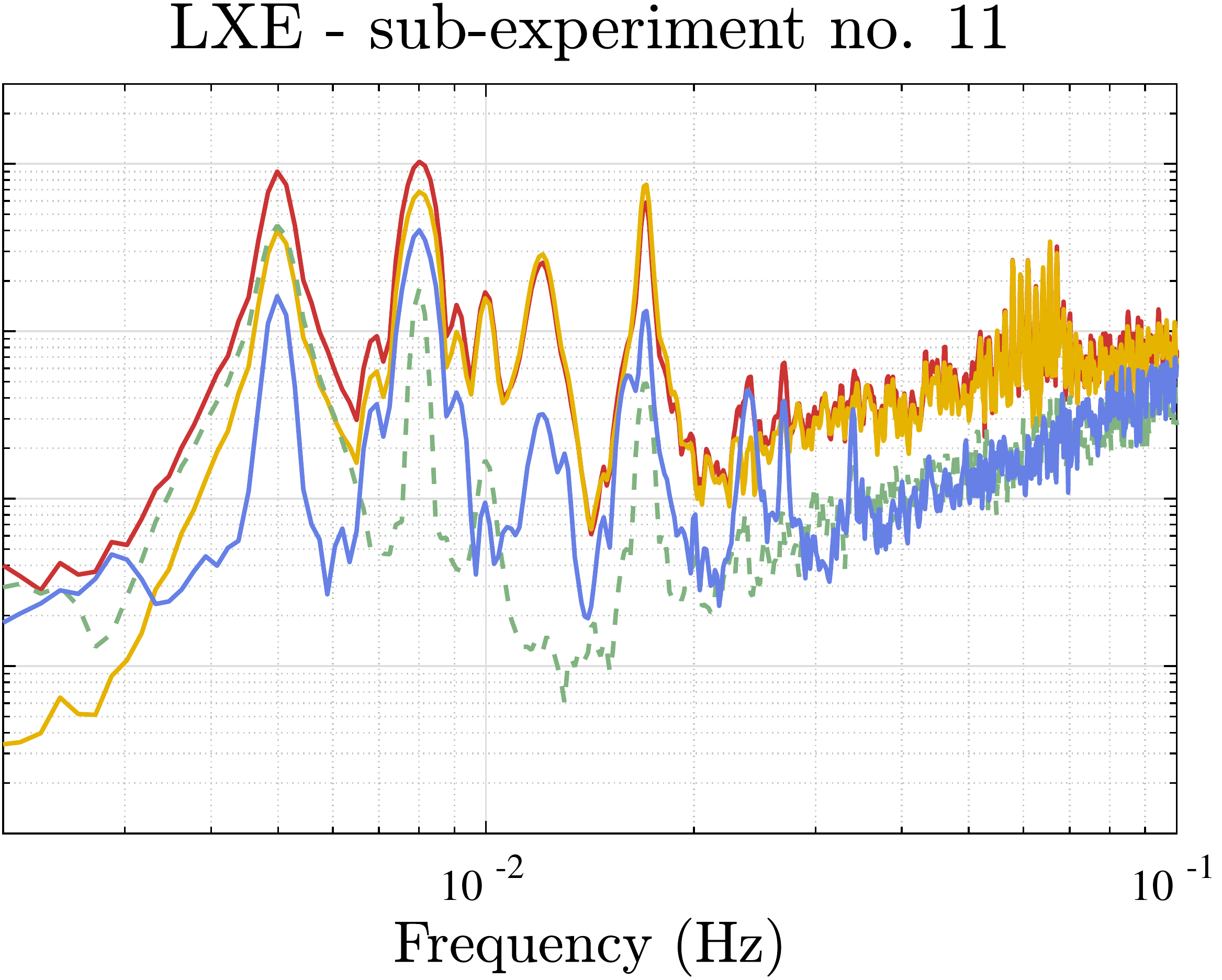}\hfill
\includegraphics[scale=0.23]{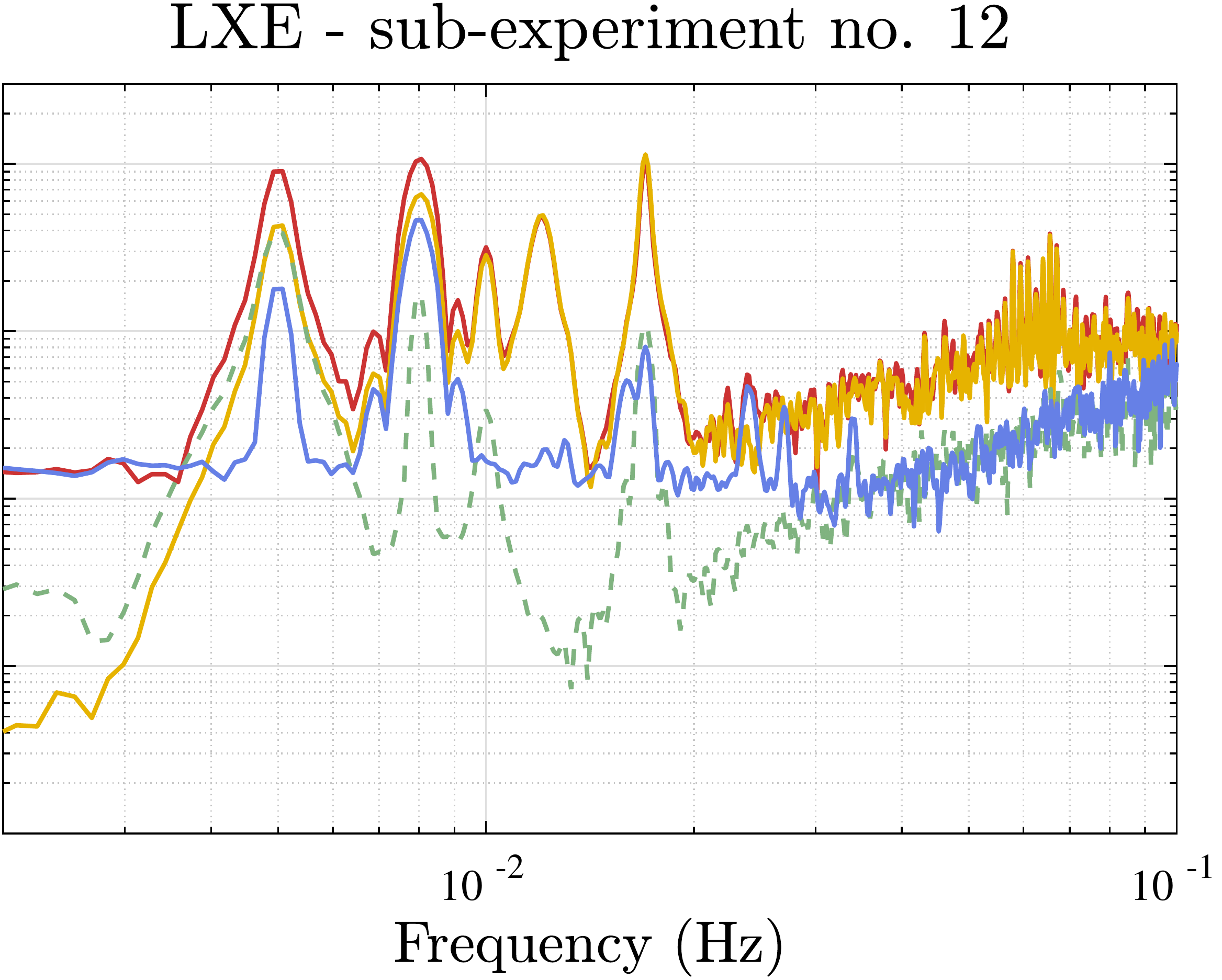} \\[2ex]
\includegraphics[scale=0.8]{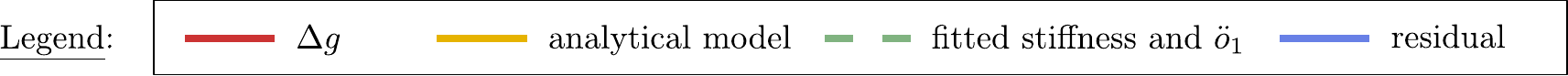}
\caption{Performance of the analytical model for the twelve sub-experiments of the LXE.
The red curve shows the \gls{ASD} of the observable $\Delta g$. The yellow curve shows the \gls{ASD} of the linear analytical model (Eq.~\eqref{eq:anamodel}). The dashed green line pictures the \gls{ASD} of the stiffness terms and the cross-coupling of residual \gls{S/C} motion along the optical axis fitted to the difference between $\Delta g$ and the analytical model. The blue curve shows the residual remaining after the subtraction of the analytical model and the fitted terms from the measurement.
The subtraction analytical model significantly reduces the noise at the full frequency range dominated by the TTL coupling, with the exception of the multiples of the injections frequencies, which are not covered by this linear model.}
\label{fig:LXE_performance_ana}
\end{figure*}

Also, the analytical model subtracted the noise at the injection frequencies well, see Fig.~\ref{fig:LXE_performance_ana}.
However, the residual at the injection frequencies is, in most cases, larger than the remaining noise after the subtracted fit model, compare e.g.\ the residuals in sub-experiment 6 in Figs.~\ref{fig:LXE_performance_fit} and~\ref{fig:LXE_performance_fit}). Still, the analytical model significantly reduced the \gls{TTL} noise in all sub-experiments.
Like the fit model, the analytical model performs slightly worse at the frequencies of the $z$-injections and naturally does not cover coupling at sums and multiples of or the differential injection frequencies.

In summary, both \gls{TTL} models, Eq.~\eqref{eq:fitmodel} and Eq.~\eqref{eq:anamodel}, allow subtracting the TTL noise sufficiently well. 
While the fit model subtracts the injected noise better, the presented analytical model explains additionally the dependency of the \gls{TTL} coupling on the \gls{TM} alignments in all sub-experiments.

\subsection{Comparison of the TTL models}
\label{sec:LXE_comparison}

The four \gls{TTL} coefficients $C_\varphi,\,C_\eta,\,C_y,$ and $C_z$
are contained by the fit and the analytical model. 
Hence, we can directly compare these coefficients with each other.
In the last section, we have seen that the coefficients provided by the fit approach yield a slightly better suppression of the \gls{TTL} noise. However, they do not provide information about the physical origin of the \gls{TTL} noise or how to suppress it.
On the other hand, the analytical coefficients directly correspond to the \gls{TM} alignments.
An agreement of both sets of coefficients within their error bars and expected variations would be a mutual confirmation of both \gls{TTL} coupling models. 

\begin{figure}
\flushright
\includegraphics[scale=0.29]{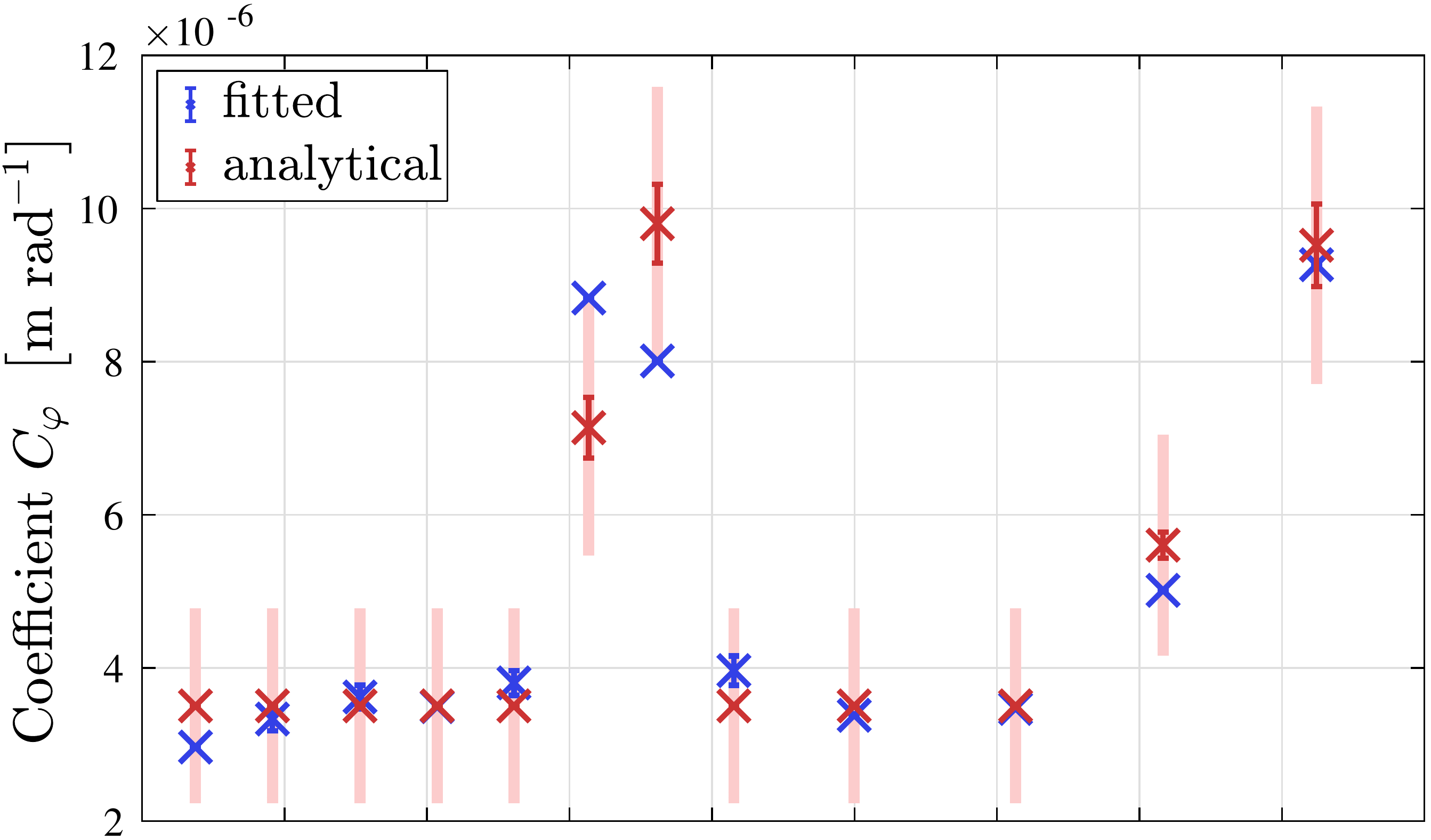}\,\,\ \\
\includegraphics[scale=0.29]{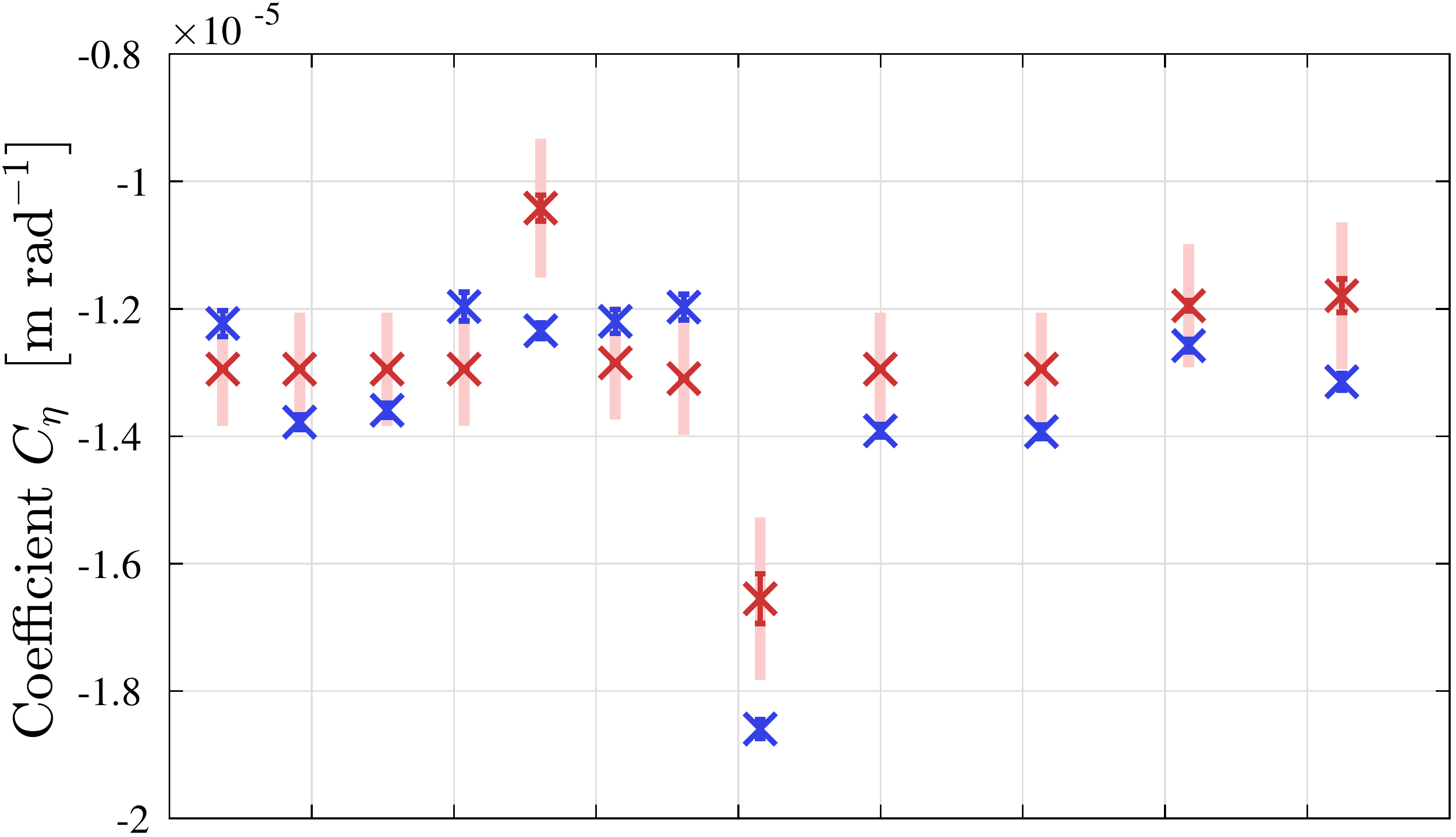}\,\,\ \\
\includegraphics[scale=0.29]{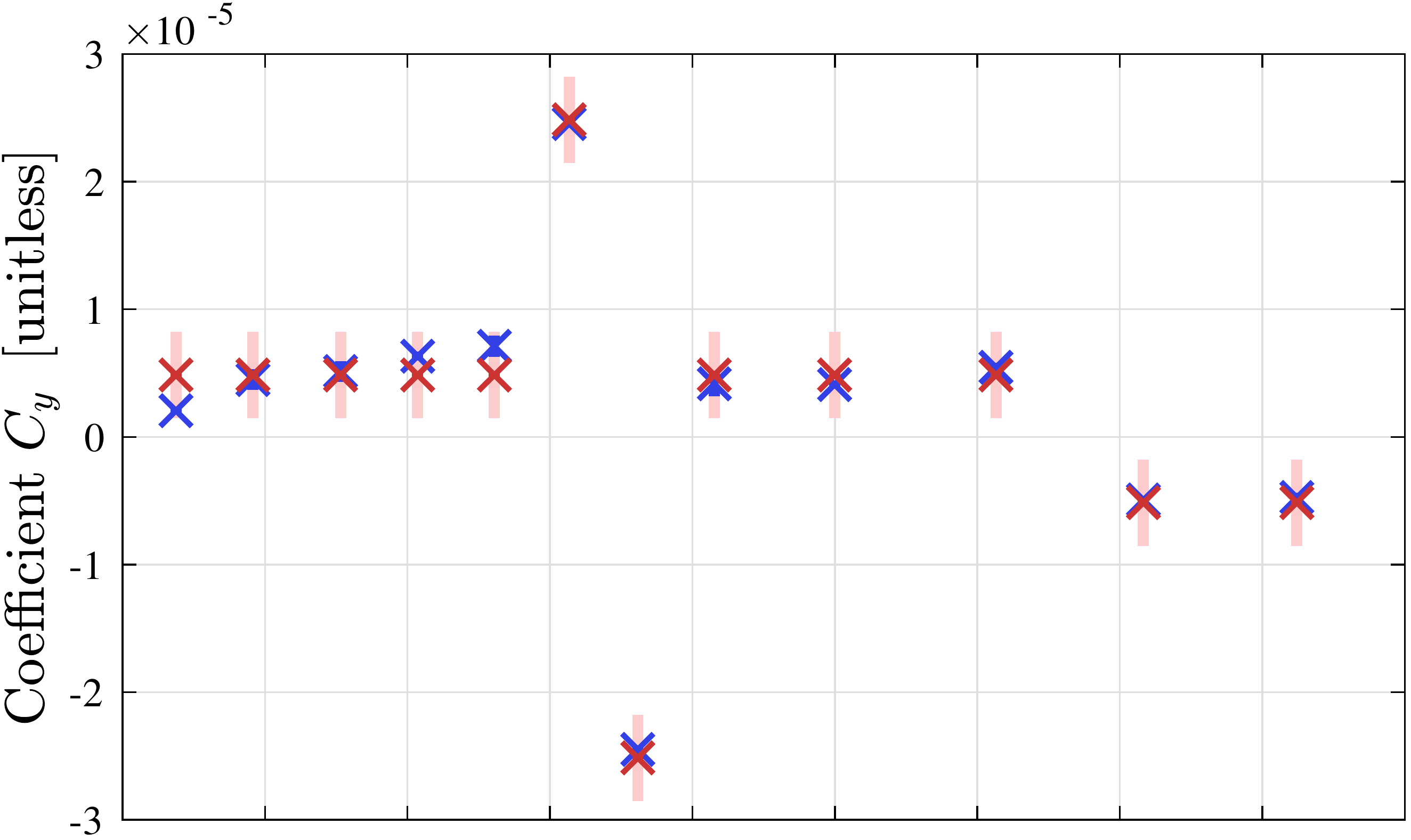}\,\,\ \\
\includegraphics[scale=0.29]{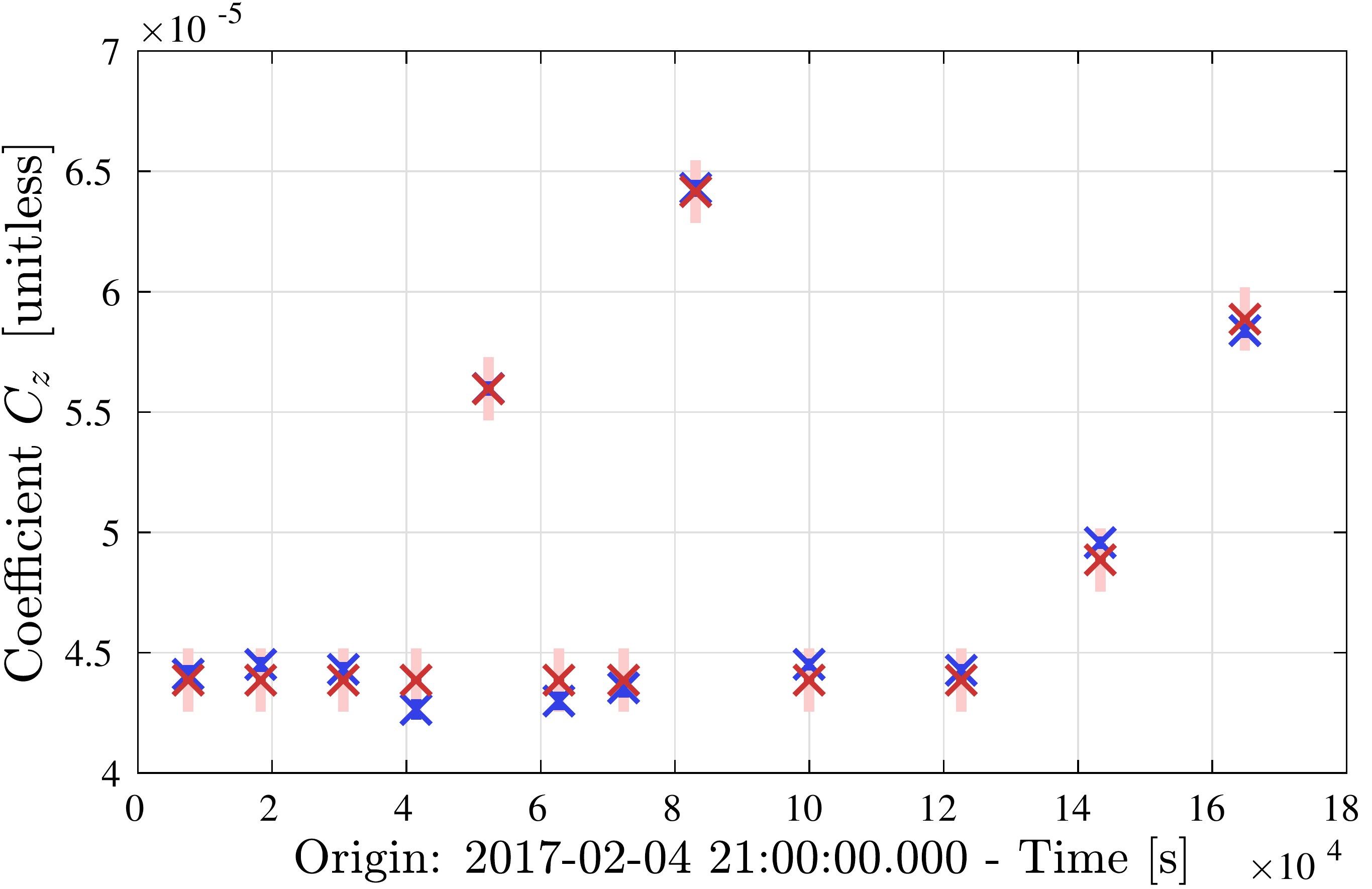} 
\caption{Comparison of the fitted (blue) and analytically derived (red) coupling coefficients for the 12 sub-experiments of the LXE. They are representatively plotted in the mid of the timespan of the experiments each. The fit was conducted in the frequency range from 2 to 70\,mHz. The error bars of the fit are plotted.
The error bars of the analytical coefficients correspond to their uncertainties given in Eq.~\eqref{eq:anamodel_C}. 
Moreover, we show the respective additional double root-mean-square of the fit deviations within a noise run to cover unmodelled instabilities (see Sec.~\ref{sec:TTLmodel_ana}) in light red.
Times are given in UTC. Most coefficients match when considering the error bars and expected variations of the fit results. Only in the case of the $C_{\eta}$ coefficient we find few mismatches.}
\label{fig:LXE_comparison}
\end{figure}

The comparison of both sets of coefficients is illustrated in Fig.~\ref{fig:LXE_comparison}.
In this figure, we see that the lateral coefficients $C_y$ and $C_z$ match to a high degree in all twelve sub-experiments.
In the case of the angular coupling coefficients, we see a deviation in some experiments.

For the $C_\varphi$ coefficient, the values of the 6th and 7th sub-experiment do not match well. At these times, the \glspl{TM} were realigned in $\varphi$. 
Compared to the fitted result, the analytical $C_\varphi$ coefficient yields a slightly smaller value in the case of the rotation of TM2 and a larger value in the case of the rotation of TM1.
The differences are small and are mostly covered by the variations of the fitted coefficients seen during the noise run mid-February. In the analytical model they could result from incorrect assumptions concerning the setup parameters or their uncertainties. For example, a longitudinal displacement of the centre of rotation by 1.4\,cm could explain these differences. This shift would be twice as big as the uncertainty assumed for this parameter. 

The differences of the $C_\eta$ coefficient are, for the most part, of another origin. 
From the analytical model (Eq.~\eqref{eq:anamodel_C}), we would have expected the coefficient to be (almost) constant in all experiments without pitch rotations of the \glspl{TM} (compare Tab.~\ref{tab:longxtalk}).
However, out of these eight experiments, we find four larger fit-coefficients in the cases where only $y$-injections were applied (the 1st, 4th, 6th and 7th sub-experiment). 
The four other experiments (the 2nd, 3rd, 9th and 10th sub-experiment) include $z$-injections, and the corresponding fit-coefficients distribute about a lower level. Additionally, we find the fit-coefficients of the other four experiments with $z$-injections but also \gls{TM} set-point in pitch (the 5th, 8th, 11th and 12th sub-experiment) below their analytical correspondence.
No such characteristic of the fit-coefficients was discovered for the $C_\varphi$ coefficients.
Thus, we suspect the $z$-injections themselves altered the level of the $C_\eta$ coefficients and investigate this in the following paragraphs.
This means that the coupling coefficient for the stimulated motion at the injection frequencies is different than the coupling in the bump (increased noise at frequencies above 20\,mHz), where the \gls{S/C} jitters freely.

\subsubsection{Adaptions of the fit settings}
\label{sec:LXE_comparison_fit-settings}

\begin{figure}
\flushright
\includegraphics[scale=0.29]{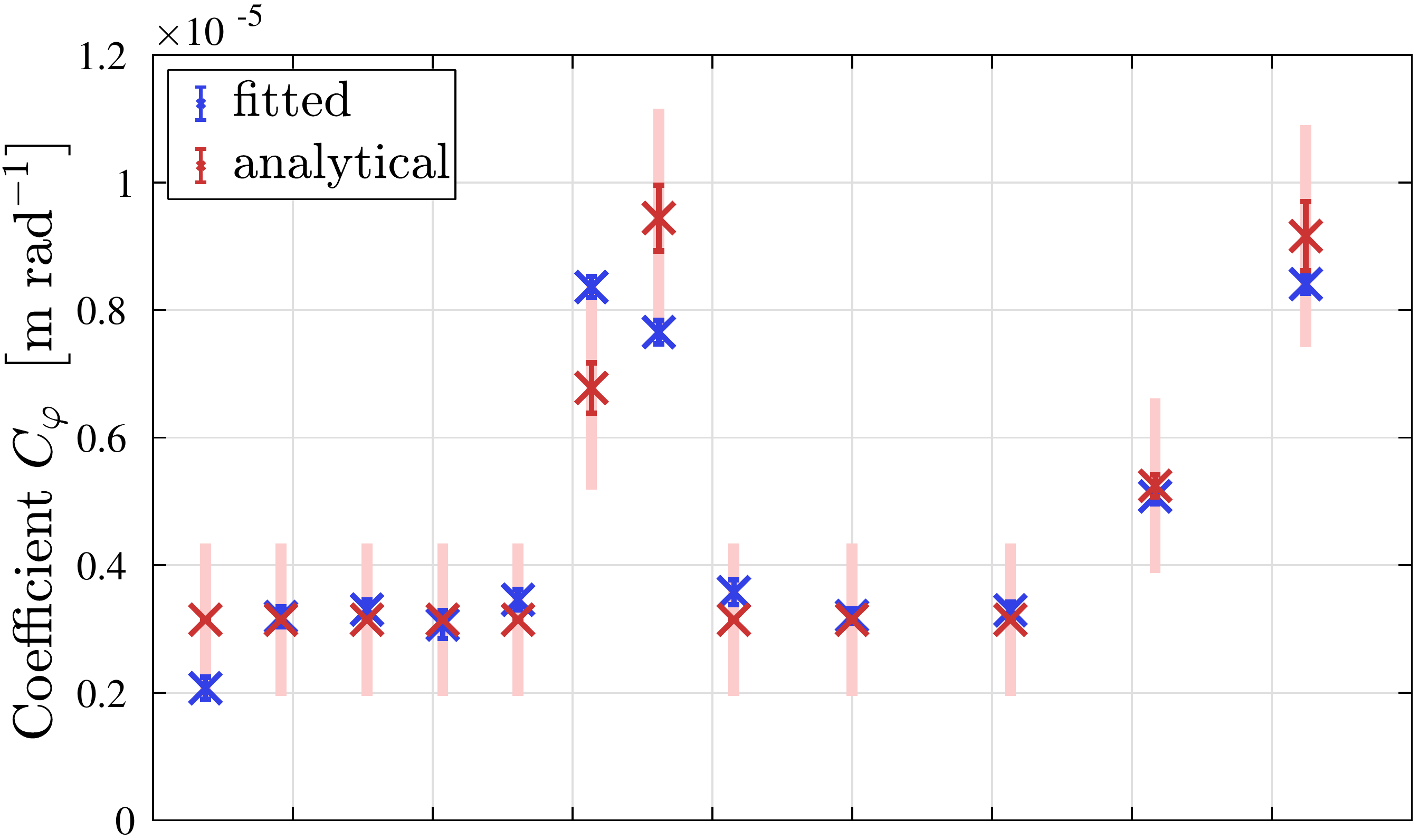}\,\,\ \\
\includegraphics[scale=0.29]{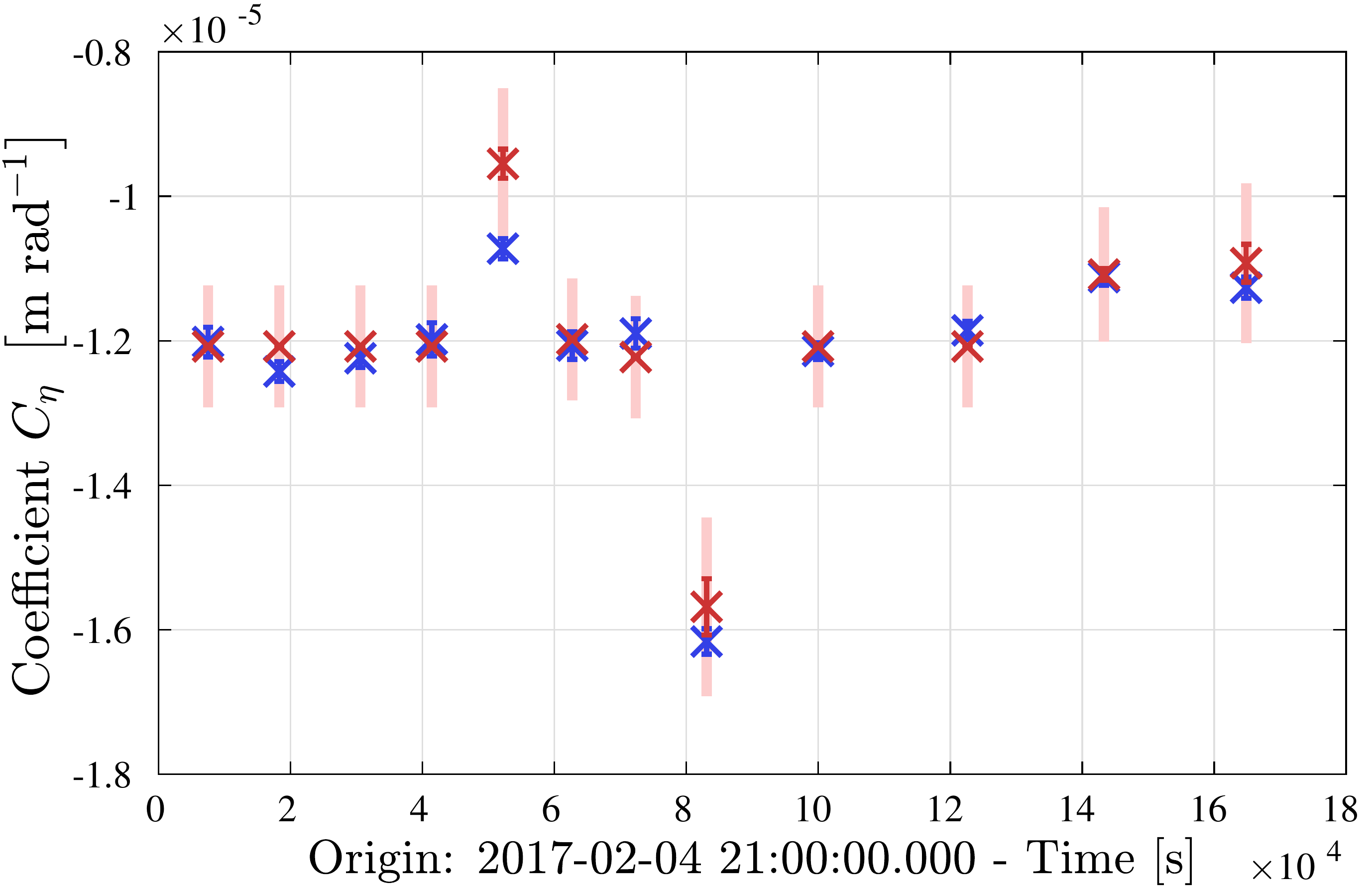} 
\caption{Same as Fig.~\ref{fig:LXE_comparison} but with the fit using a frequency range restricted to the bump only (i.e. 20 to 70\,mHz).
The results of this fit is shown in blue. 
The distribution of the analytical coefficients is again plotted in red. Their offset was adjusted to the new mean of the fitted coefficients in experiments in which they are not expected to (significantly) change.
Compared to Fig.~\ref{fig:LXE_comparison}, the fitted $C_\eta$ coefficients do not split into two levels. Therefore, the fitted and analytical coefficients match better.} 
\label{fig:LXE_comparison_bump}
\end{figure}

For comparison, we repeated the fit restricting the considered frequency span to the bump frequencies, i.e.\ 20\,mHz to 70\,mHz (the original upper bound).
This changed the lateral coefficients only negligibly. The changes of the angular coupling coefficients are shown in Fig.~\ref{fig:LXE_comparison_bump}.
In this figure, the fitted $C_\eta$ coefficients do not distribute about two levels and almost match the analytical predictions, which confirms our assumption that the $z$-injections affect the fit result.
The absolute difference of the $C_\eta$ coefficients, fitted for frequency regimes with and without the injection frequencies, is almost 2\,$\mu$m/rad in experiments with $z$-injections.

\begin{figure}
\flushright
\includegraphics[scale=0.29]{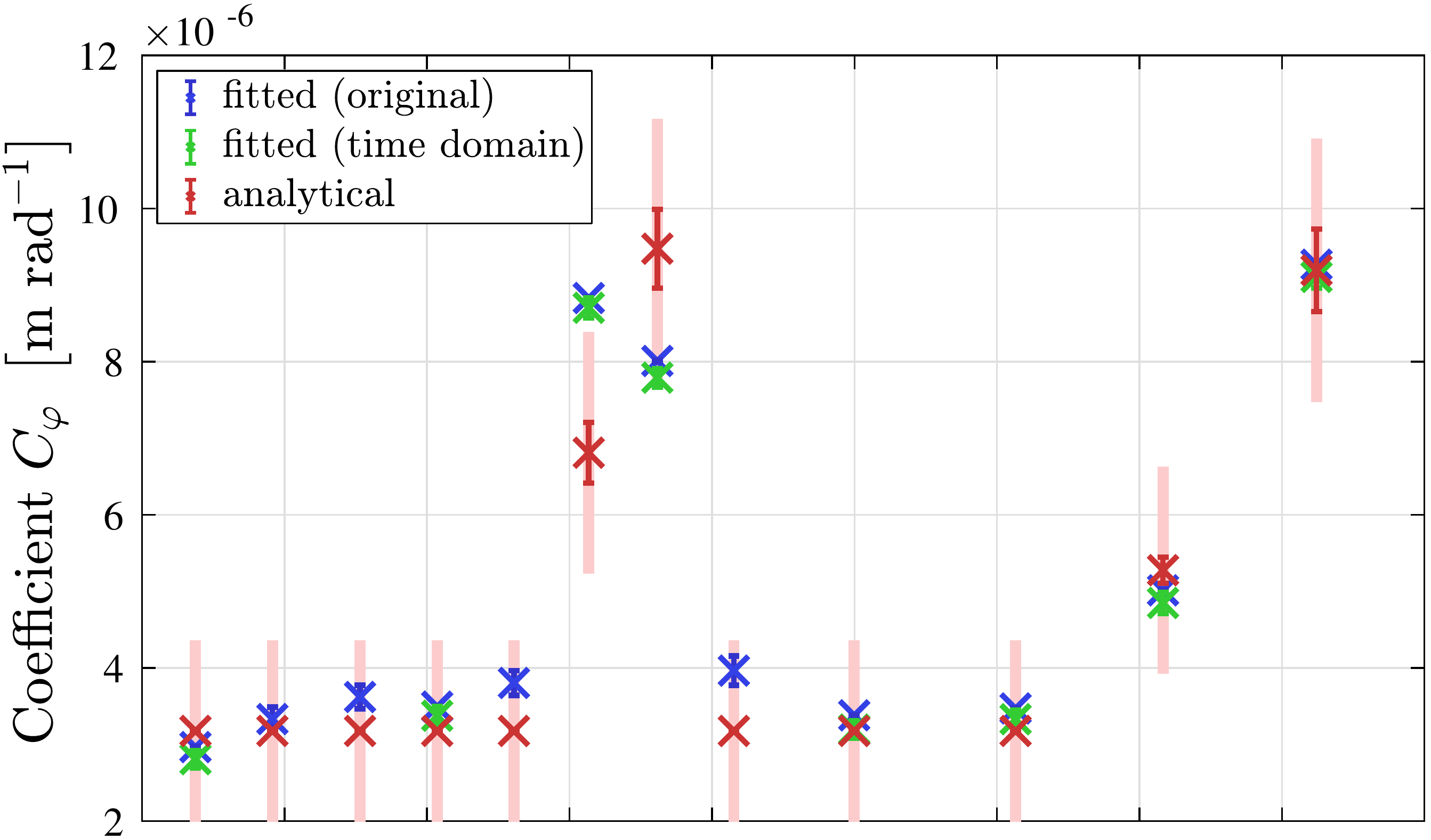}\,\,\ \\
\includegraphics[scale=0.29]{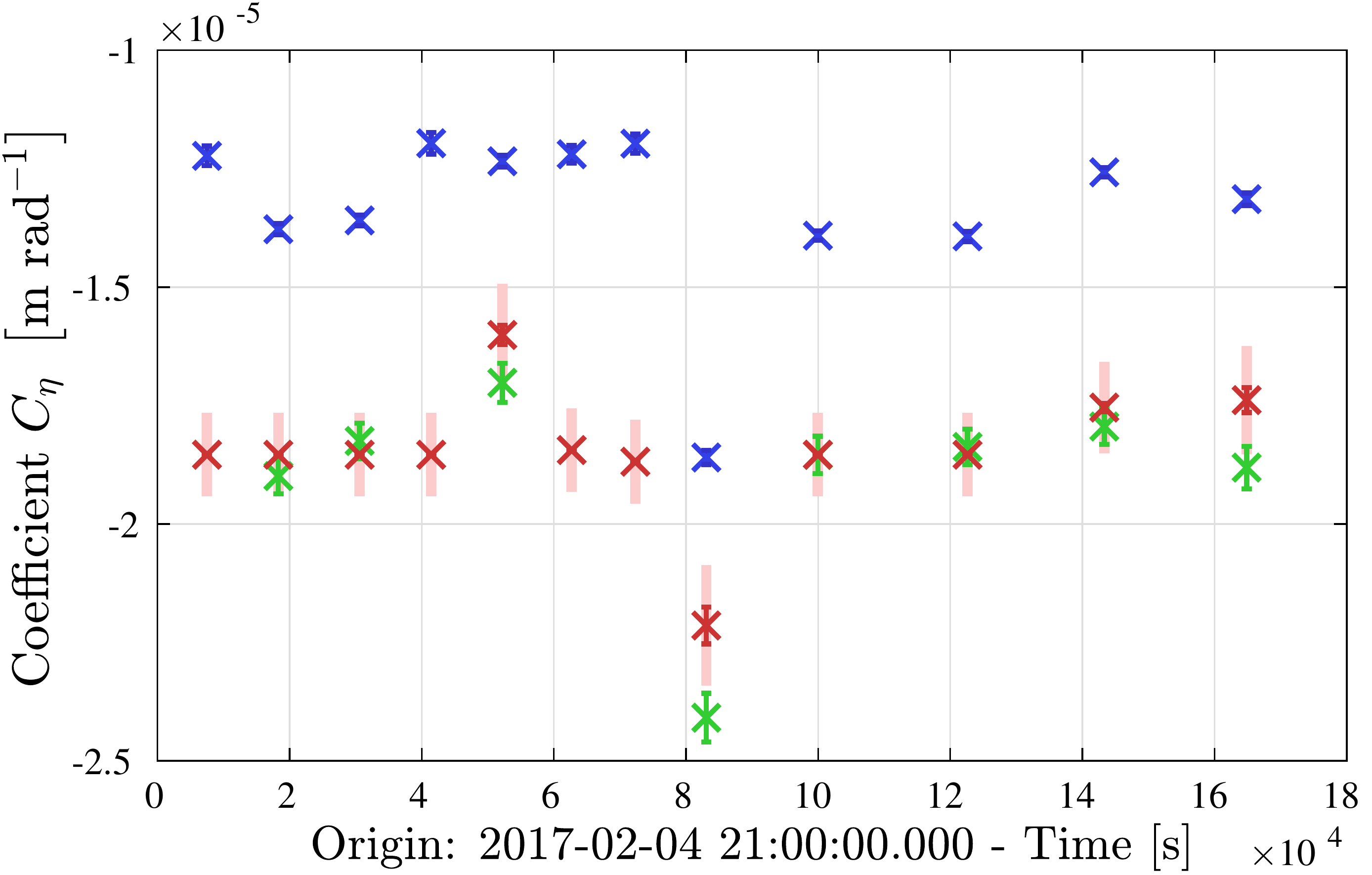} 
\caption{Comparison of the fitted (injections only, green) and analytically derived (red) coupling coefficients for the 12 sub-experiments of the LXE. 
The time domain fit for the injection frequencies is further described in Sec.~\ref{sec:LXE_comparison_fit-settings}.
Compared to Fig.~\ref{fig:LXE_comparison_bump}, the resulting $C_\eta$ coefficients distribute about a significantly lower level, while the $C_\varphi$ coefficients did not significantly change.
The distribution of the analytical coefficients is plotted in red. Their offset was adjusted to the mean of the fitted coefficients (green) in experiments in which they are not expected to (significantly) change.
The blue coefficients correspond to the original fit result shown in Fig.~\ref{fig:LXE_comparison}.} 
\label{fig:LXE_comparison_inj}
\end{figure}

We further compared these results with the numbers we find when fitting the coefficients 
for the injection frequencies only. 
We chose to fit the oscillations in the time domain (least-squares routine) by finding a model
\begin{align}
f(t) 
= C_s\,\sin(\Omega_\text{inj}\,t)
+ C_c\,\cos(\Omega_\text{inj}\,t)
+ \sum_{i=0}^4 c_i\,t^i \,,
\label{eq:model_timedomain}
\end{align}
with the time $t$ and the injection frequency $\Omega_\text{inj}$, for the measurements of $\Delta g$ and the exceeded coupling parameter $\ddot{\bar{j}}$, $j\in\{\varphi,\,\eta,\,y,\,z\}$.
The coupling coefficients correspond to the quotient of the two fitted $C_s$'s.
The results are plotted in Fig.~\ref{fig:LXE_comparison_inj} together with the fit results in the original frequency domain and the analytical computations. We find that the $C_\eta$ coupling coefficients for the injection and bump frequencies differ even by approximately  6\,$\mu$m/rad.
This is in the same order of magnitude as the fitted coefficients themselves.

We do not see a comparable splitting into different levels in the case of the $C_\varphi$ coefficients (compare $C_\varphi$-graphs in Figs.~\ref{fig:LXE_comparison} and~\ref{fig:LXE_comparison_bump}).
Thus, the split of the $C_\eta$ coefficients into two levels must be related to a characteristic difference between the jitter coupling in the $xy$- and the $xz$-plane.
We discuss these in detail in the following subsection.

\subsubsection{Origin of the different $C_\eta$ coupling levels}
\label{sec:LXE_comparison_Ceta_discussion}

First, it was shown in \cite{LPFana22} that the lateral coupling coefficients almost only depend on the differential angular alignment of the two \glspl{TM} in \gls{LPF}.
Thus, we cannot only compute the coupling coefficients from the \gls{TM} alignment angles but also derive the differential \gls{TM} angles from the coupling coefficient fits. 
This is discussed in further detail in \cite{LPFstabi23}. 
Using the mean levels of the fitted coefficients in Fig.~\ref{fig:LXE_comparison} and their relation to the \gls{TM} alignments in Eqs.~\eqref{eq:anamodel_Cy} and~\eqref{eq:anamodel_Cz}, we find an approximate differential pitch angle $\Delta \eta = \eta_2-\eta_1 \approx -45\,\mu$rad, which is nine times as big as the differential yaw angle $\Delta \varphi = \varphi_2-\varphi_1 \approx 5\,\mu$rad. 
Therefore, any \gls{TTL} coupling differences for the injection and bump frequency span that depend on the \gls{TM} alignments were different for yaw and pitch jitter.
A level change of the $C_\varphi$ coefficient of a ninth of the magnitude of the observed $C_\eta$ coefficient changes is smaller than the alternations due to the fit itself. Hence, we would not see it in Fig.~\ref{fig:LXE_comparison}, which could explain the absence of levels in the $C_\varphi$ case.

The equations presented in this paper show the \gls{TTL} coupling effect for a \gls{S/C} rotation about its centre of rotation.
This is representative for the \gls{TTL} noise measured during noise runs.
However, in the case of the applied injections during the \gls{LXE}, the \gls{S/C} rotated about a pivot defined by the control loops. This pivot was the geometrical centre of the optical bench.
Both centres of rotation differ by less than a millimetre in $x$- and $y$-direction. However, in $z$-direction, the \gls{S/C} centre of mass lies several centimetres below the optical bench.
If computing the differences of the angular coupling coefficients \cite[suppl.\ material]{LPFana22}, we find the \gls{TM} alignment dependent terms of $C_\varphi$ and $C_\eta$ to change only in the third relevant digit.
Using the numbers of the differential \gls{TM} alignment angles presented above, we deduce that the different centres of rotation can explain at most 10\,\% of the coupling coefficient differences. 

\gls{S/C} jitter in $\eta$ about the lowered centre of rotation at the bump frequencies would also make the \gls{S/C} jitter along the $x$-axis. Thus, we further investigated whether a correlation between the $\eta$- and $o_1$-jitter caused the $C_\eta$ coefficient changes. 
Therefore, we repeated the fit using the frequency range of the bump without the $C_{o_1}$ coefficients. 
This attempt yielded only minimal coefficient changes.

To preserve the stability of the \gls{LPF} satellite, its control scheme \cite{Schleicher2018} not only commanded the \gls{S/C} response to \gls{TM} displacements but also rotated the \glspl{TM} with respect to the \gls{S/C}. 
This control loop was implemented at low frequencies. Thus, it was too slow to play a role in the analysis of \gls{TTL} coupling at the frequencies of the bump.
However, for the injections, we can expect \gls{TM} rotations of approximately 10\,\% of the measured \gls{S/C} jitter but of inverse sign. 
The \gls{DWS} measurements were the sum of both rotations. 
These \gls{TM} rotations would alter the angular coupling coefficients for the injections compared to the bump.
The exact change depends on the absolute alignment angles of the \glspl{TM}, which are unknown.
We estimated that the mechanism could only explain the $C_\eta$ coefficient levels if the \glspl{TM} were nominally rotated by more than $100\,\mu$rad in pitch but not in yaw. These large angles are very unlikely.
Therefore, \gls{TM} rotations were not the only reason for the $C_\eta$ coefficient levels but could explain parts of it.

As we will show in Sec.~\ref{sec:LXE_noise-contributors}, the stiffness terms of the fit model (Eq.~\eqref{eq:fitmodel}) contribute significantly at the injection frequencies but are almost negligible at the bump frequencies. 
Thus, they potentially altered the fit results of the coefficient in the same plane if injections were applied.
However, the relevance of the stiffness terms in our analysis remains unknown since we cannot study them on their own: If injections were applied, the terms $\bar{y},\,\bar{z}$ oscillate at the same frequency as their second derivatives $\ddot{\bar{y}},\,\ddot{\bar{z}}$.

Fig.~\ref{fig:LXE_performance_fit} showed that the linear fit model did not cover the noise peaks at the sums of the injections frequencies, which result from higher-order \gls{TTL} coupling. Since these peaks are within the fit regime, the question arises whether they alter the fit results.
We discuss higher-order coupling in Sec.~\ref{sec:LXE_2nd-order}.
There, we show that extensions of the fit model to higher-order coefficients do not significantly alter the linear coupling coefficients.
Thus, the coefficient changes were not related to the higher-order effects.

In summary, we have learned that different coupling coefficients explain the $C_\eta$ jitter at the injection and the bump frequencies.
Also, we found a few mechanisms that yield a deviation of the $C_\eta$ coupling coefficient fit for these frequencies.
Neither of these mechanisms could explain the coefficient changes we found alone. Therefore, we assume that a combination of several effects sums up to the observed deviations.
\\

All together, we could show the fitted and analytical \gls{TTL} coupling coefficients match to a high degree in the absence of injections, i.e.\ during noise runs.
We make use of this result in Sec.~\ref{sec:TMalignments}.

\subsection{Noise contributors}
\label{sec:LXE_noise-contributors}

\begin{figure*}
\centering
\includegraphics[scale=0.23,valign=t]{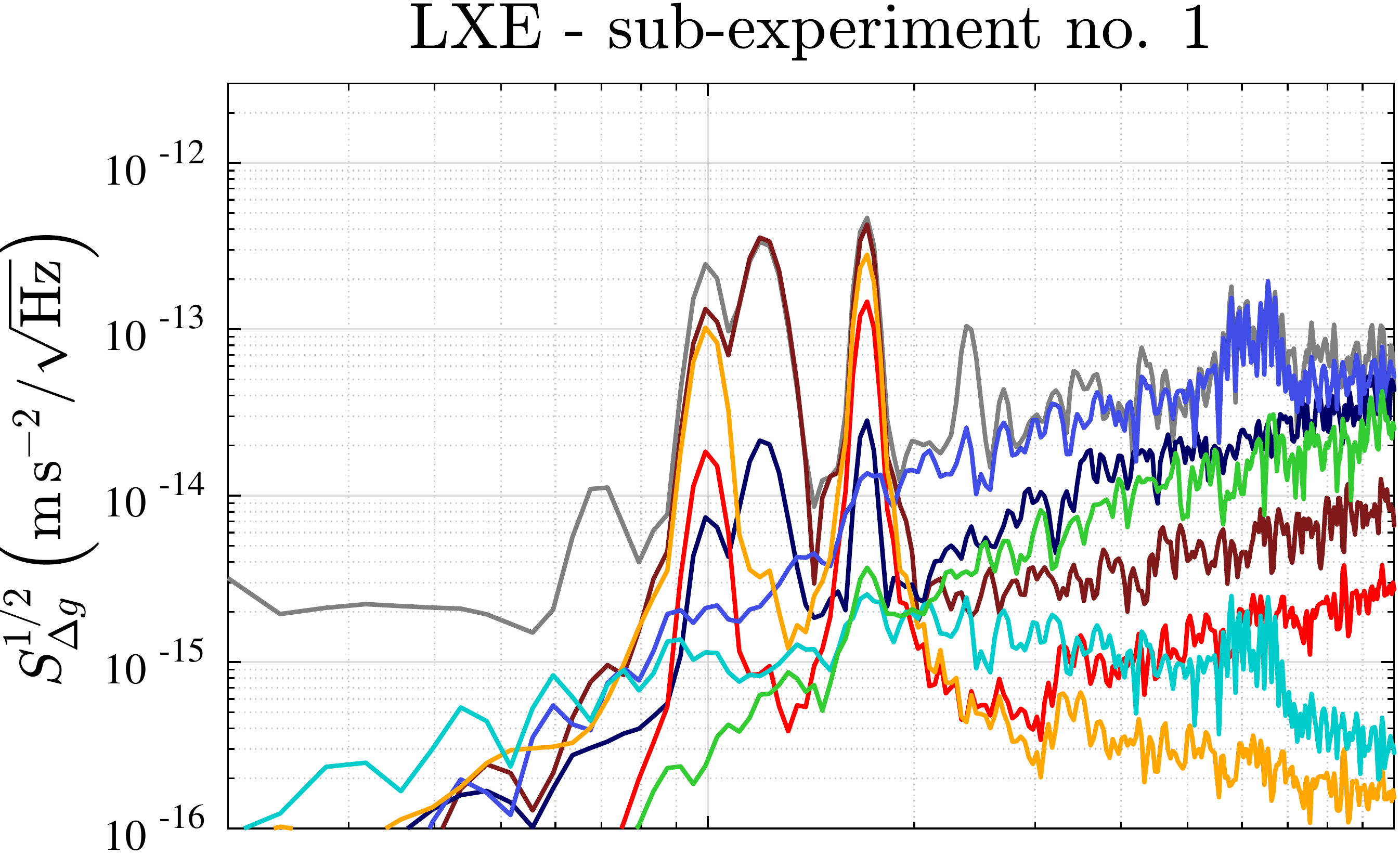}\,\quad
\includegraphics[scale=0.23,valign=t]{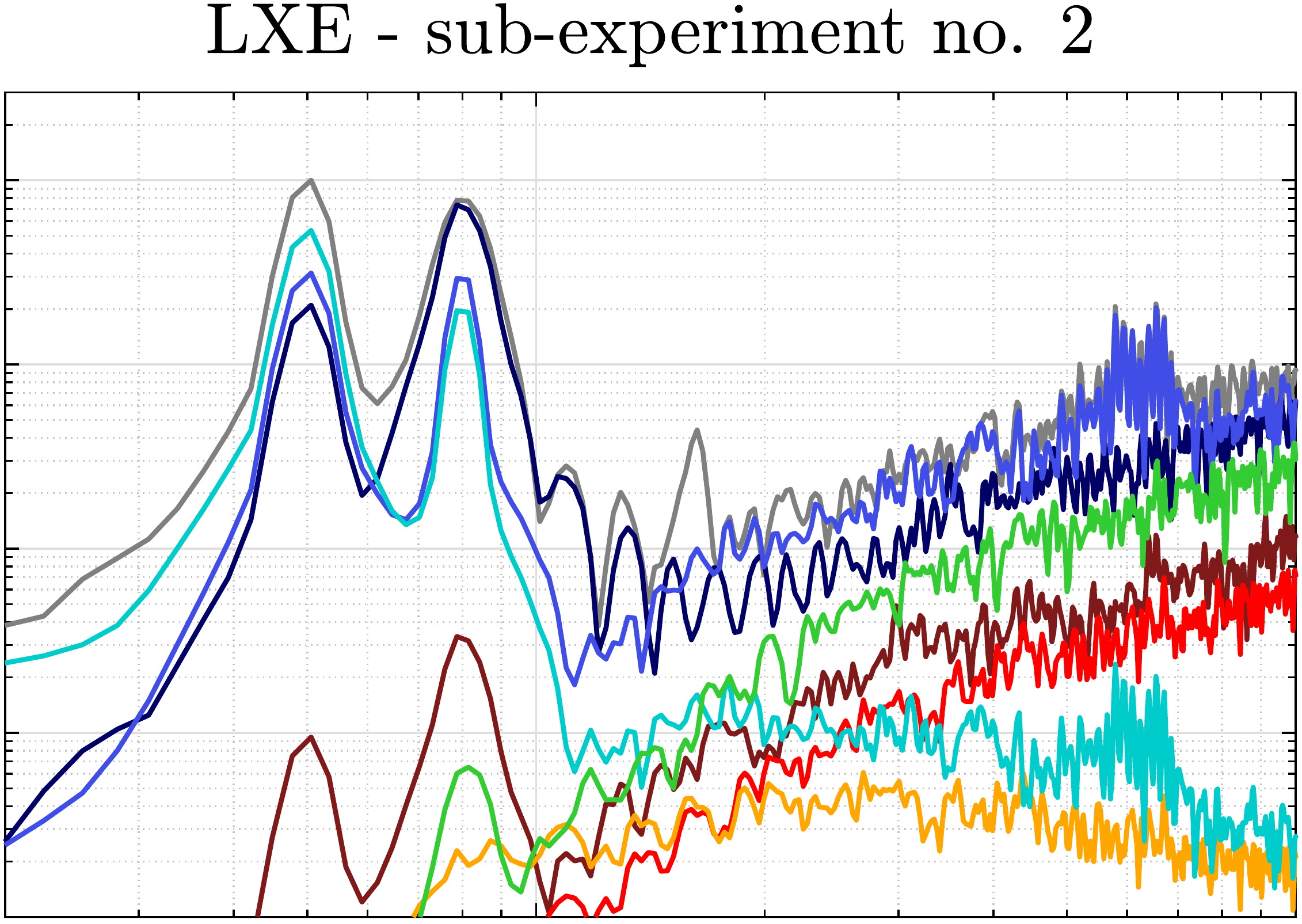}\quad
\includegraphics[scale=0.23,valign=t]{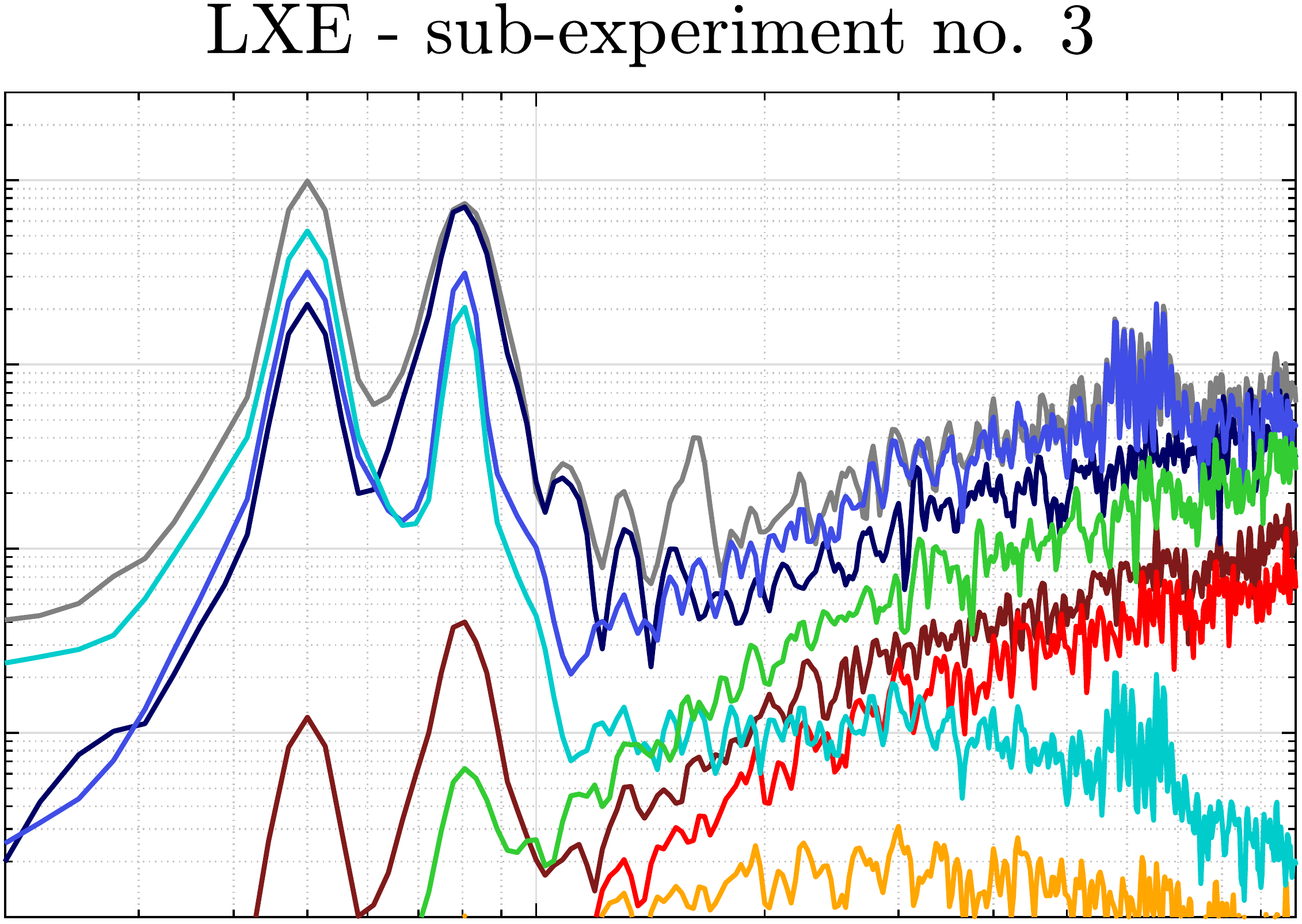} \\[1ex]
\includegraphics[scale=0.23,valign=t]{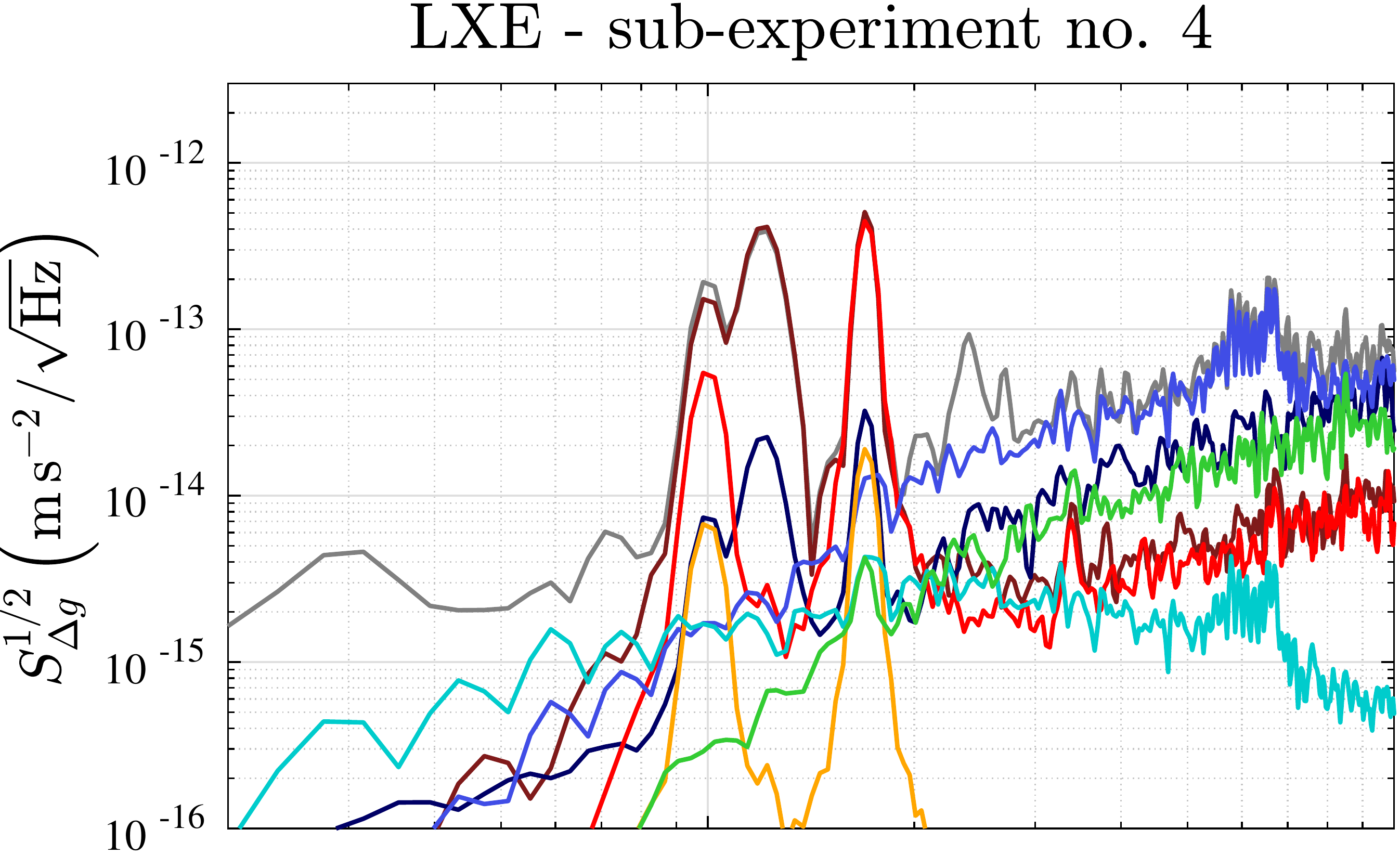}\,\quad
\includegraphics[scale=0.23,valign=t]{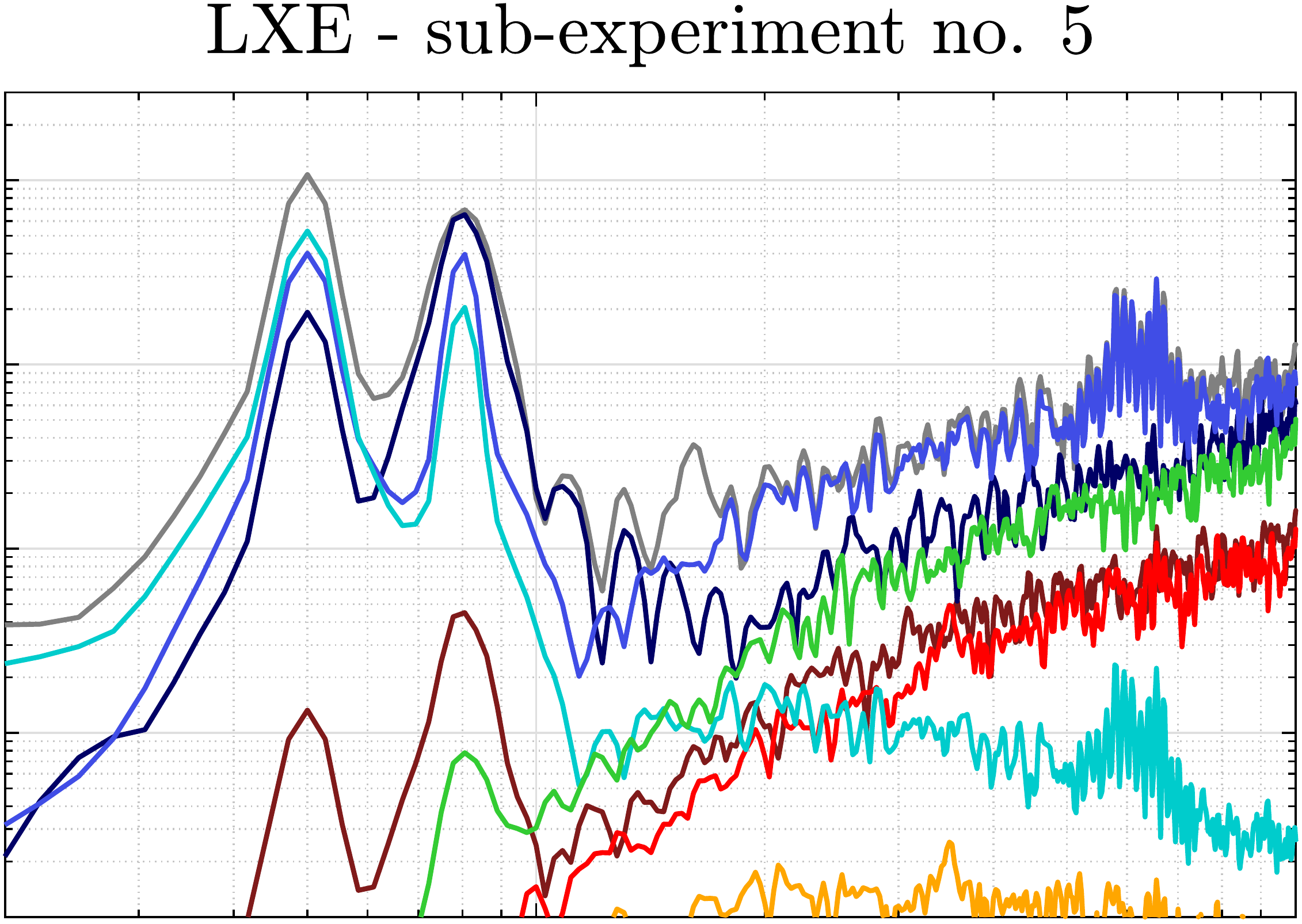}\quad
\includegraphics[scale=0.23,valign=t]{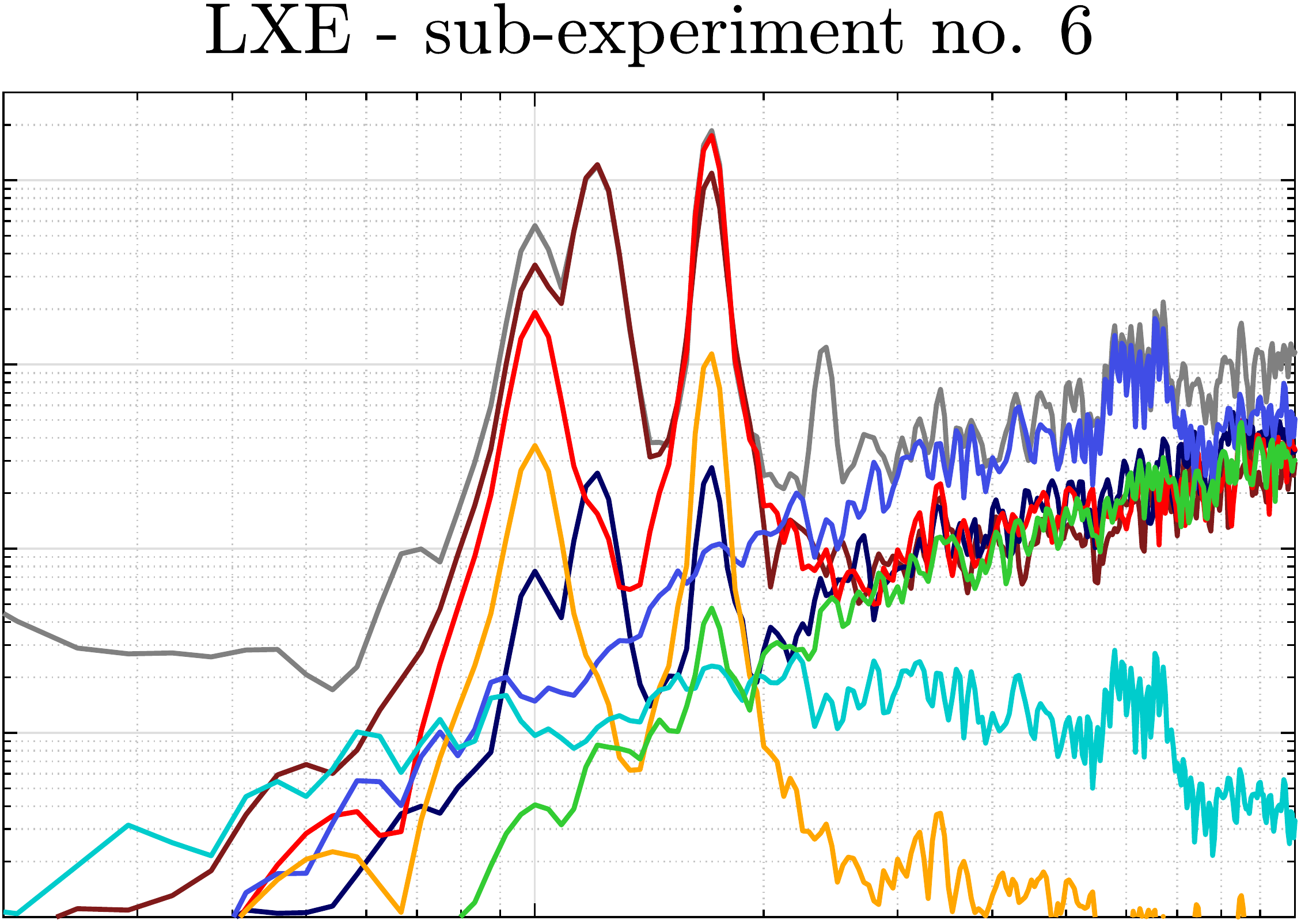} \\[1ex]
\includegraphics[scale=0.23,valign=t]{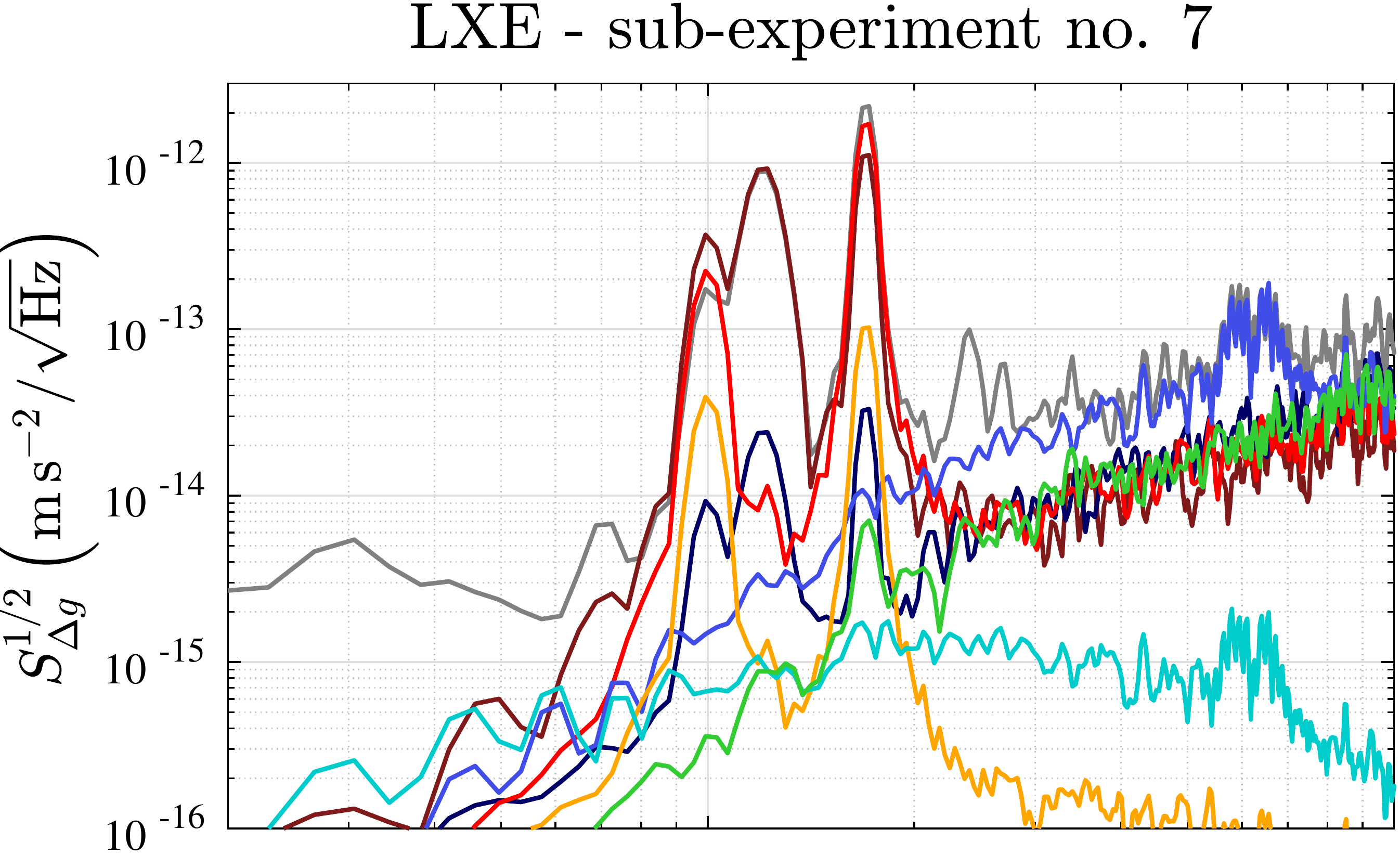}\,\quad
\includegraphics[scale=0.23,valign=t]{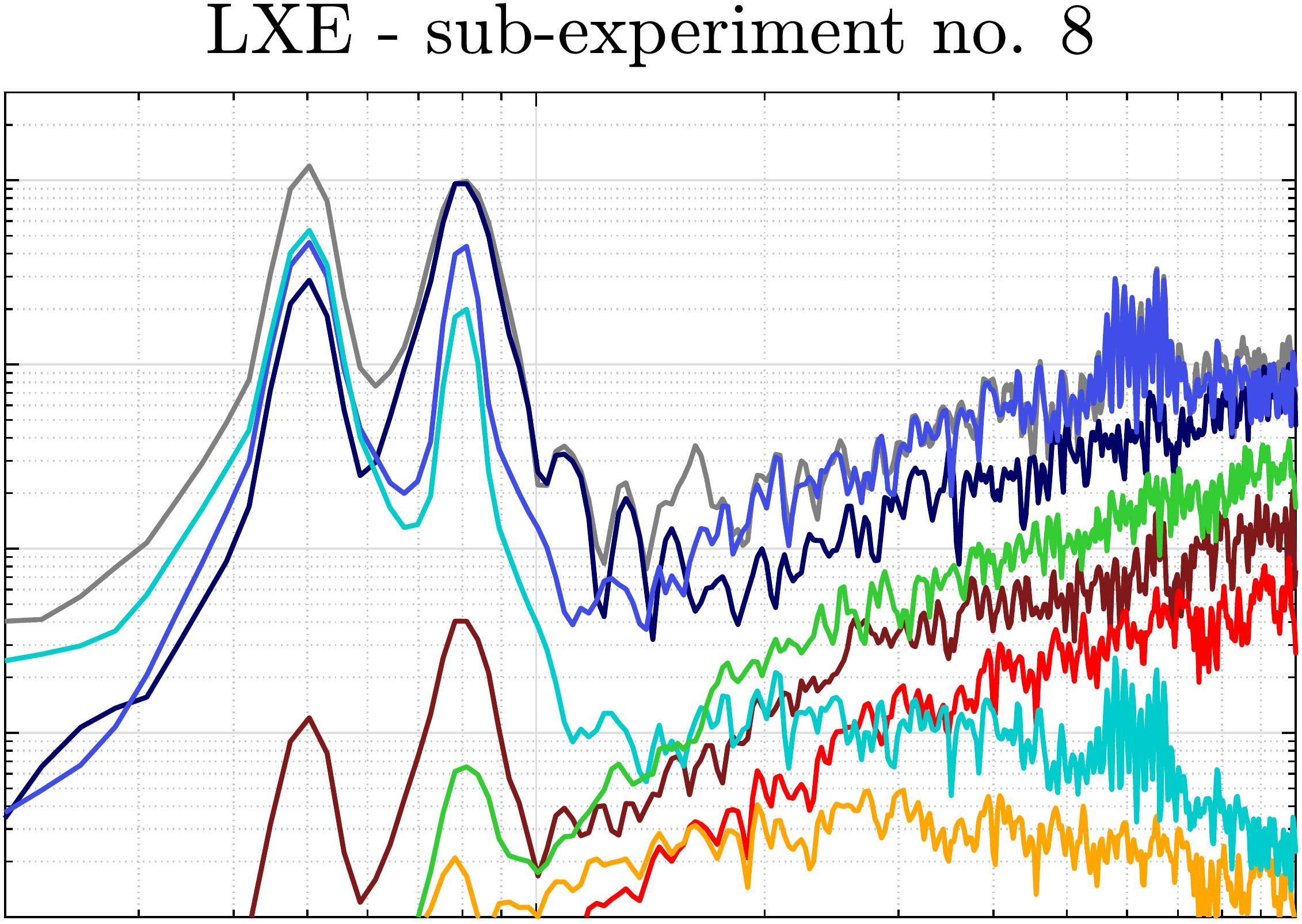}\quad
\includegraphics[scale=0.23,valign=t]{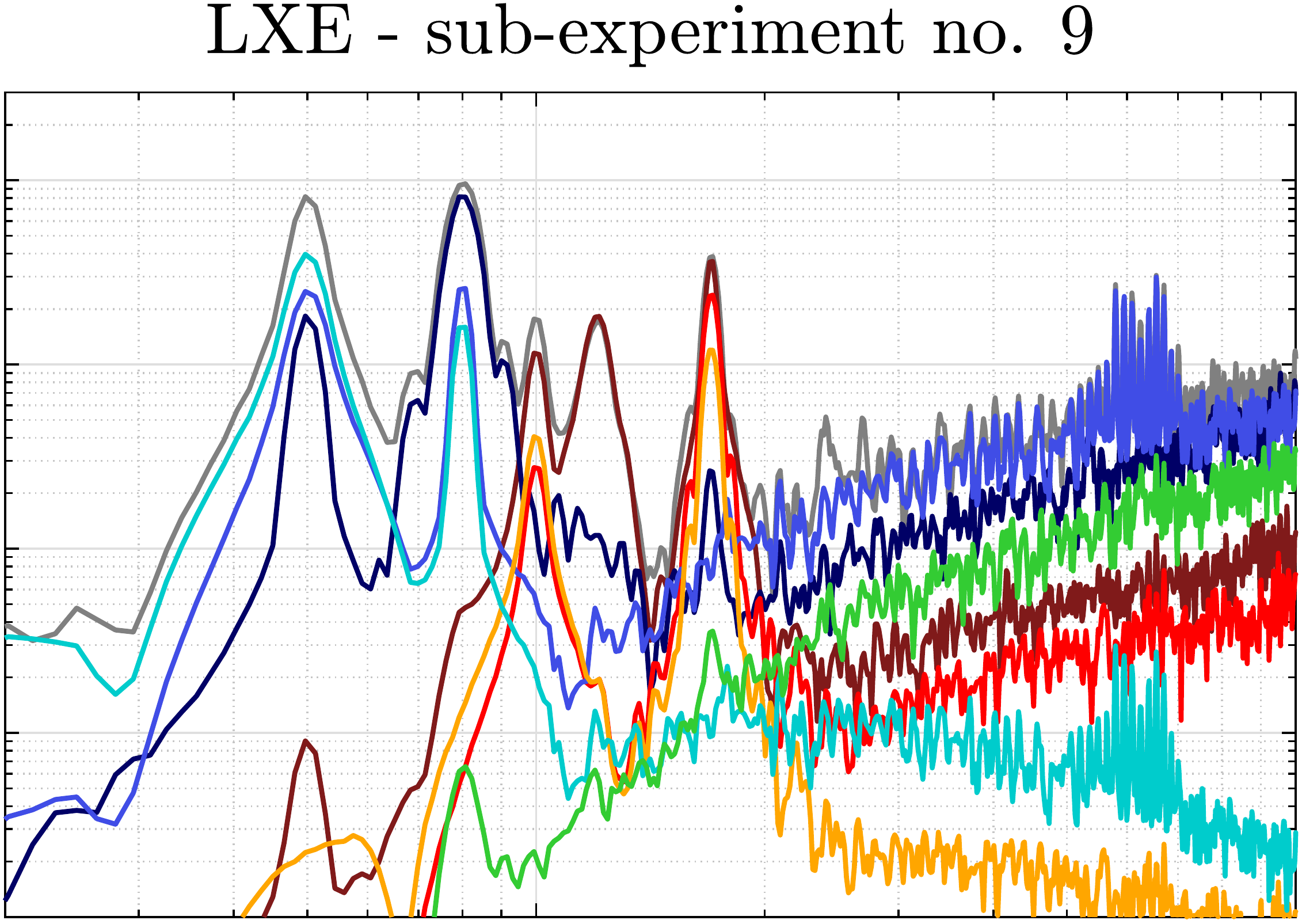} \\[1ex]
\,\ \includegraphics[scale=0.23,valign=t]{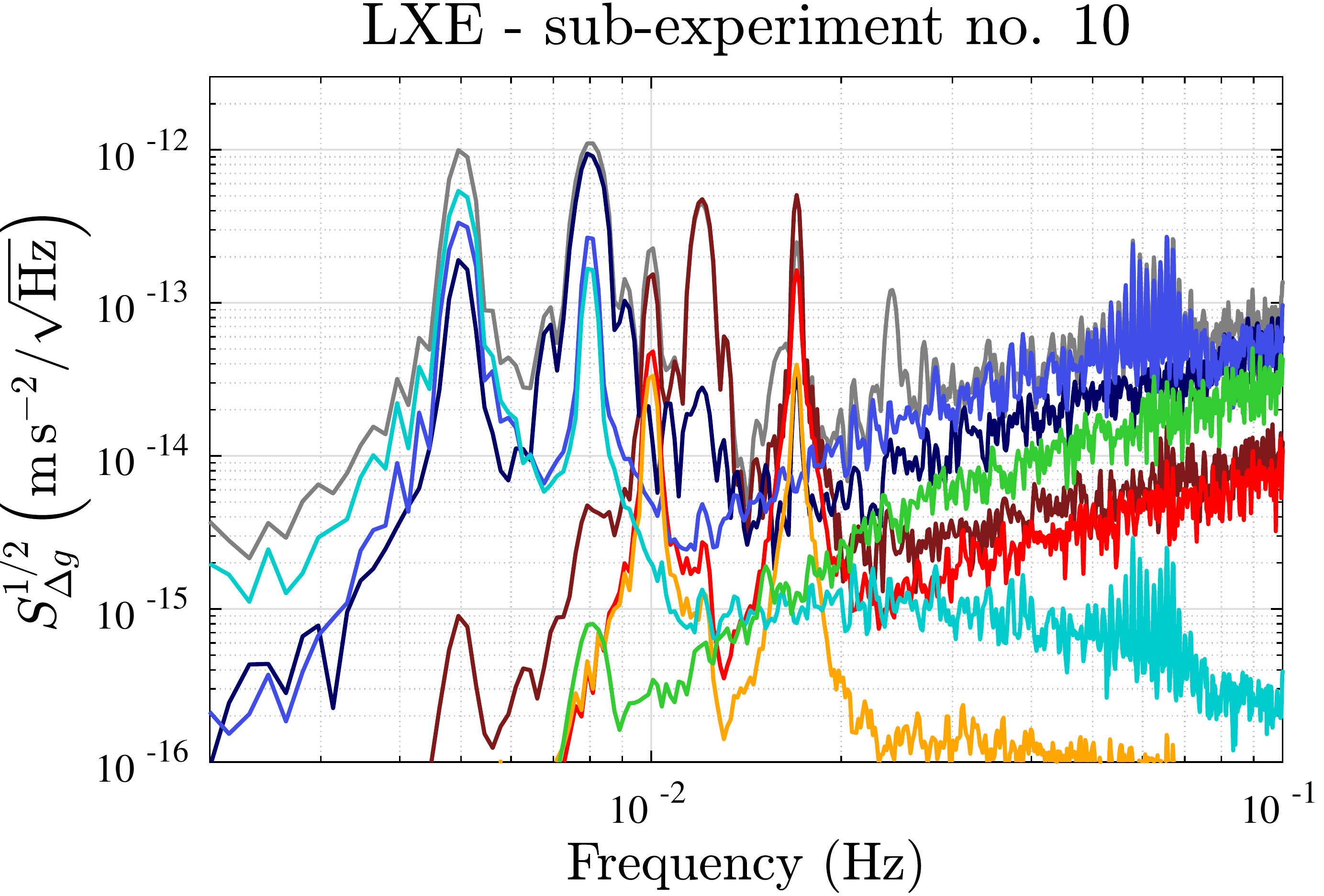}\,\,\
\includegraphics[scale=0.23,valign=t]{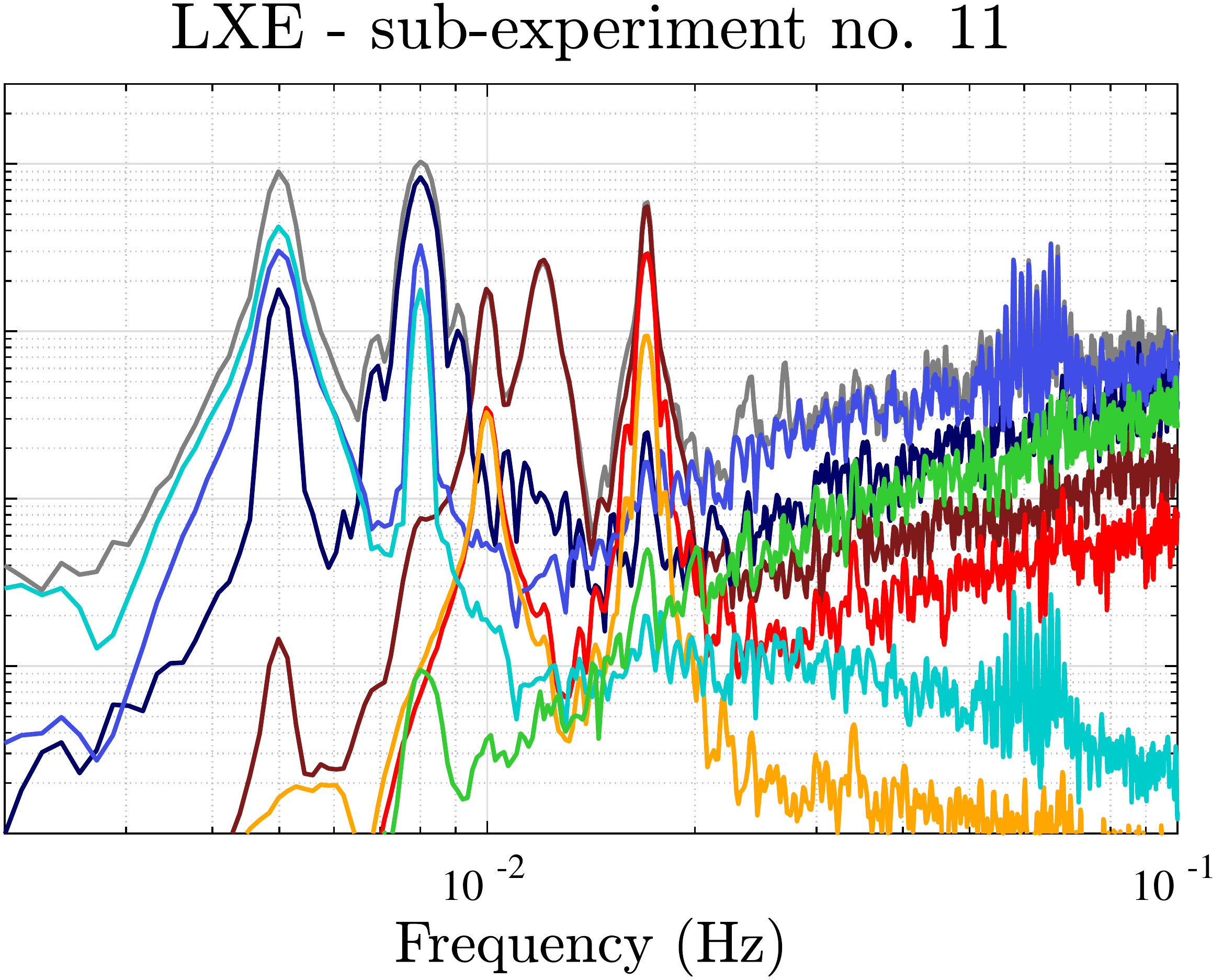}\hfill
\includegraphics[scale=0.23,valign=t]{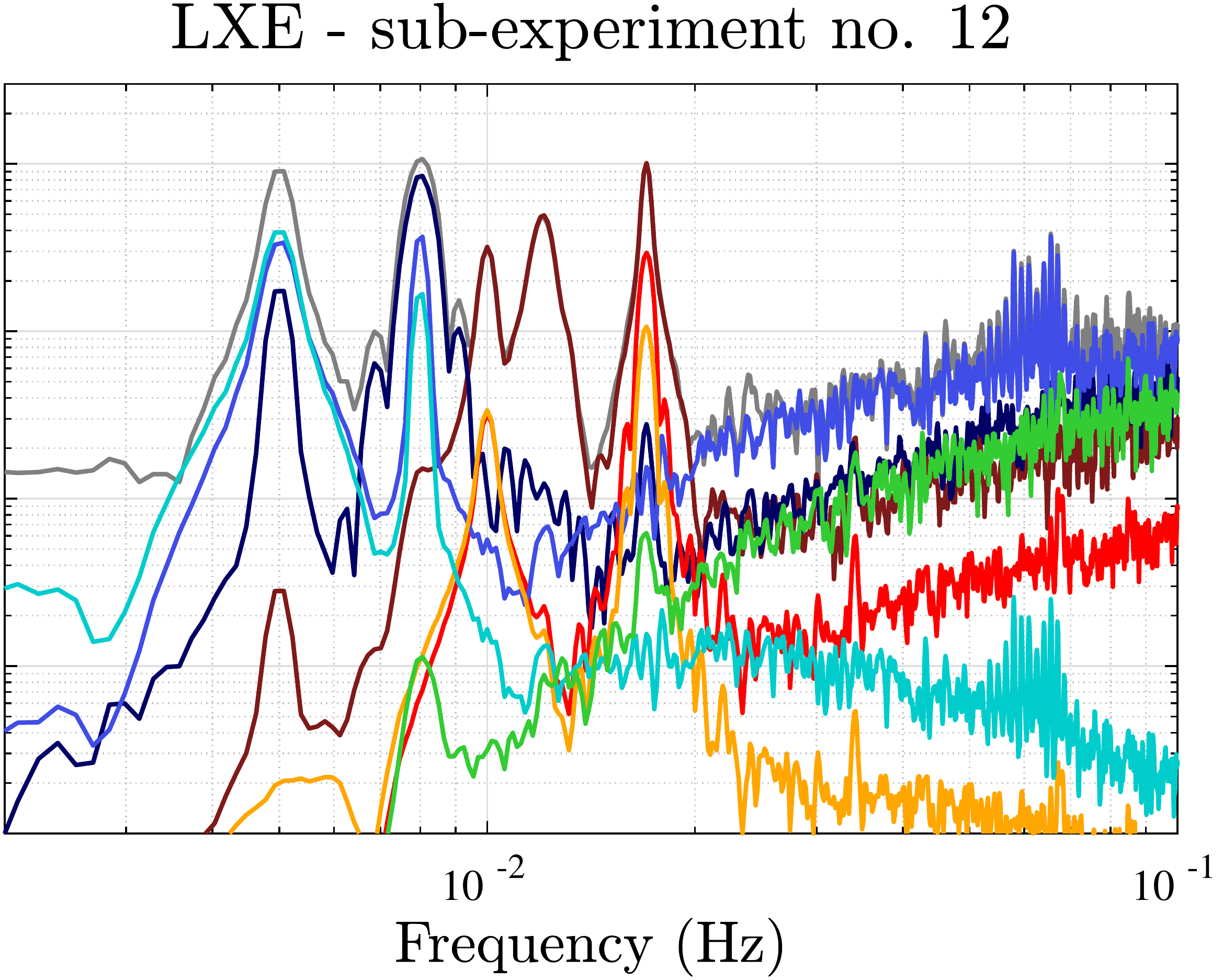} \\[2ex]
\includegraphics[scale=0.8]{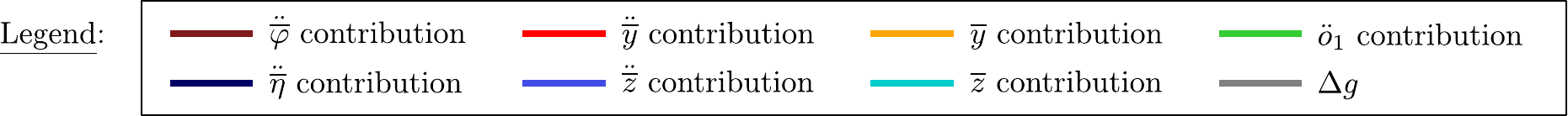}
\caption{The \glspl{ASD} of the noise contributors during the sub-experiments of the \gls{LXE}. The observable $\Delta g$ is given in grey. The coloured curves show the measurements in the respective \gls{DoF} scaled by their fitted coupling coefficients. In all experiments, the acceleration noise in the $xz$-plane is dominant at TTL noise bump. The most dominant is the $\ddot{\bar{z}}$ noise contribution, which almost only explains the bulge between 55 and 70\,mHz.}
\label{fig:LXE_performance_contributors}
\end{figure*}

We have shown in the Sec.~\ref{sec:LXE_comparison_fit-settings} that we find different $C_\eta$ coefficients if fitting either only the bump or only the injections.
When applying the fit to the full frequency regime (2-70\,mHz), the resulting coupling coefficient is a compromise between both single coupling coefficients. 
The impact of either of the coefficients on the combined result was the higher, the higher the noise in the respective frequency regime.
Here, we investigate the noise levels of the single \gls{TTL} coupling contributors. These are illustrated in Fig.~\ref{fig:LXE_performance_contributors}.
We find that the jitter in $\ddot{\bar{\eta}}$ was one of the main noise contributors at the bump frequencies. 
Thus, it significantly contributes to the $C_\eta$ fit in the combined frequency regime.
Consequently, when using the resulting compromise fit coefficient for the noise subtraction, we expect considerable residuals at the $z$-injection frequencies (compare Fig.~\ref{fig:LXE_performance_fit}). 

In general, Fig.~\ref{fig:LXE_performance_contributors} shows that the \gls{TTL} coupling jitter terms add with very different levels to the full noise.
We found that the stiffness coupling was large at the injection frequencies (if injections were applied in the corresponding plane) but negligible at the bump frequencies.
The accelerations along $y$ and $\varphi$ were small if no realignments in $\varphi_1$ or $\varphi_2$ were applied (6th, 7th, 11th and 12th sub-experiment).
In these cases, the acceleration along $z$ and $\eta$ were the largest \gls{TTL} noise contributors to the \gls{TTL} noise bump. %
Particularly, the $z$-jitter explains almost completely the bulge between 55 and 70\,mHz.

We show in Sec.~\ref{sec:TMalignments} that the large $\ddot{\bar{\eta}}$- and $\ddot{\bar{z}}$-jitter noise at high frequencies resulted partially from their insufficient suppression by \gls{TM} realignments. Their reduction was worse compared to the jitter in the other plane. 
Further, we discuss in \cite{LPFstabi23} that additional \gls{TTL} coupling changes resulted from drifts of the coefficients (largest for the $C_z$-coefficient).

\subsection{Second-order models}
\label{sec:LXE_2nd-order}

We have seen in Figs.~\ref{fig:LXE_performance_fit} and~\ref{fig:LXE_performance_ana} that neither of the linear models subtract the noise peaks at sums and multiples of, or the differential injection frequencies.
These would, for sinusoidal injections, be covered by a higher-order noise model according to trigonometric addition theorems.

When extending the accelerations in $\varphi$,\,$\eta$,\,$y$ and $z$ to the second-order noise contributions within the fit model, this model would contain ten additional coupling coefficients. 
The resulting large fit model would be computationally expensive and largely affected by correlations between the fit quantities.

However, the evaluation of the analytical \gls{TTL} coupling model \cite{LPFana22} has shown that most second-order noise contributors are negligibly small.
Based on this analysis, we can reduce the second-order model to
\begin{align}
\begin{split}
\Delta g_\text{xtalk,2nd} &= \Delta g_\text{xtalk} \\
&+ C_{\varphi2} \left[2\left(\overline{\varphi}\,\ddot{\overline{\varphi}}+\dot{\overline{\varphi}}^2\right)\right]
+ C_{\eta2} \left[2\left(\overline{\eta}\,\ddot{\overline{\eta}}+\dot{\overline{\eta}}^2\right)\right] \,.
\end{split}
\label{eq:2ndordermodel}
\end{align}

The two additional second-order coupling coefficients do not depend on the \gls{TM} alignments. Therefore, they are invariant to the realignments at the beginning of the mission and the \gls{TM} set-points during the \gls{LXE}.
Moreover, the analytical model derivation shows that these coefficients originate mainly from geometric \gls{TTL} coupling \cite{G21} and are very stable. 
Therefore, the analytical derivation of these coefficients is assumed to hold for the entire mission duration. 

The analytical model yields:
\begin{subequations}
\begin{align}
C_{\varphi2}^\text{ana} &= 0.196^{+0.001}_{-0.002}\,\frac{\text{m}}{\text{rad}^2} \,,
\label{eq:anamodel_2nd_phi}\\
C_{\eta2}^\text{ana}    &= 0.193^{+0.001}_{-0.002}\,\frac{\text{m}}{\text{rad}^2} \,.
\label{eq:anamodel_2nd_eta}
\end{align}
\label{eq:anamodel_2nd}
\end{subequations}

\begin{figure*}
\centering
\includegraphics[scale=0.23,valign=t]{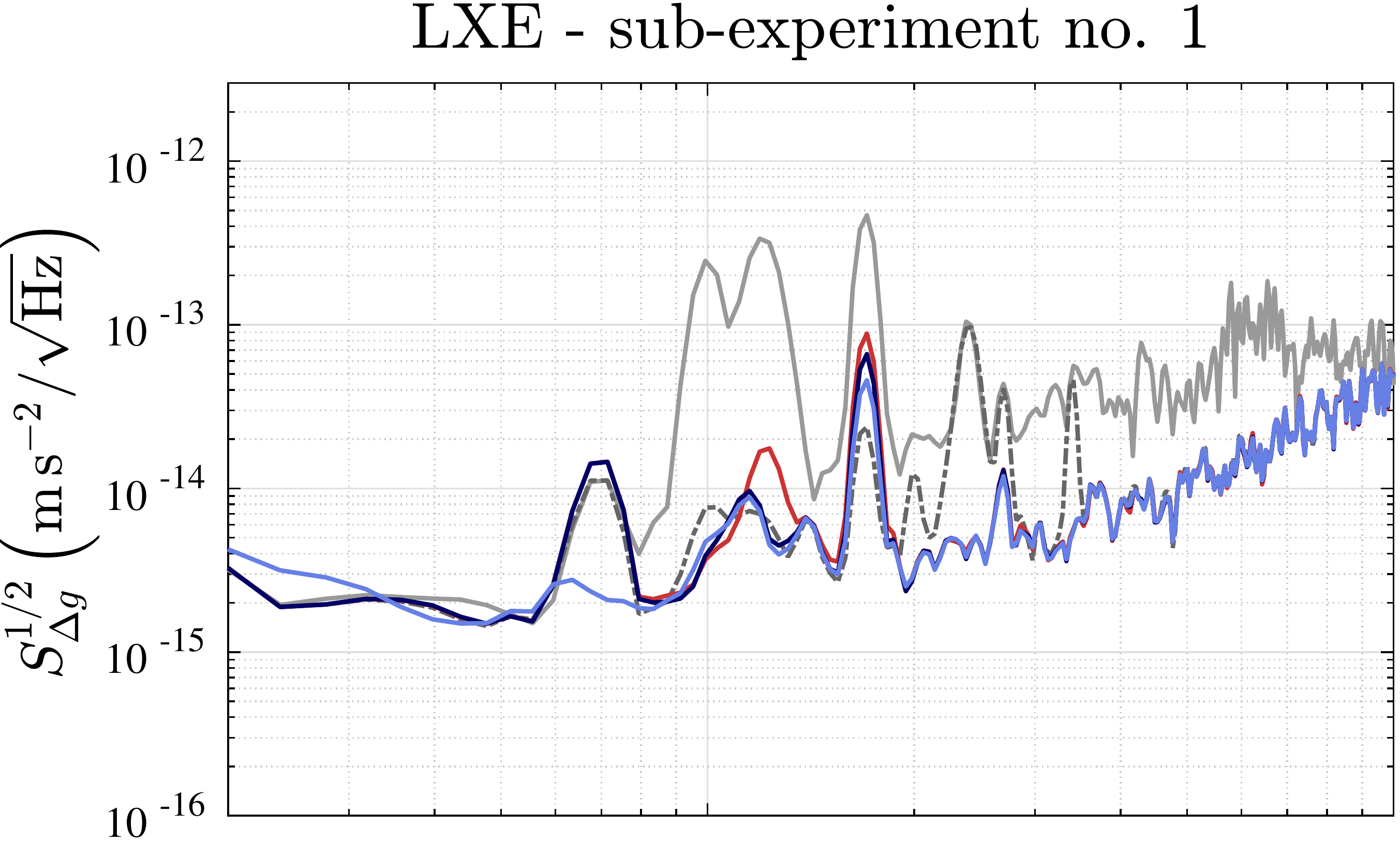}\,\quad
\includegraphics[scale=0.23,valign=t]{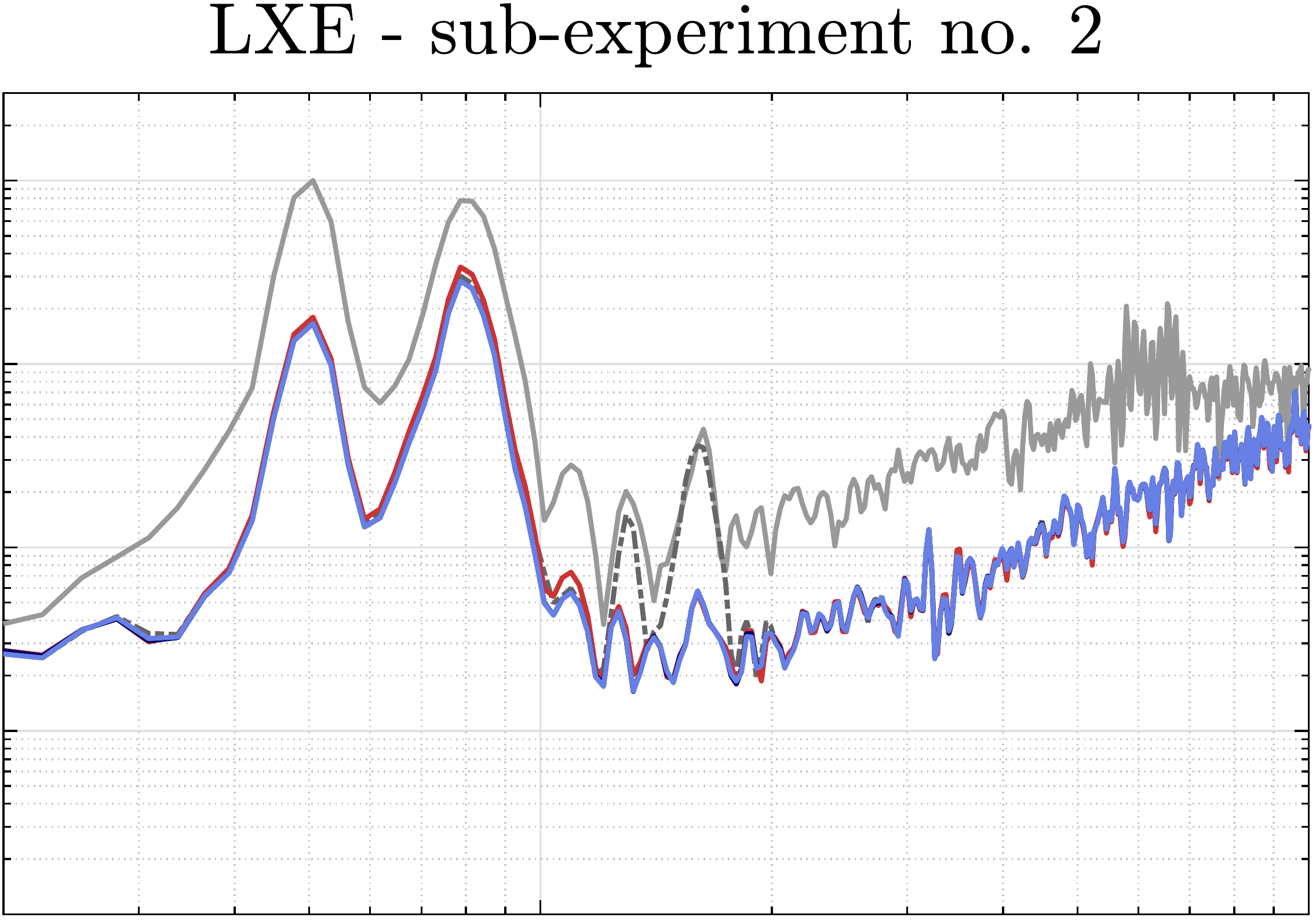}\quad
\includegraphics[scale=0.23,valign=t]{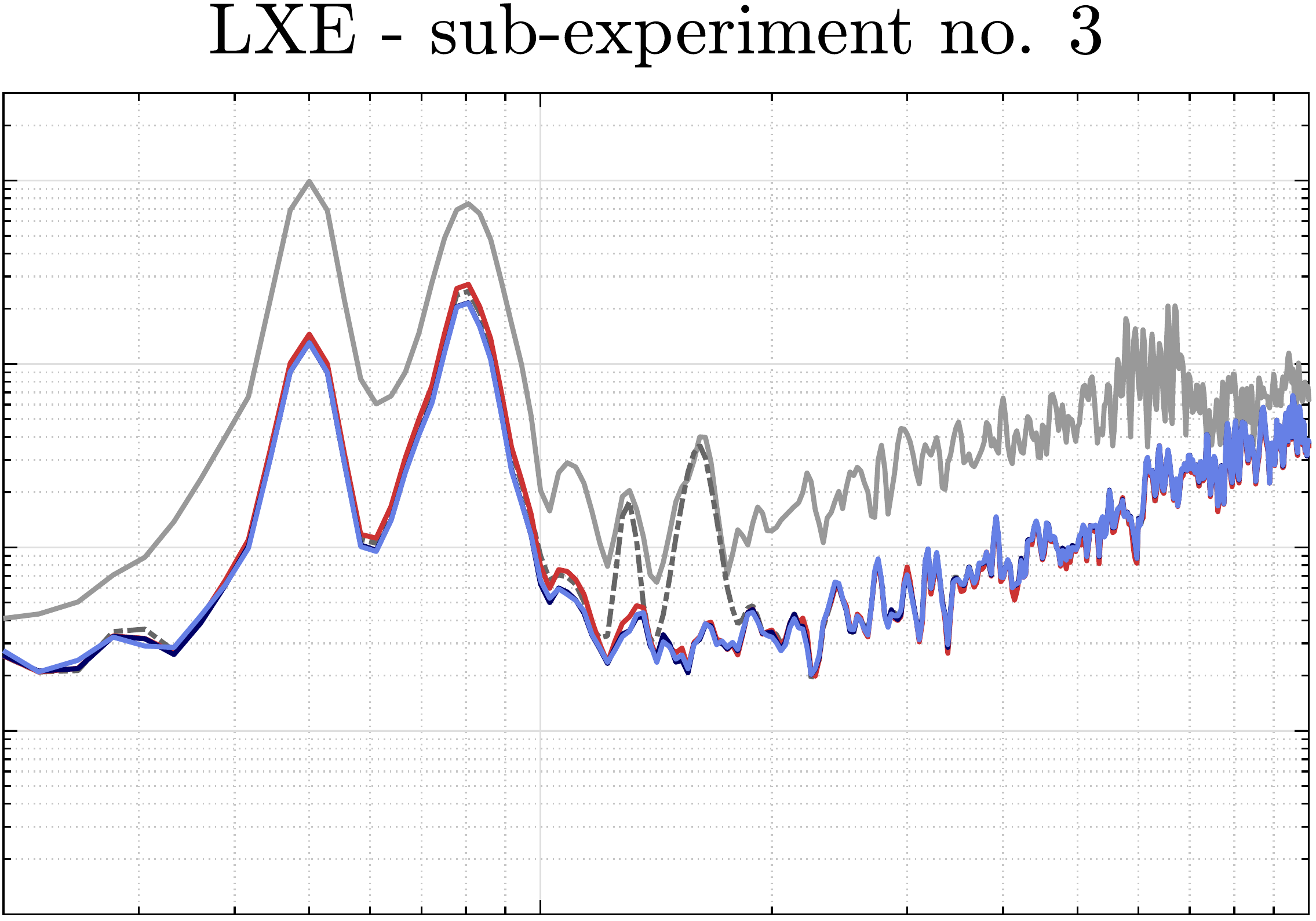} \\[1ex]
\includegraphics[scale=0.23,valign=t]{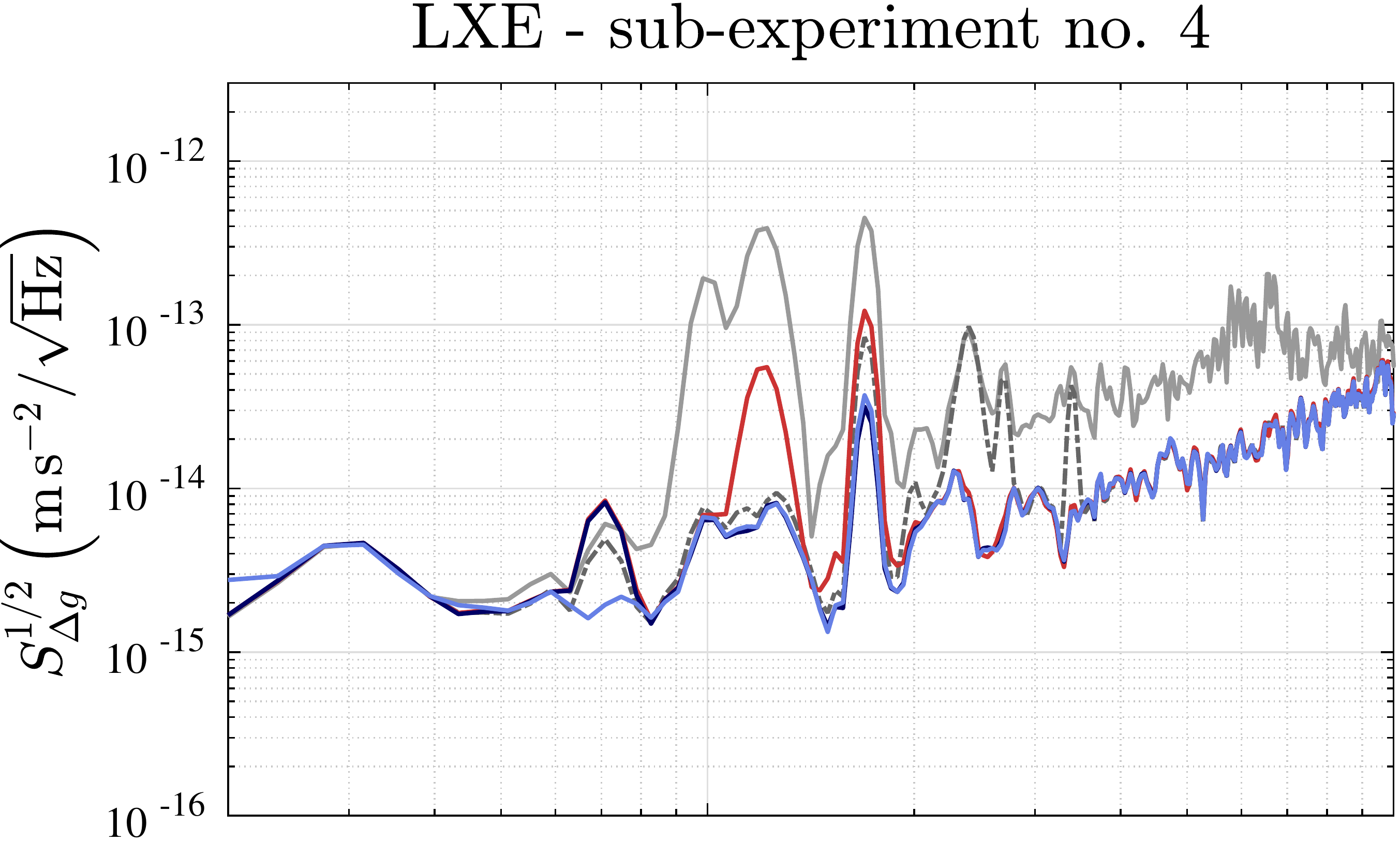}\,\quad
\includegraphics[scale=0.23,valign=t]{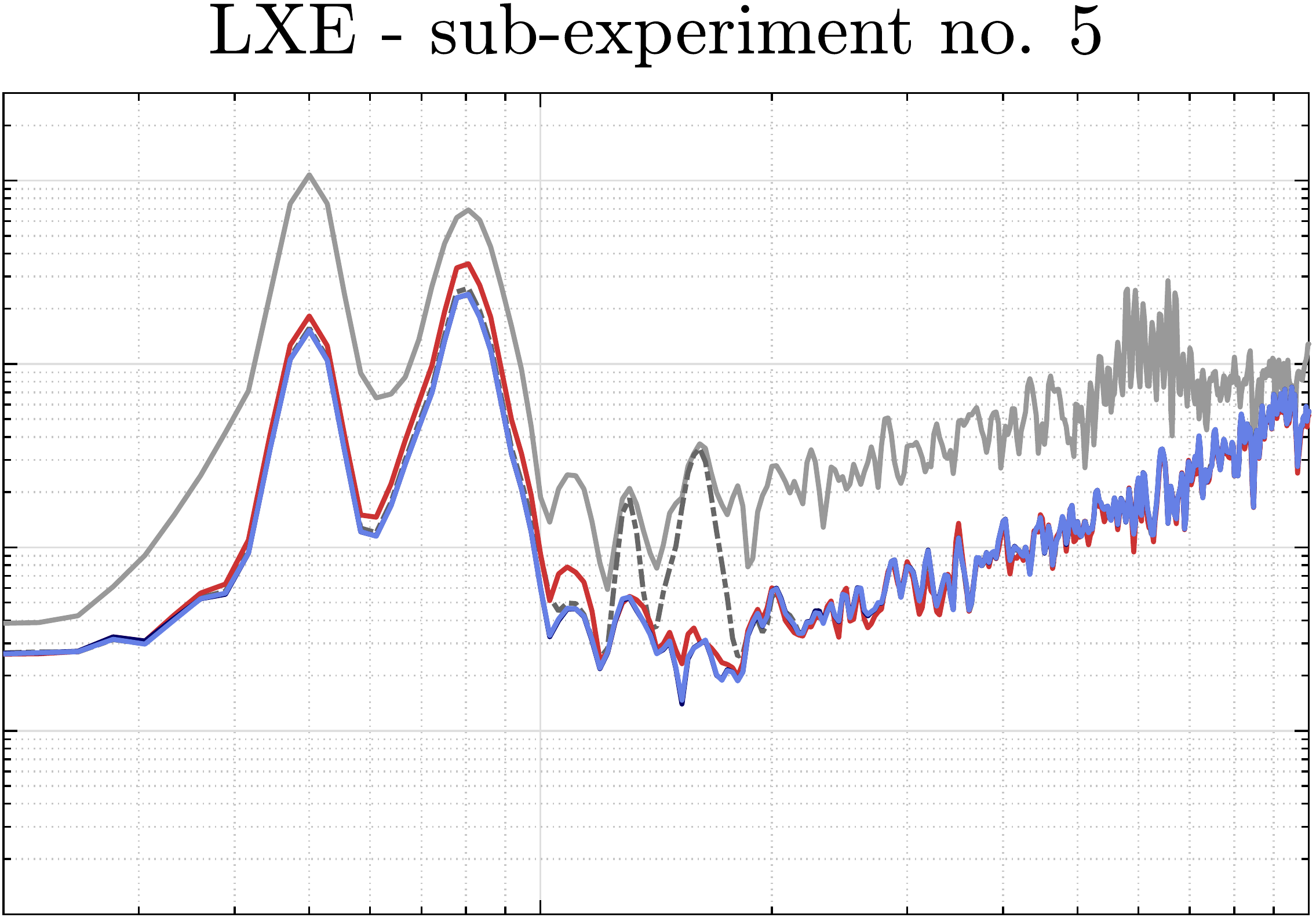}\quad
\includegraphics[scale=0.23,valign=t]{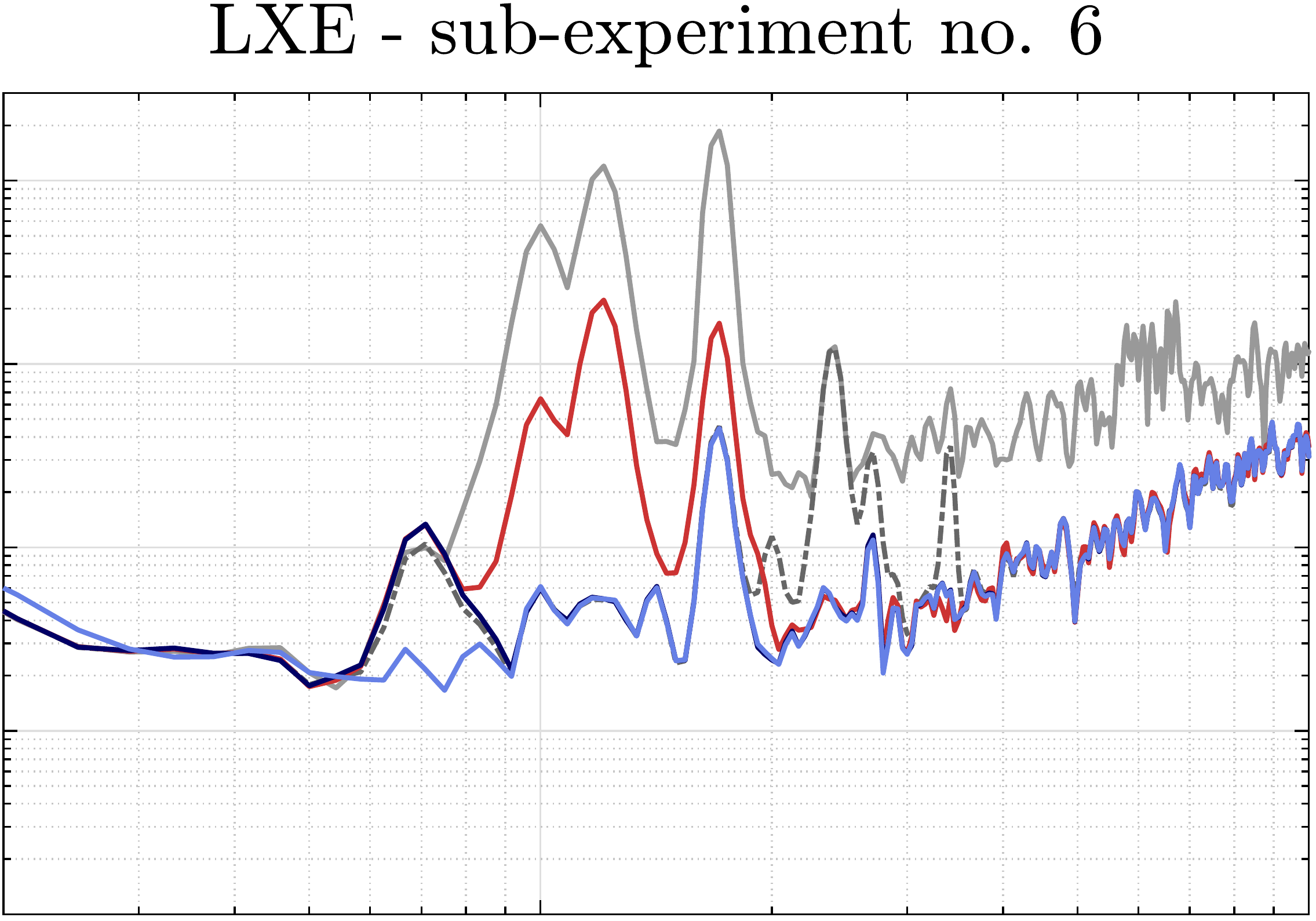} \\[1ex]
\includegraphics[scale=0.23,valign=t]{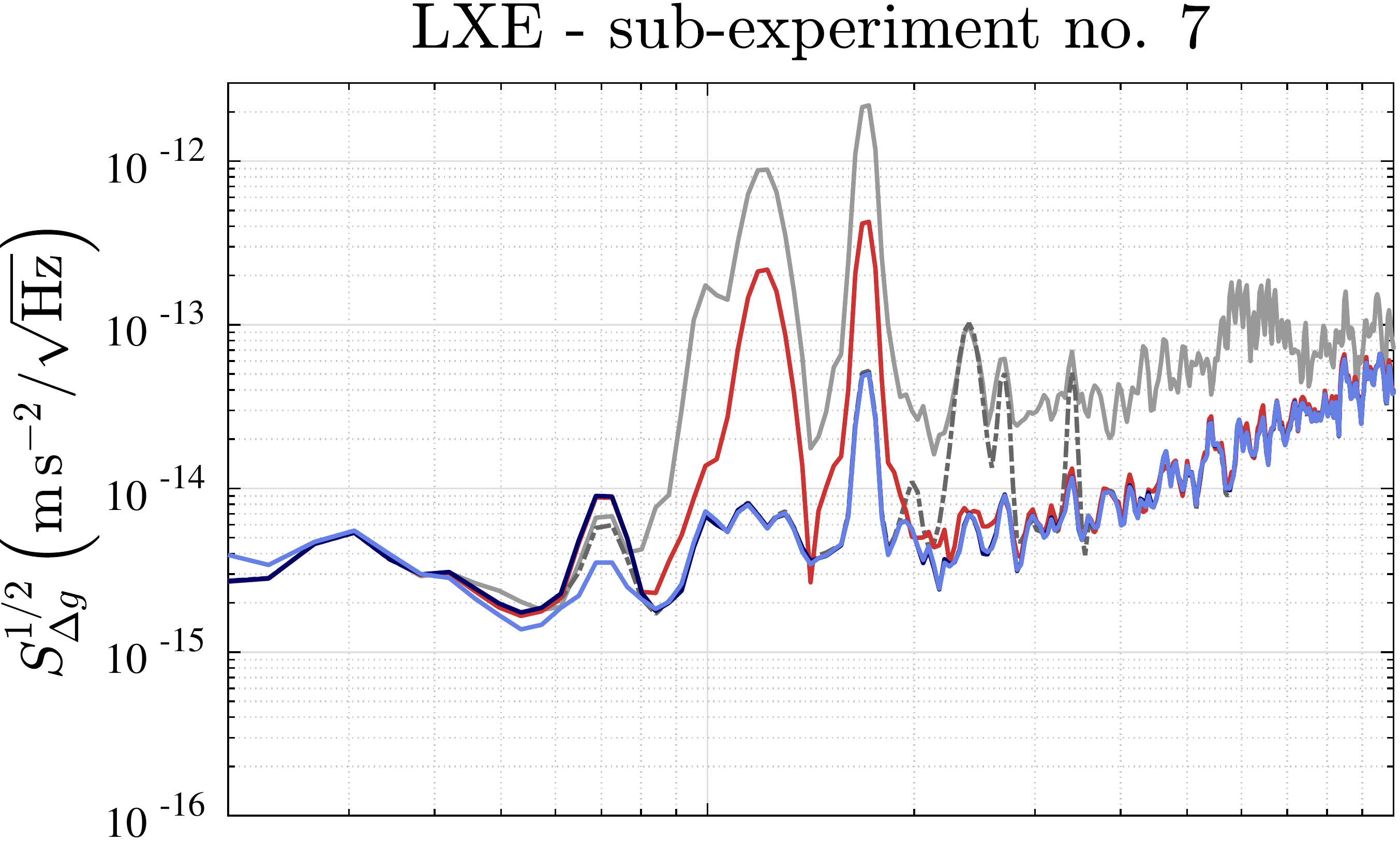}\,\quad
\includegraphics[scale=0.23,valign=t]{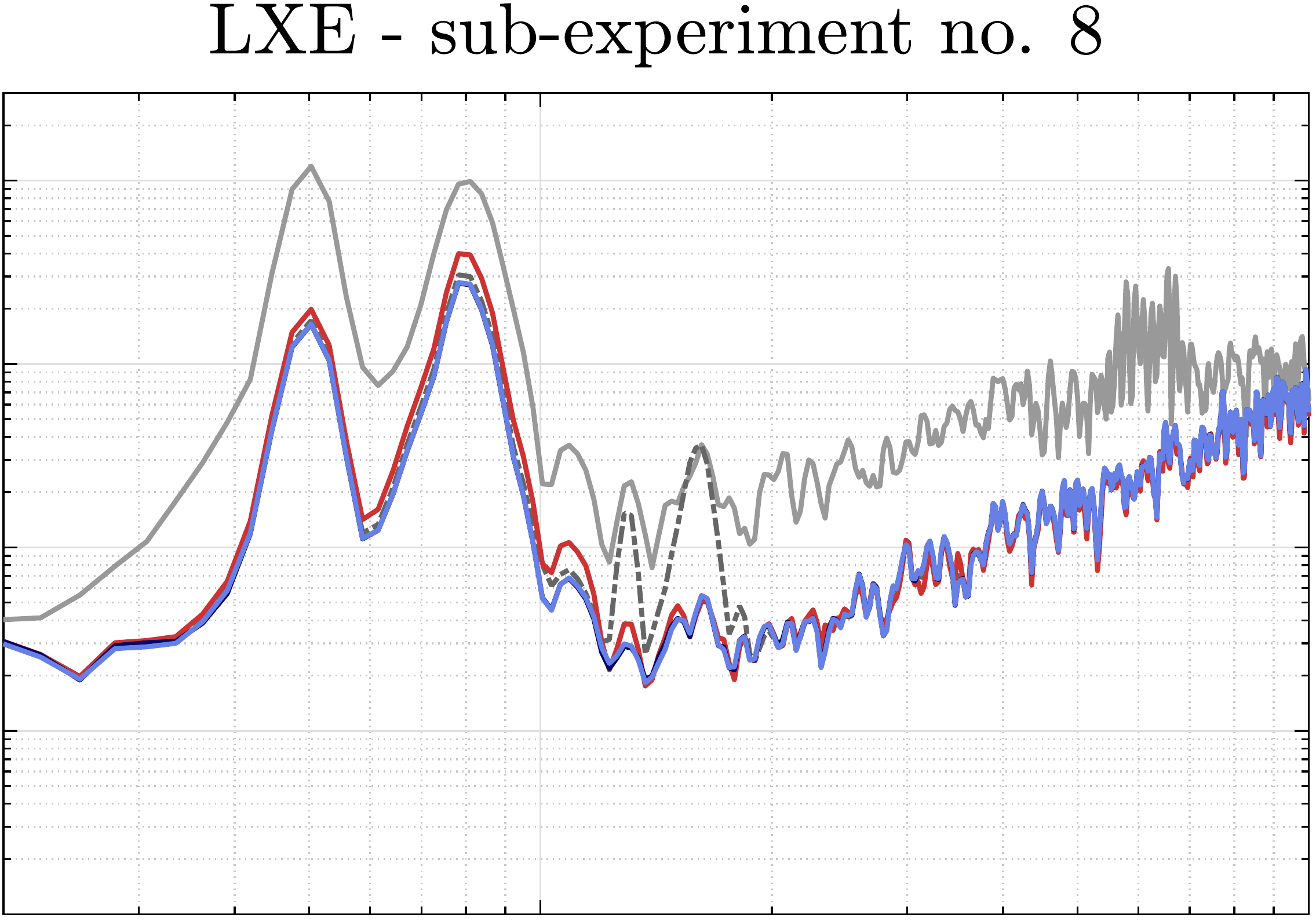}\quad
\includegraphics[scale=0.23,valign=t]{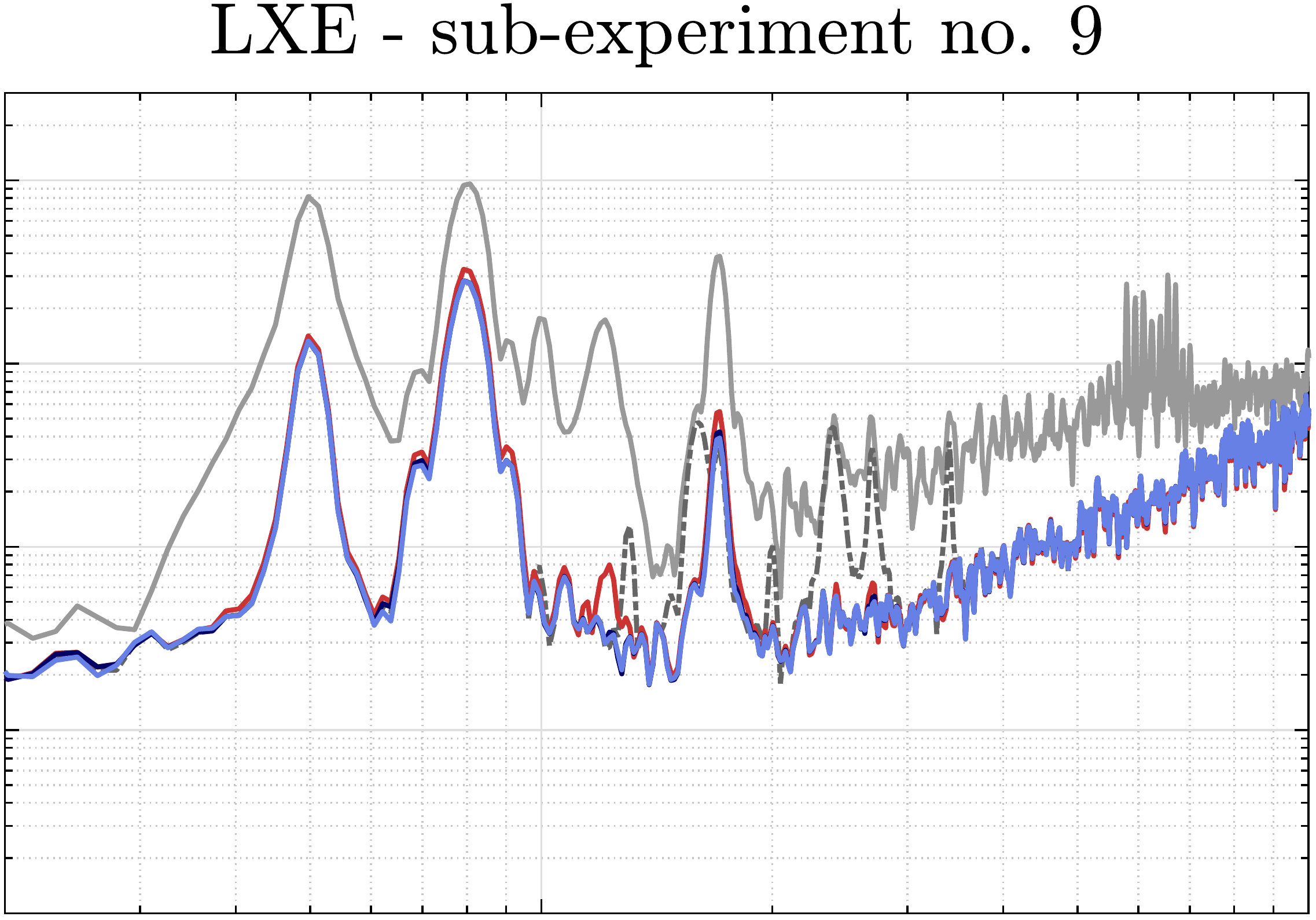} \\[1ex]
\,\,\includegraphics[scale=0.23,valign=t]{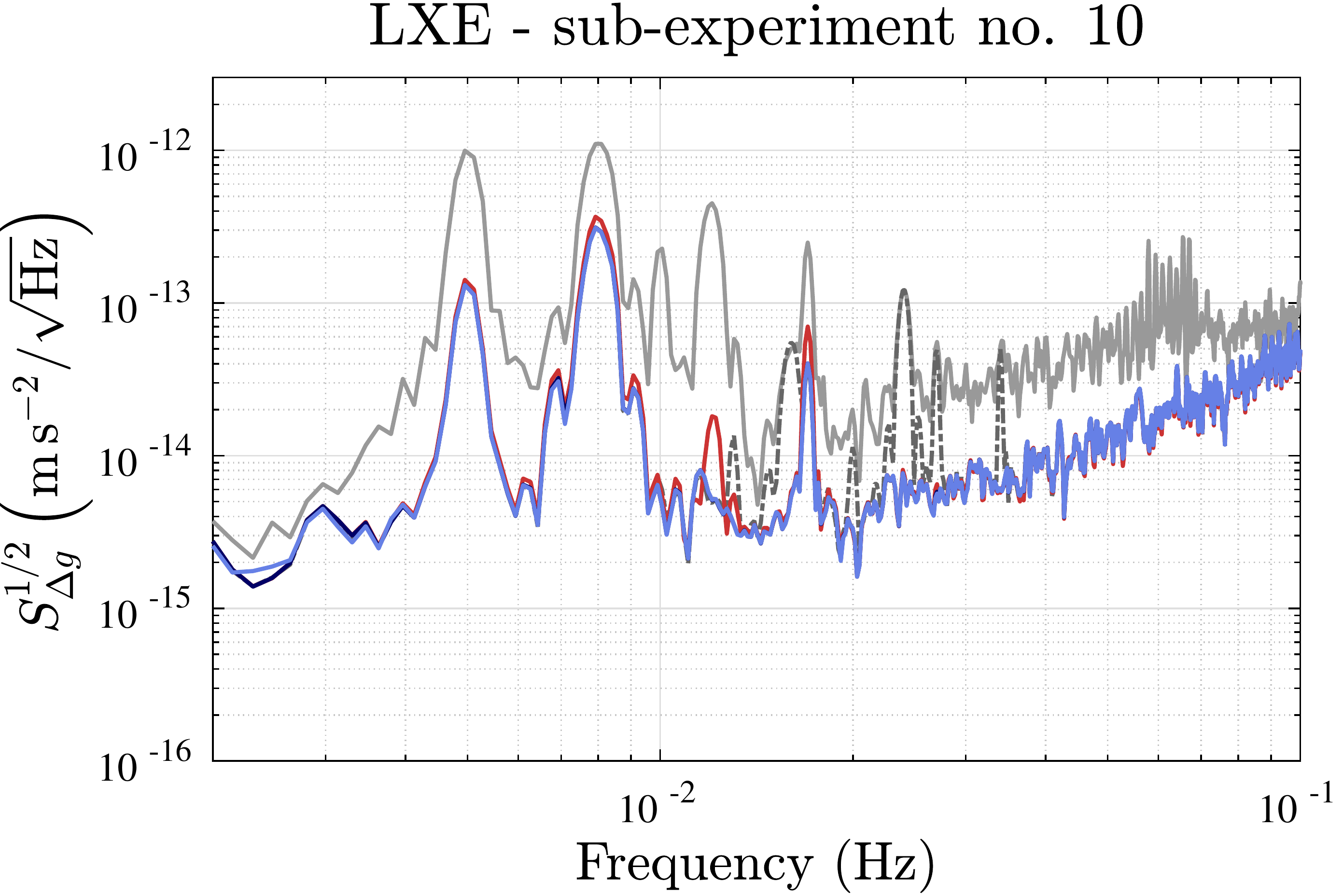}\,\,
\includegraphics[scale=0.23,valign=t]{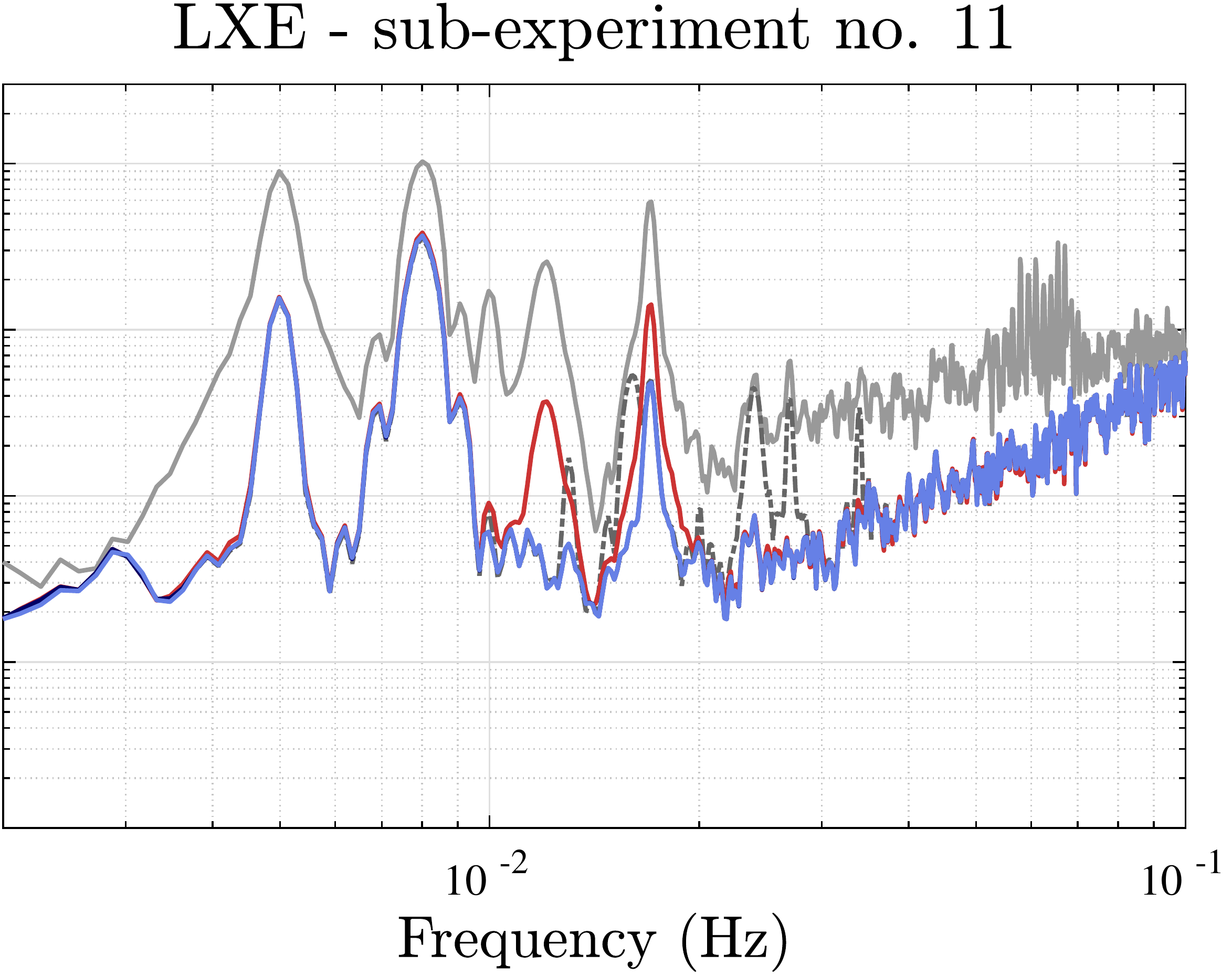}\hfill
\includegraphics[scale=0.23,valign=t]{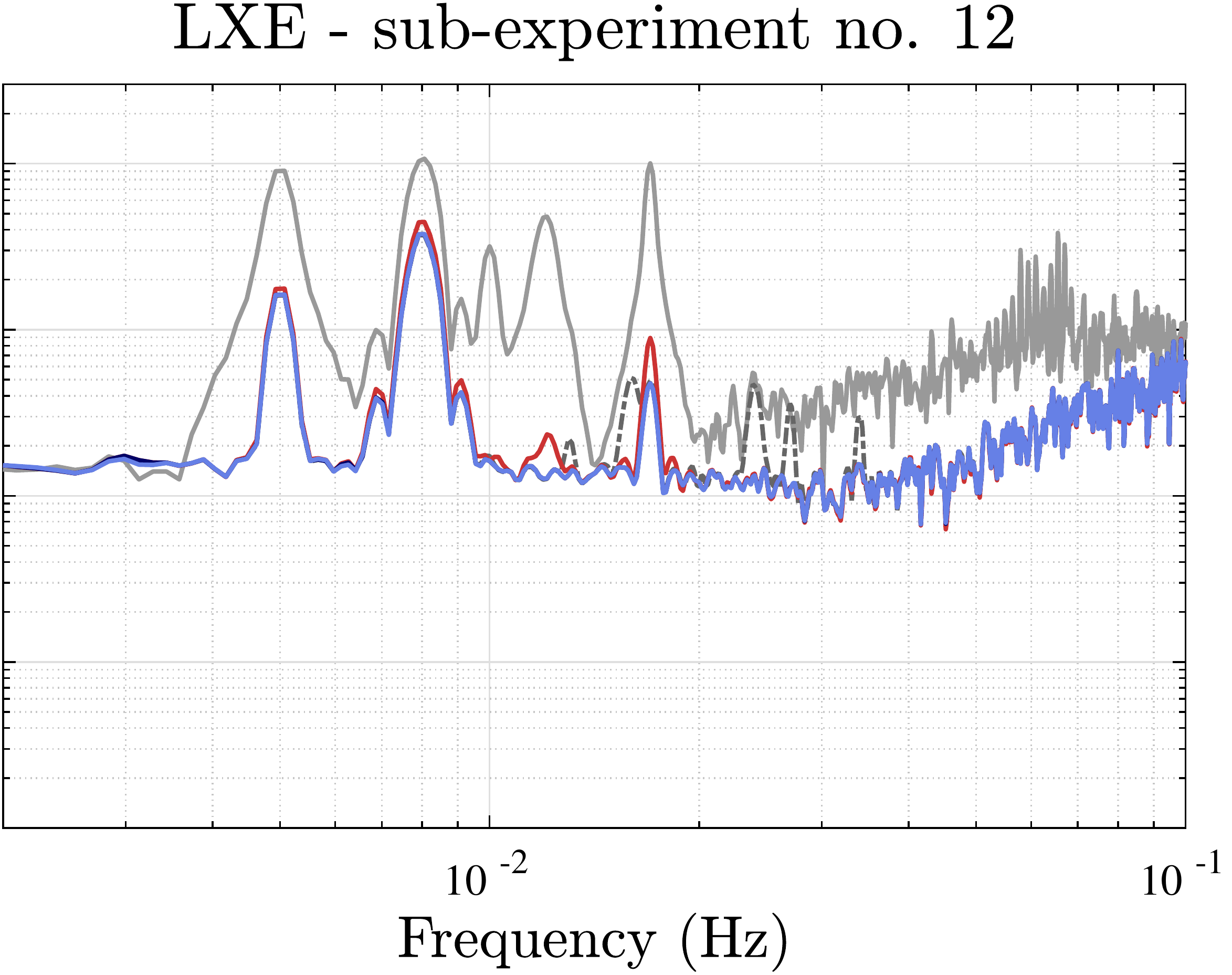} \\[2ex]
\includegraphics[scale=0.8]{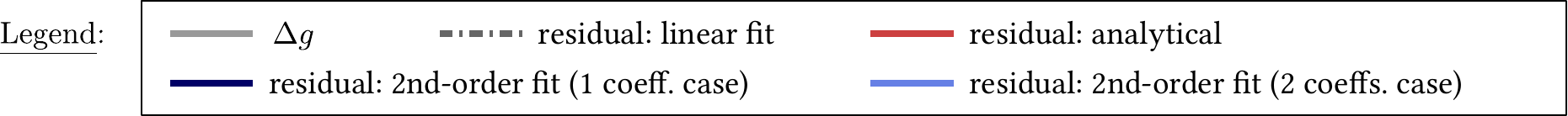}
\caption{Residuals after subtracting the second-order models for the twelve sub-experiments of the LXE.
The continuous grey curve shows the \gls{ASD} of the observable $\Delta g$.
The colored curves show the ASD of the residuals after subtracting the second-order models from $\Delta g$.
Red: second-order analytical model plus the fitted stiffness and $o_1$ terms.
Dark blue: fit model with one additional angular second-order coefficient in each plane. 
Blue: fit model with two additional angular second-order coefficients in each plane.
For comparison, we show the ASD of the residual after subtracting the linear fit model (dashed grey).
All three models subtract the peaks at the multiples of the injections frequencies that were not subtracted by the linear models.
However, the extensions change very little at the injection frequencies. 
The fit model with two second-order coefficients in each plane subtracts the noise at the differential injection frequencies better (see 7\,mHz peaks).}
\label{fig:LXE_performance_2nd_fit_residual}
\end{figure*}

The extended analytical model additionally subtracts the noise peaks at the sums of, but not the differential injection frequencies (e.g.\ 17\,mHz$-$10\,mHz=7\,mHz), see the red residual in Fig.~\ref{fig:LXE_performance_2nd_fit_residual}.

Next, we fit the \gls{TTL} coupling using the model Eq.~\eqref{eq:2ndordermodel} with $\Delta g_\text{xtalk}$ being replaced by Eq.~\eqref{eq:fitmodel}.
Like in the case of the analytical model, this second-order fit model subtracts the peaks at the sums of the injection frequencies nicely but not at the differential frequencies, see dark blue curve in Fig.~\ref{fig:LXE_performance_2nd_fit_residual}.
Also, the fit yielded second-order \gls{TTL} coefficients comparable to the analytical results (Eq.~\eqref{eq:anamodel_2nd}):
\begin{subequations}
\begin{align}
C_{\varphi2}^\text{fit} &= 0.194\,\frac{\text{m}}{\text{rad}^2} \,,
\label{eq:fitmodel_2nd_phi}\\
C_{\eta2}^\text{fit}    &= 0.193\,\frac{\text{m}}{\text{rad}^2} \,,
\label{eq:fitmodel_2nd_eta}
\end{align}
\label{eq:fitmodel_2nd}
\end{subequations}
These numbers correspond to the mean of the second-order coefficients in the eight sub-experiments with injections in the respective plane.
For both coefficients, the root-mean-square of their deviations from this mean is small, i.e.\ 0.005\,m/rad$^2$.
We neglected the other four fitted coefficients since the higher-order coupling was too small in these cases yielding large error bars. 

It is evident that neither the second-order analytical nor the fit model subtract the peaks at the differential frequencies (e.g., visible at 7\,mHz in the first sub-experiment). 
We found that these could be covered by a set of two additional second-order coefficients. 
Therefore, we fit a coefficient for the two second-order terms in both planes, each, i.e.,
\begin{align}
\begin{split}
\Delta g_\text{xtalk,2nd*}^\text{fit} &= \Delta g_\text{xtalk}^\text{fit} \\
&+ 2\,C_{\varphi2A}^\text{fit} \,\overline{\varphi}\,\ddot{\overline{\varphi}}
 + 2\,C_{\varphi2B}^\text{fit} \,\dot{\overline{\varphi}}^2 \\
&+ 2\,C_{\eta2A}^\text{fit} \,\overline{\eta}\,\ddot{\overline{\eta}}
 + 2\,C_{\eta2B}^\text{fit} \,\dot{\overline{\eta}}^2 \,.
\end{split}
\label{eq:2ndordermodel_extended}
\end{align}
Subtracting this model from the $\Delta g$ measurements subtracted additionally the differential peaks, see blue curve in Fig.~\ref{fig:LXE_performance_2nd_fit_residual}.
While, in general, more fit terms could provide a more accurate fit, we see no physical explanation for the extension of the second-order model Eq.~\eqref{eq:2ndordermodel} to the model Eq.~\eqref{eq:2ndordermodel_extended}.

\subsection{Extraction of a TTL model from the fit results}
\label{sec:LXE_minimiser}

So far, we have discussed two \gls{TTL} coupling models:
First, we have introduced the fit model, which successfully subtracts the \gls{TTL} coupling in noise runs as well as during the \gls{LXE} but does not provide a physical explanation of the cross-coupling.
Second, there is the analytical model, which is the first model explaining how the \gls{TTL} coupling depends on the system parameters, but its derivation is based on a lengthy analysis. 
Here, we show that we can build a model that, like the analytical model, describes how the \gls{TM} alignments change the \gls{TTL} coupling coefficients and is based only on the \gls{LXE} data.

We find this system by searching for coefficients that describe the dependency of the fitted coupling coefficients on the set-points best. 
Mathematically speaking, we compute the coefficients $k_{j}^\alpha$ that minimise the systems of equations
\begin{align}
0 = - C^\text{fit}_{ji} + \sum_\alpha k_{j}^\alpha \cdot \alpha_{i}\,,\qquad\ i=1,...,12 \,,
\label{eq:minimiser}
\end{align}
where
\begin{itemize}
\item $C^\text{fit}_{ji}$, $i=1,...,12$, are the coupling coefficients fitted in the bump frequency range for each of the twelve sub-experiments of the \gls{LXE} and for each \gls{S/C} degree of freedom $j\in\{\varphi,\eta,y,z\}$
\item $\alpha\in\{\varphi_1,\varphi_2,\eta_1,\eta_2,y_1,y_2,z_1,z_2,1\}$ are the \gls{TM} set-points chosen in each experiment, see Tab.~\ref{tab:longxtalk}, plus an additional entry accounting for coupling coefficient offsets $C_{j,0}$.
\end{itemize}
Note that we intentionally do not discard any of the $k_j^\alpha$'s that are assumed to be negligible based on the analytical modelling.
Showing that the minimising routine also yields small results for these coefficients is a further confirmation of our model Eq.~\eqref{eq:anamodel}.

We ran the \gls{LTPDA} minimiser `lscov' (least-squares algorithm) on Eqs.~\eqref{eq:minimiser} 
for each coefficient yielding
\begin{subequations}
\begin{align}
\begin{split}
C_\varphi^\text{min} &= \frac{0.15}{\text{rad}^2}\varphi_1+\frac{0.25}{\text{rad}^2}\varphi_2+\frac{0.02}{\text{rad}^2}\eta_1-\frac{0.01}{\text{rad}^2}\eta_2 \\
&+\frac{0.03}{\text{rad}}y_1+\frac{0.02}{\text{rad}}z_1+\frac{0.01}{\text{rad}}z_2+\frac{3.21\text{e-6m}}{\text{rad}} \,,
\label{eq:minimizer_Cphi}
\end{split} \displaybreak[3]\\ 
\begin{split}
C_\eta^\text{min} &= \frac{0.01}{\text{rad}^2}\varphi_1+\frac{0.01}{\text{rad}^2}\varphi_2+\frac{0.13}{\text{rad}^2}\eta_1+\frac{0.19}{\text{rad}^2}\eta_2 \\
&-\frac{0.01}{\text{rad}}y_1-\frac{0.01}{\text{rad}}y_2-\frac{0.01}{\text{rad}}z_2-\frac{12.26\text{e-6m}}{\text{rad}} ,
\label{eq:minimizer_Ceta}
\end{split} \displaybreak[3]\\ 
\begin{split}
C_y^\text{min} &= -\frac{0.97}{\text{rad}}\varphi_1+\frac{1.06}{\text{rad}}\varphi_2+\frac{0.15}{\text{rad}}\eta_1+\frac{0.01}{\text{rad}}\eta_2 \\
&-\frac{0.06}{\text{m}}y_1+\frac{0.01}{\text{m}}y_2-\frac{0.13}{\text{m}}z_2+5.97\text{e-6} \,,
\label{eq:minimizer_Cy}
\end{split} \displaybreak[3]\\ 
\begin{split}
C_z^\text{min} &= -\frac{0.02}{\text{rad}}\varphi_1-\frac{0.06}{\text{rad}}\varphi_2+\frac{0.97}{\text{rad}}\eta_1-\frac{1.01}{\text{rad}}\eta_2 \\
&+\frac{0.03}{\text{m}}y_2-\frac{0.04}{\text{m}}z_1+\frac{0.07}{\text{m}}z_2+43.53\text{e-6} \,.
\label{eq:minimizer_Cz}
\end{split}
\end{align}
\label{eq:minimizer_C}
\end{subequations}

\begin{table}
\begin{tabular}{c|llll}
\toprule
term & $C_\varphi$ & $C_\eta$ & $C_y$ & $C_z$ \\ 
\midrule
$\varphi_1$ & 0.02m\,/rad$^2$ & 0.00\,m/rad$^2$ & 0.01/rad & 0.01/rad \\
$\varphi_2$ & 0.02m\,/rad$^2$ & 0.00\,m/rad$^2$ & 0.02/rad & 0.02/rad \\
$\eta_1$    & 0.04m\,/rad$^2$ & 0.01\,m/rad$^2$ &  0.02/rad & 0.03/rad  \\
$\eta_2$    & 0.02m\,/rad$^2$ & 0.00\,m/rad$^2$ &  0.02/rad & 0.02/rad  \\
$y_1$       & 0.01\,m/rad & 0.00\,m/rad & 0.01 & 0.01 \\
$y_2$       & 0.02\,m/rad & 0.00\,m/rad & 0.01 & 0.02 \\
$z_1$       & 0.02\,m/rad & 0.00\,m/rad & 0.01 & 0.02 \\
$z_2$       & 0.03\,m/rad & 0.01\,m/rad & 0.02 & 0.03 \\
1           & 0.21e-6\,m/rad & 0.04e-6\,m/rad & 0.14e-6 & 0.19e-6 \\ 
\bottomrule
\end{tabular}
\caption{Rounded absolute errors of the coefficients in Eq.~\eqref{eq:minimizer_C} found by the minimising routine.}
\label{tab:minimiser_errors}
\end{table}

\begin{figure}
\flushright
\includegraphics[scale=0.29]{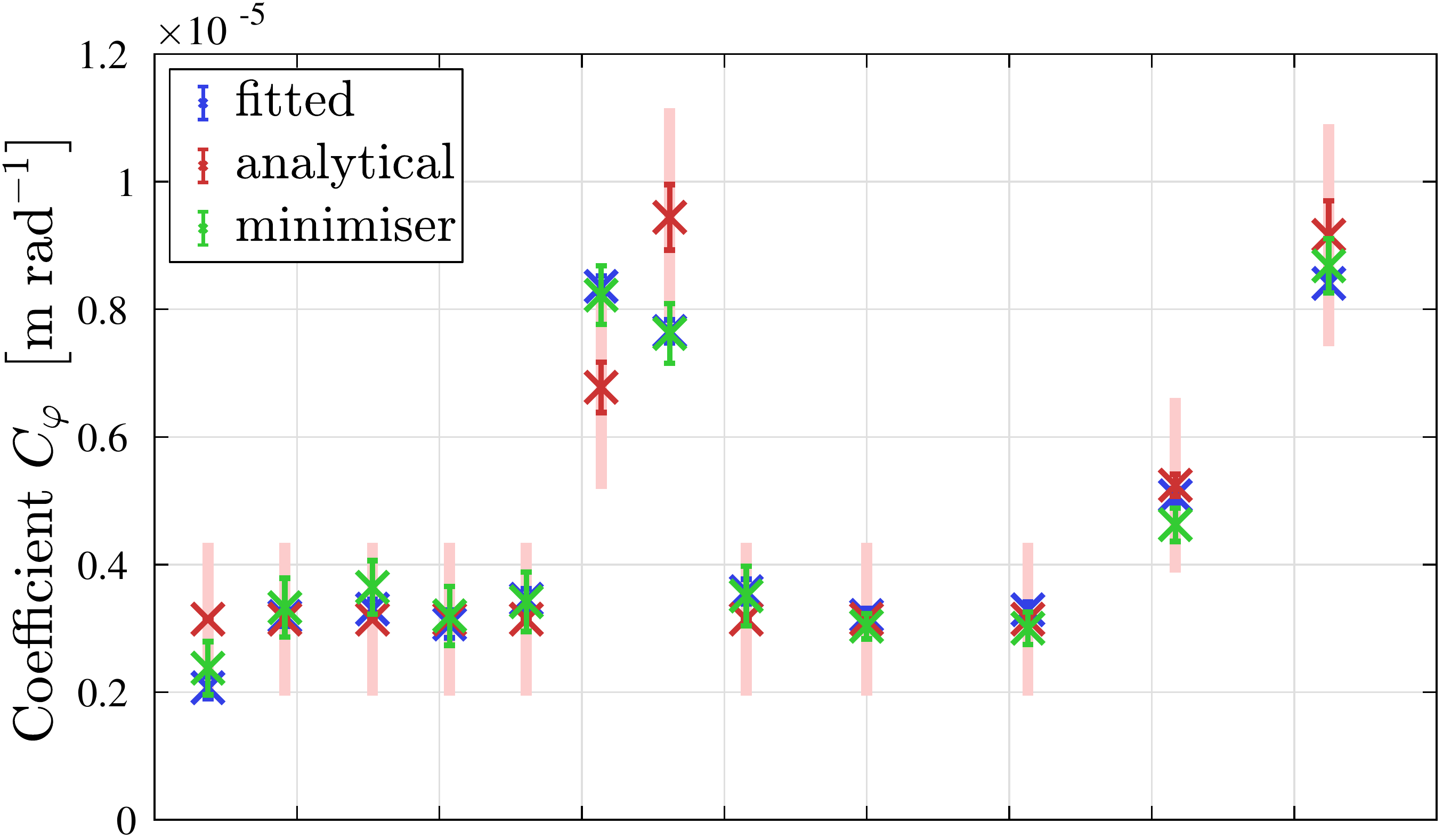}\,\,\ \\
\includegraphics[scale=0.29]{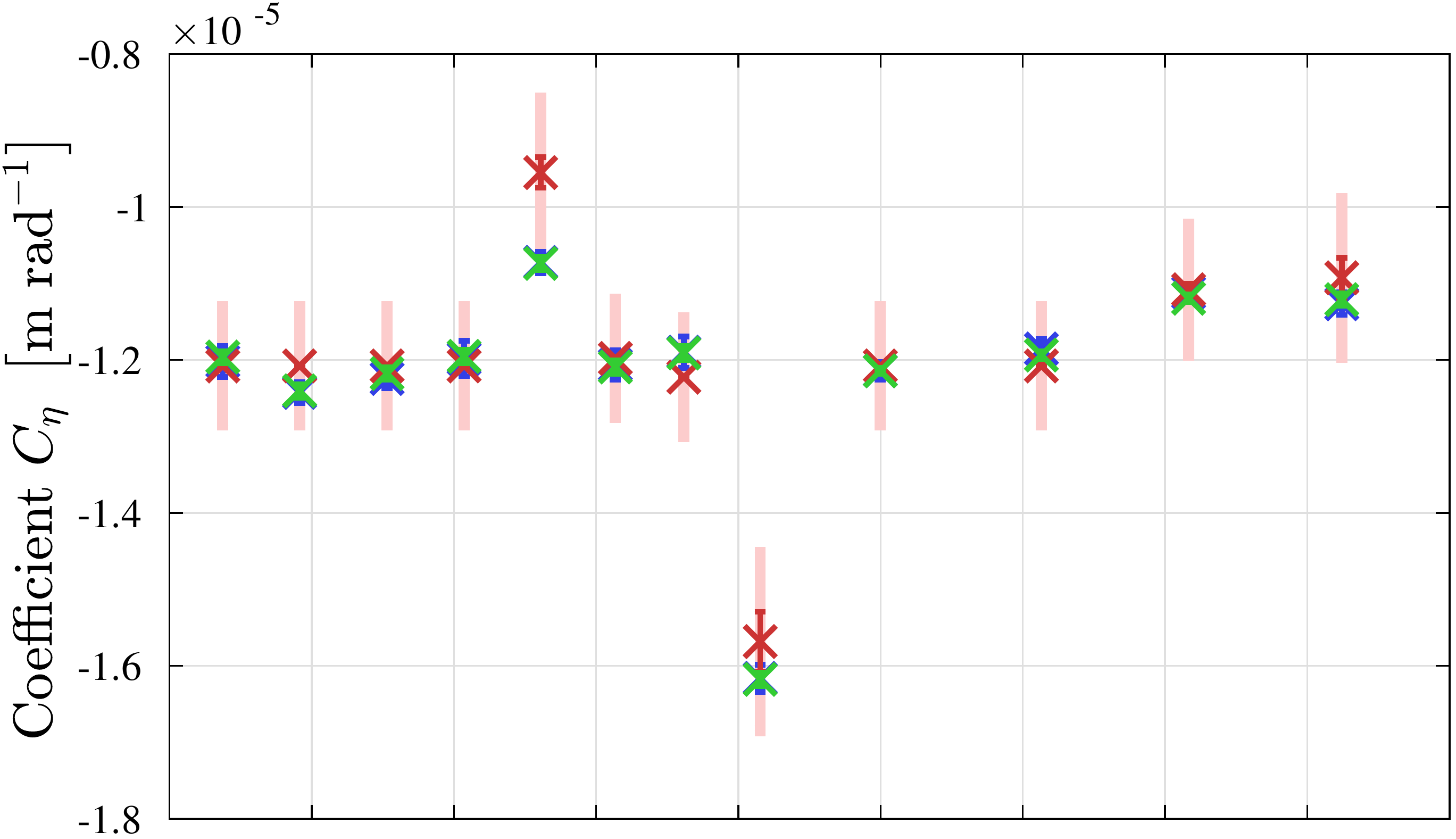}\,\,\ \\
\includegraphics[scale=0.29]{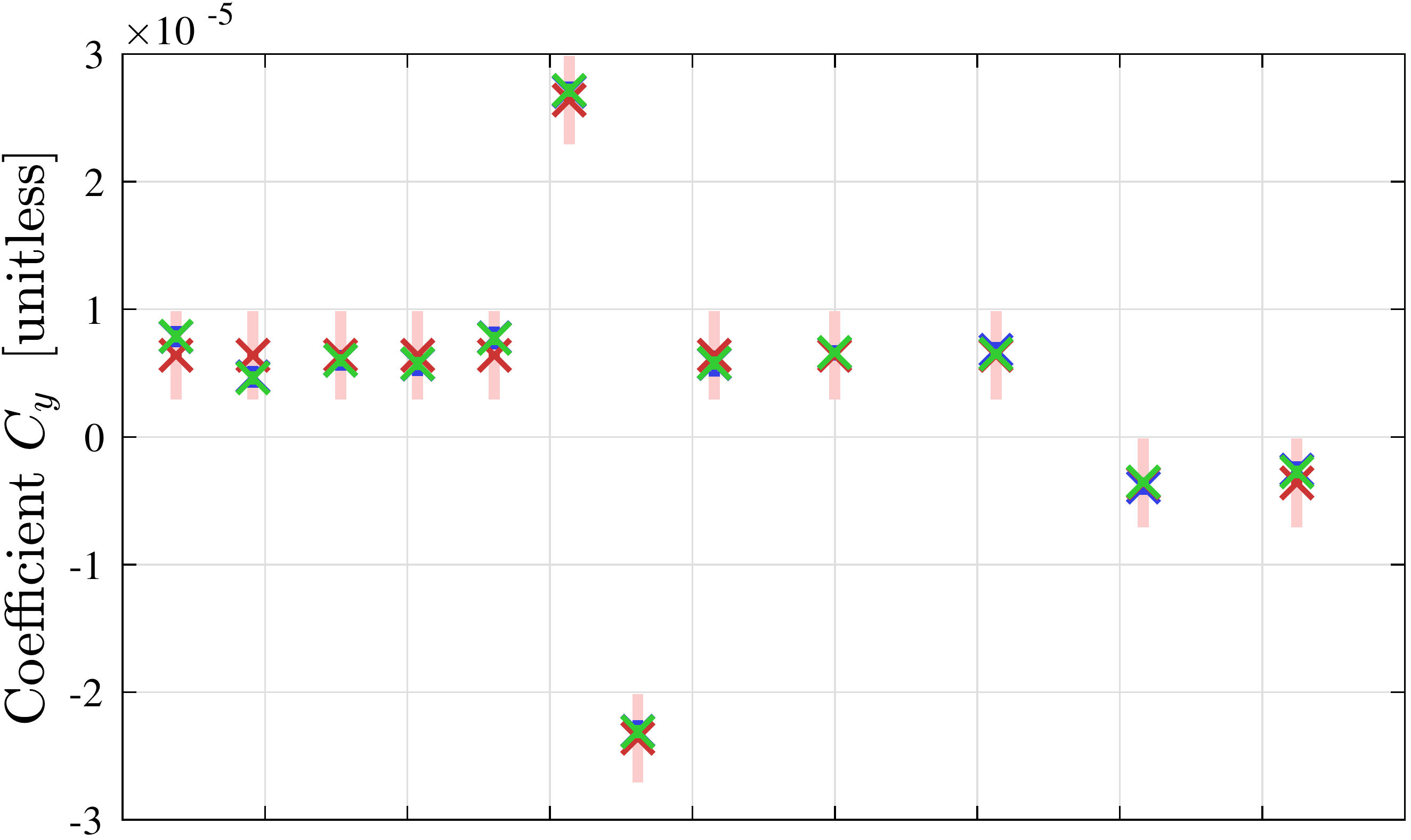}\,\,\ \\
\includegraphics[scale=0.29]{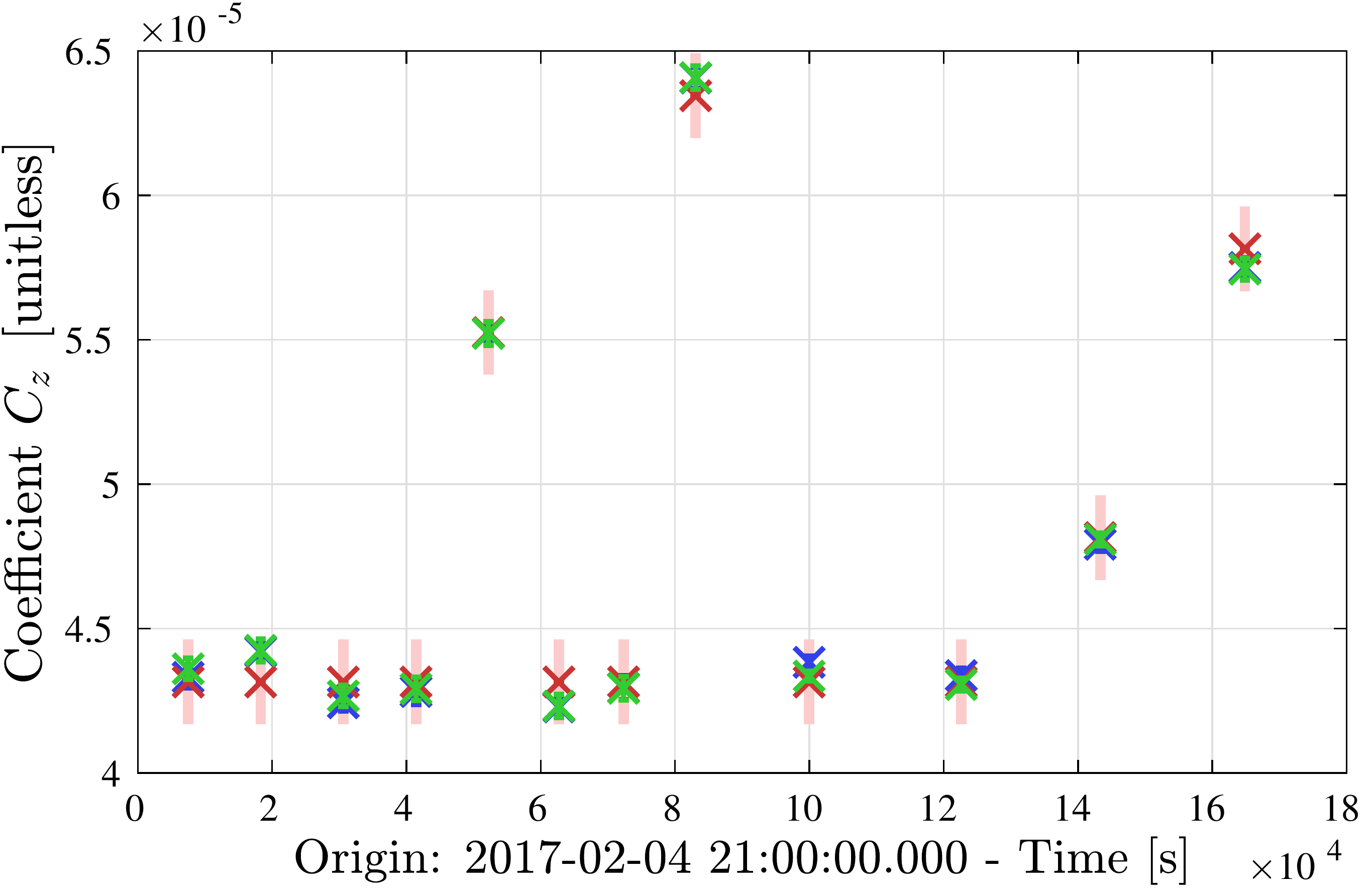} 
\caption{Comparison of the coefficients derived via the minimising routine with the fitted (fit range 20 to 70\,mHz) and analytically derived coupling coefficients. The minimiser and fit coefficients match well. The coefficient changes due to TM realignments show the same characteristic behaviour as analytically predicted with the deviations discussed in Sec.~\ref{sec:LXE_comparison}. The time is given in UTC.}
\label{fig:LXE_minimiser}
\end{figure}

The coefficients obtained via the minimising routine agree with the fitted coefficients within their error bars, see Fig.~\ref{fig:LXE_minimiser}. 

The coefficients found via the minimiser (Eq.~\eqref{eq:minimizer_C}) show similarities to the corresponding analytically derived coefficients (Eq.~\eqref{eq:anamodel_C}).
This is particularly true for the lateral coefficients $C_y$ and $C_z$, which mainly depend on the differential angular \gls{TM} alignment in the same plane. All factors that scale with other alignment parameters are comparatively small. 
Also, the angular coefficients $C_\varphi$ and $C_\eta$ show the most significant dependency on the same-plane angular \gls{TM} alignments. Here, the factors scaling this dependency differ for the analytical and the minimiser result by about 25\,\% but show the same characteristic behaviour.

We conclude that by varying set-points in a laser interferometric setup, a model showing the dependency of the \gls{TTL} coupling on the alignment parameters can potentially be found computationally.
This is particularly interesting in the case of \gls{LISA}, where the setup is too complex to evaluate a detailed \gls{TTL} model analytically.

\section{Test Mass Realignments}
\label{sec:TMalignments}

After analysing the \gls{LXE} in the previous section, we investigate here the three \gls{TM} realignments introduced in Sec.~\ref{sec:realingments_description}.
We show here how the realignments changed the \gls{TTL} coupling.
Also, we show that the observed \gls{TTL} noise changes can be explained with our new analytical model.

\subsection{Engineering Days}
\label{sec:TMalignments_ED}

At the beginning of the mission phase, \gls{TTL} was the largest noise contributor between 20 and 200\,mHz. 
In order to optimise the \gls{TM} orientations for \gls{TTL} suppression, injections at different set-points were applied to the \glspl{TM} during the \gls{ED}, which took place from the 14th to the
17th of March 2016 \cite{dlr2018,dlr2020}.
Based on these experiments, the \glspl{TM} were rotated by the angles shown in Tab.~\ref{tab:realignments}, which reduced the \gls{TTL} coupling by a factor of two but did not fully mitigate it \cite{Wanner2017}.

\begin{figure*}
\centering
\includegraphics[scale=0.29]{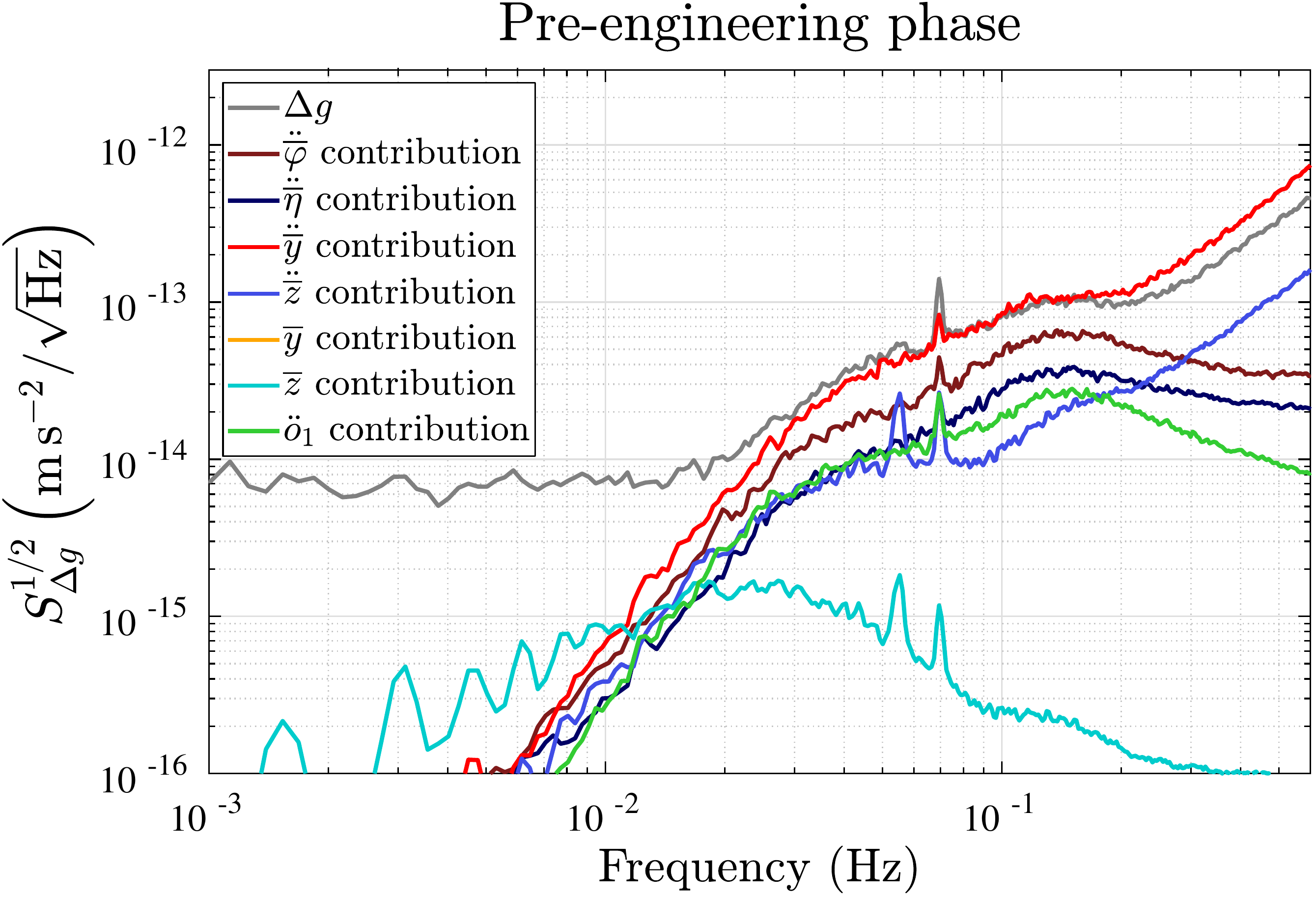}\quad\quad\
\includegraphics[scale=0.29]{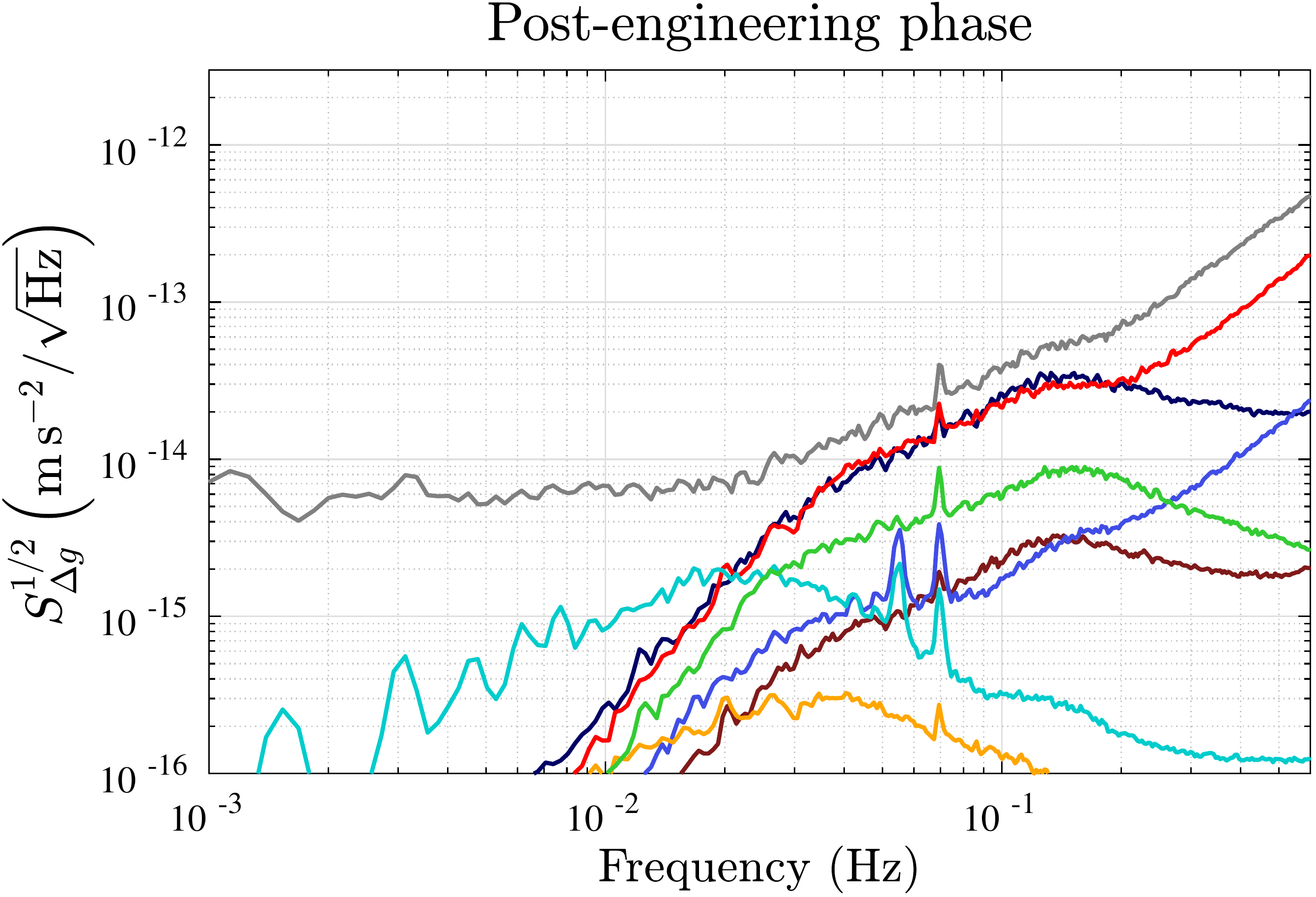} %
\caption{Noise contributors in the pre- and post-engineering phase of the ED. The noise contributions due to accelerations in $\varphi$, $y$, $z$ and $o_1$ were reduced due to the realignments. The largest noise contributors after the realignments were the accelerations in $\eta$ and $y$.}
\label{fig:ED_fit_contributors}
\end{figure*}

We start here with investigating how the \gls{TTL} coupling contributors changed due to this first realignment.
Therefore, we compare the contributing jitter noise before (\textit{pre-engineering phase}) and after (\textit{post-engineering phase}) the realignments. This is shown in Fig.~\ref{fig:ED_fit_contributors}. We see significant differences.
Due to the new alignment, the coupling of the accelerations along $z$ (blue curve) was well suppressed. 
The same applied to the coupling of the accelerations along $\varphi$ (brown) and $o_1$ (green). 
The main noise contributors after the realignment were the accelerations along $y$ (red), which was also dominant before and decreased approximately by a factor of three, and $\eta$ (dark-blue), which was almost not affected by the realignments. 

For the pre-engineering phase, we observe that the $\ddot{\bar{y}}$ noise contributor appears to be larger than the measured $\Delta g$ above 20\,mHz (left plot of Fig.~\ref{fig:ED_fit_contributors}). 
However, this noise is not part of the $\Delta g$ measurement.
The \gls{GRS} measurement of the lateral accelerations at frequencies above 0.2\,Hz are dominated by displacement readout noise \cite{Armano2017}. This measurement noise adds up to the true accelerations in $y$ and $z$ and is scaled by the corresponding coupling coefficient. Due to the large coupling of the $y$-accelerations in the pre-engineering phase (compare Tab.~\ref{tab:realignments_ana_C}), the high-frequency behaviour of the red curve in the left plot of Fig.~\ref{fig:ED_fit_contributors} is largely affected by this noise.
Note that we add this noise to $\Delta g$ when we subtract the \gls{TTL} coupling.  

For the comparison of the realignment strategy applied during the mission and the predictions of our analytical model (Eq.~\eqref{eq:anamodel}), we computed the optimal \gls{TM} alignment angles for cross-coupling suppression with this new model.
Therefore, we replaced the $C_{i,0}$ in Eq.~\eqref{eq:anamodel_C} by the fitted coupling coefficients for the pre-engineering noise (Tab.~\ref{tab:realignments_ana_C}) and solved Eq.~\eqref{eq:anamodel_C} for the \gls{TM} alignments for which the analytical coefficients $C_i$ would be zero.
By this, we found the angles shown in Tab.~\ref{tab:realignments_ana}.
The comparison with the rotations applied during the mission shows a large difference in the case of the angles $\varphi_2$, $\eta_1$ and $\eta_2$. However, the $\varphi_1$ angles as well as the differential $\eta$ angles (considered in the computation of the coupling coefficient $C_z$) are approximately the same. 
Hence, $C_z$ was well suppressed (compare Fig.~\ref{fig:ED_fit_contributors}).

\begin{table}
\begin{tabular}{ll|ccc}
\toprule
coeff. & [unit] & pre-ED (fit) & post-ED (fit) & post-ED (ana.) \\ 
\midrule
$C_\varphi$ & [$\upmu$m/rad] &  14.3 &  -0.7 &  -1.8 \\
$C_\eta$    & [$\upmu$m/rad] &  -7.5 &  -6.7 &  -7.5 \\
$C_y$       & [10$^{-6}$]    & -50.2 & -13.6 & -12.5 \\
$C_z$       & [10$^{-6}$]    &   7.6 &   1.1 &   0.6 \\
\bottomrule
\end{tabular}
\caption{TTL coupling coefficients during the pre- and post-engineering phase. The first two columns show the fitted coefficients. The last column provides the coefficients computed by the analytical TTL noise model (Eq.~\eqref{eq:anamodel}) for the post-engineering phase based on the pre-engineering fit-results. These predictions are close to the fit results.}
\label{tab:realignments_ana_C}
\end{table}

\begin{table}
\begin{tabular}{cc|cc}
\toprule
DoF & [unit] & mission & analytical \\ 
\midrule
$\varphi_1$ & [$\upmu$rad] & -59.25 & -60.3  \\
$\varphi_2$ & [$\upmu$rad] & -21.35 & -10.3  \\
$\eta_1$    & [$\upmu$rad] &  -3.5  &  15.8  \\
$\eta_2$    & [$\upmu$rad] &   3.5  &  23.4  \\
\bottomrule
\end{tabular}
\caption{TM realignments for TTL suppression at the time of the ED derived via the analytical model (Eq.~\eqref{eq:anamodel}) compared to the alignments performed at that time of the mission. These angles largely deviate.
Only $\varphi_1$ and the differential $\eta$ angle are comparable.}
\label{tab:realignments_ana}
\end{table}

\begin{figure*}
\includegraphics[scale=0.335]{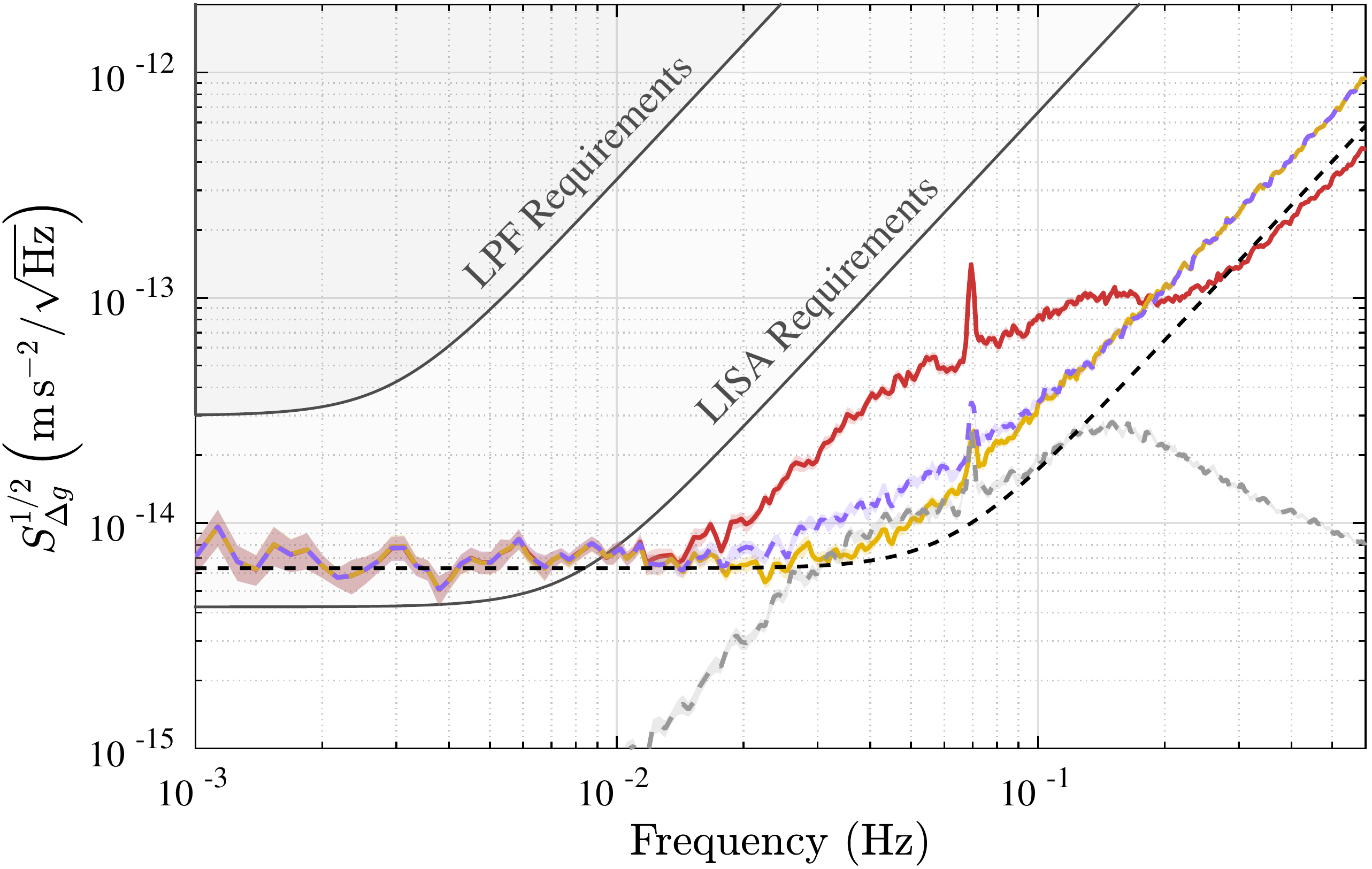}\hfill
\includegraphics[scale=0.335]{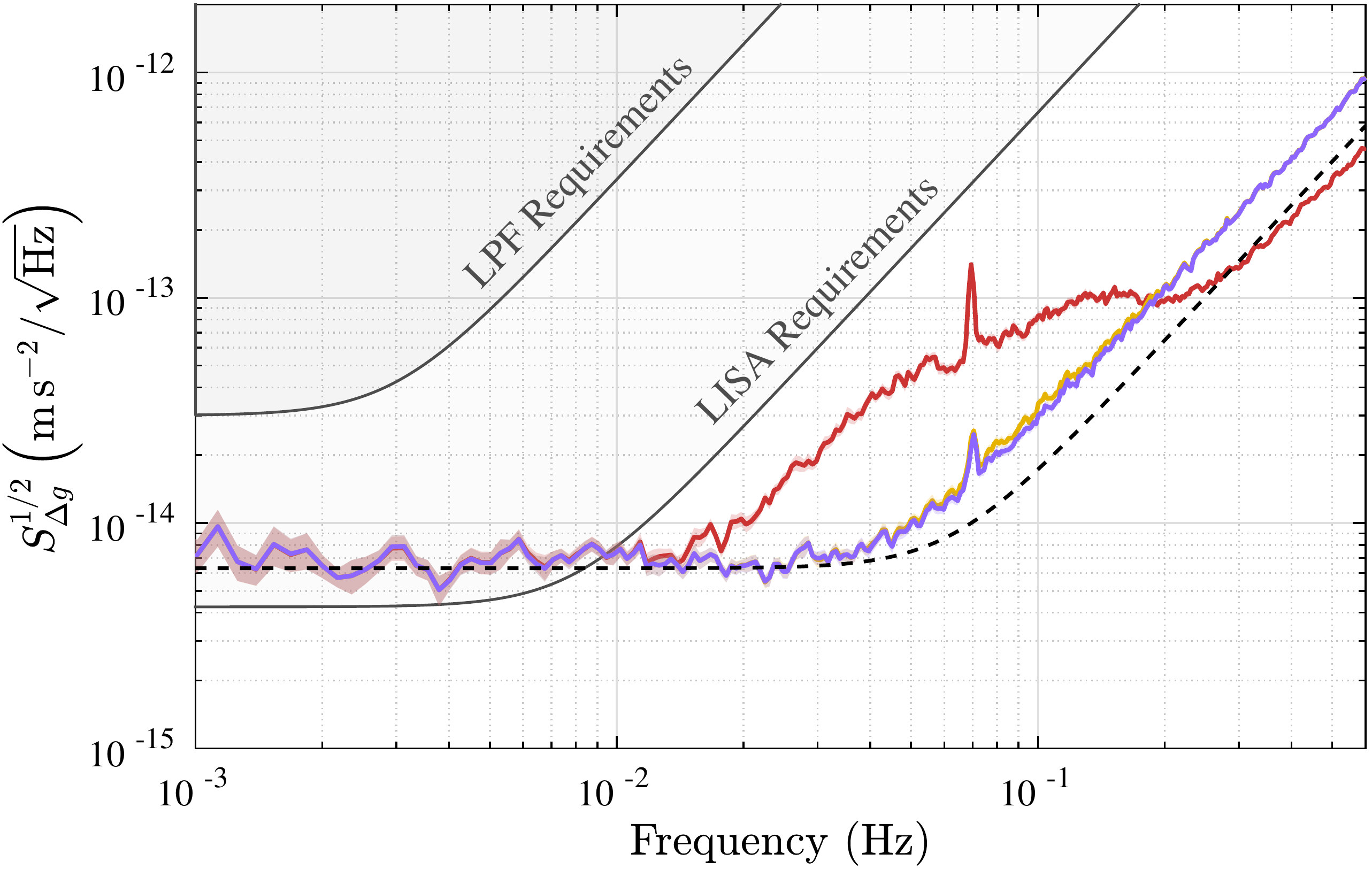}\\[2ex]
\flushright{
\includegraphics[scale=0.8]{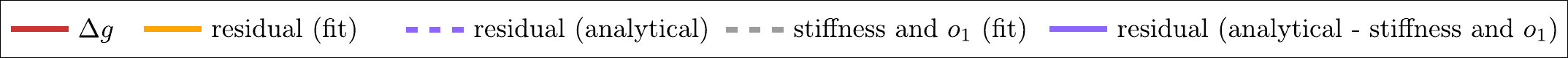}}
\caption{
  Analytical modelling of the expected noise after realigning the TMs in accordance to the analytically derived angles (right hand side of Tab.~\ref{tab:realignments_ana}).
  The results validate the analytical model (Eq.~\eqref{eq:anamodel}).
  Red curves: measured $\Delta g$ before the realignment (pre-engineering data). 
  Yellow curve: residual after the subtraction of the fit model (Eq.~\eqref{eq:fitmodel}) from the measured $\Delta g$ before the realignment.
  Dashed black line: LPF performance model.
  Left figure: The analytical model (Eq.~\eqref{eq:anamodel}) for the analytically derived realignment angles (Tab.~\ref{tab:realignments_ana}) added to the pre-engineering noise (dashed purple line), and the sum of the fitted stiffness and $o_1$-noise contributions (dashed grey line).
  Right figure: The analytical model added to the pre-engineering noise minus the fitted stiffness and $o_1$-noise contributors (purple line). 
  The overlap of the purple and the yellow curve indicates that a full TTL noise mitigation via realignment could have been achieved via the 
  analytically derived TM angles. 
   The noise added above 200\,mHz originates from the measurement noise of the GRS scaled by the TTL coefficients. 
  It would not be present if performing the realignment.}%
\label{fig:ED_ana_subtract}
\end{figure*}

Next, we reproduced the \gls{TTL} noise suppression we would have achieved with the analytical results.
Since we could not repeat the measurements themselves, we computed the expected residual noise by adding the analytical model to the pre-engineering measurements. 
Note that this addition effectively reduces the noise as the analytical model explains how the coupling coefficients would change.
For this, we inserted the analytically derived realignment angles into the analytical model (Eq.~\eqref{eq:anamodel_C} with $C_{i,0}=0$) and and added it to the pre-engineering $\Delta g$.
We then find the residual noise provided by the dashed purple curve in the left-hand side of Fig.~\ref{fig:ED_ana_subtract}.
It exceeds the noise floor we gain when subtracting the \gls{TTL} noise via the fit model~\eqref{eq:fitmodel} (yellow curve) between 20 and 90\,mHz. This noise almost only originates from the coupling of \gls{S/C} accelerations along the optical axis ($o_1$-term).
While we have no model on hand describing this noise contribution, it showed a decrease when the overall \gls{TTL} coupling decreases (e.g., Fig.~\ref{fig:ED_fit_contributors}).
Therefore, we subtract it additionally from the data (together with the small stiffness terms), yielding the solid purple curve at the right-hand side of Fig.~\ref{fig:ED_ana_subtract}.
This purple curve now coincides with the noise we get when subtracting the \gls{TTL} coupling via the fit model~\eqref{eq:fitmodel}. 
This shows that we could have achieved the after-fit sensitivity only by a realignment of the \glspl{TM} in accordance with the new analytically predicted angles. 
Thereby, the increased noise at higher frequencies was only added due to the addition of the analytical model, which is affected by the \gls{GRS} sensing noise. This noise increase does not show up when actually performing the realignments of the \glspl{TM} by the analytically derived angles. 

\begin{figure*}
\includegraphics[scale=0.335]{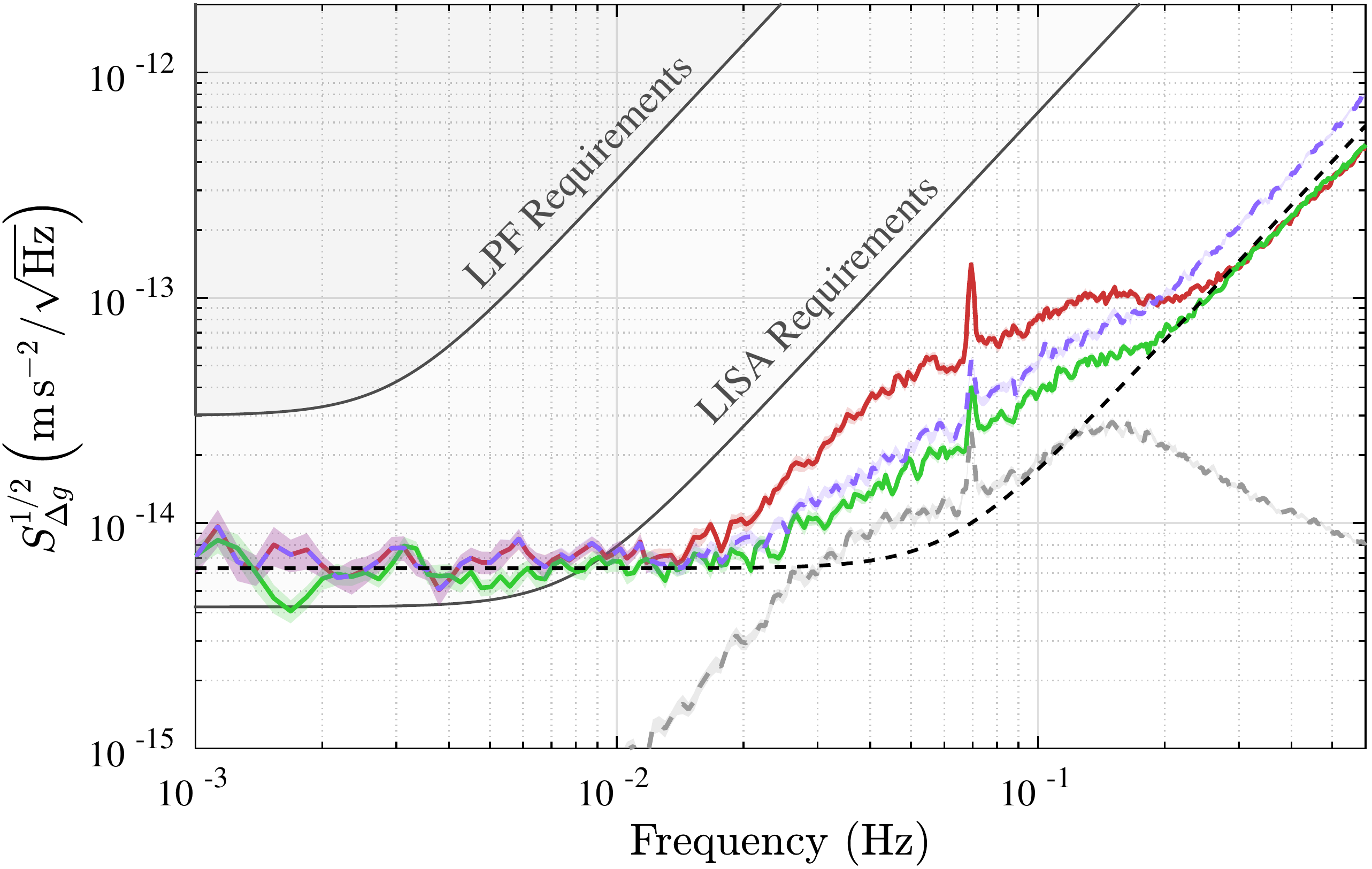}\hfill
\includegraphics[scale=0.335]{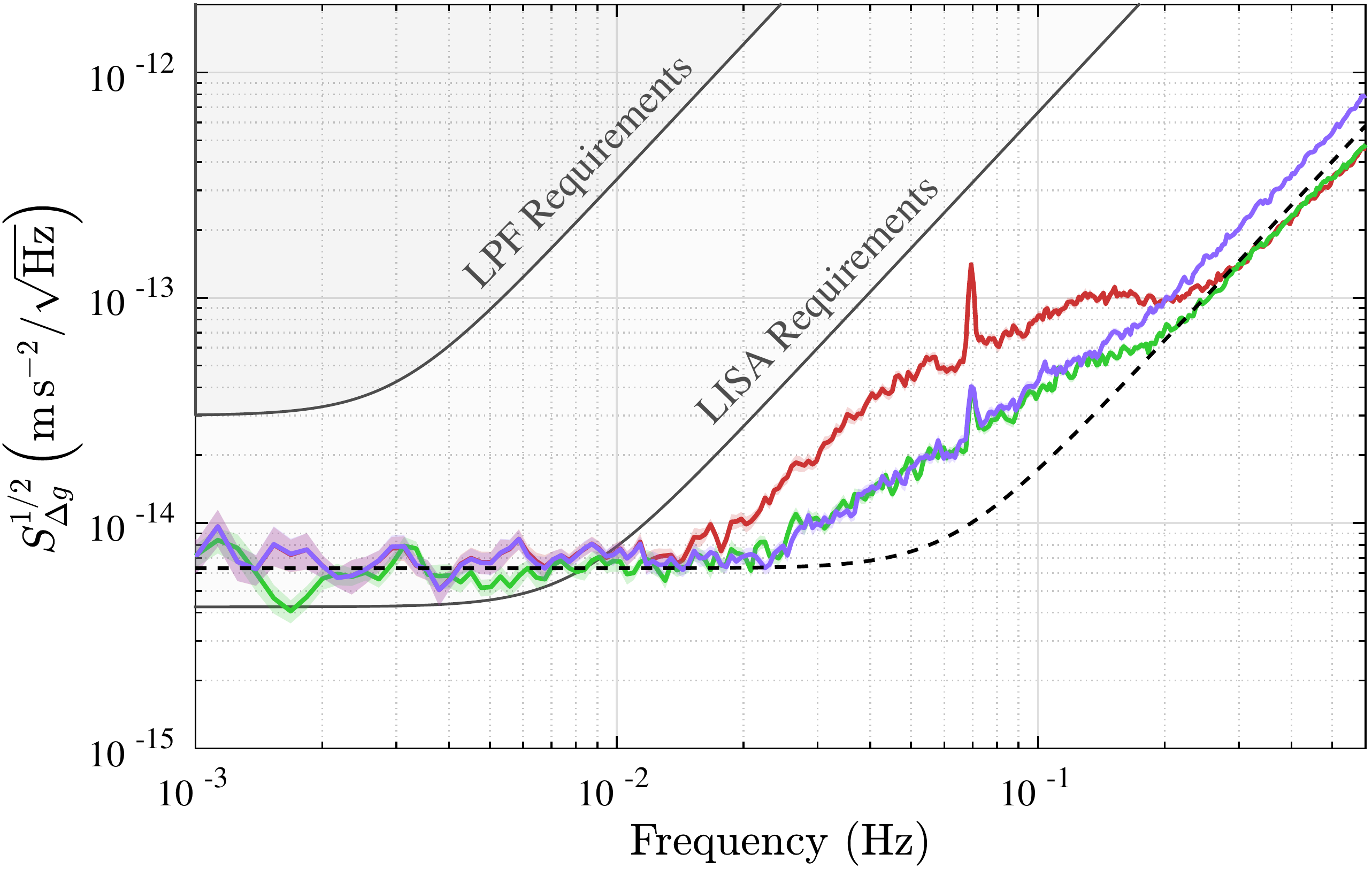} \\[2ex]
\flushright{
\includegraphics[scale=0.8]{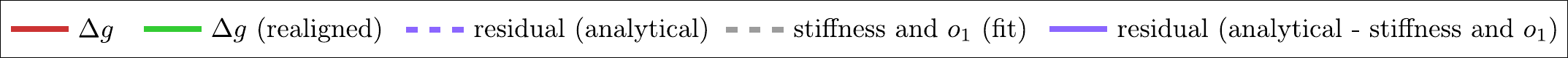}}
\caption{Analytical modelling of the expected noise after realigning the TMs to the angles defined during the mission (left hand side of Tab.~\ref{tab:realignments_ana}).
  The results further validate the analytical model (Eq.~\eqref{eq:anamodel}).
  Red curves: measured $\Delta g$ before realignment. 
  Green curve: measured $\Delta g$ after the realignments. 
  Dashed black line: LPF performance model.
  Left figure: The analytical model~\eqref{eq:anamodel} evaluated for the realignments applied during the engineering days added to the pre-engineering noise (dashed purple line), and the sum of the fitted stiffness and $o_1$-noise contributions (dashed grey line).
  Right figure: Analytical model added to the pre-engineering noise minus the fitted stiffness and $o_1$-noise contributors (purple line). 
  The overlap of the purple and the green curve indicates that the observed noise decrease due to the commanded realignment could have been predicted with the analytical model.
  In comparison to Fig.~\ref{fig:ED_ana_subtract} we see that not the realignment angles defined during the mission but the new analytically derived ones would have yielded a suppression of TTL noise.}
\label{fig:ED_ana_realign}
\end{figure*}

After having demonstrated that we could have suppressed the \gls{TTL} coupling noise using the new analytically predicted angles (Fig.~\ref{fig:ED_ana_subtract}), we confirmed our analytical model further with a second simulation:
We reproduced the \gls{TTL} coupling measured during the post-engineering phase using the analytical model. 
Therefore, we inserted the alignment angles applied during the mission into Eq.~\eqref{eq:anamodel} (setting the $C_{i,0}=0$) and add it to the pre-engineering noise (see Fig.~\ref{fig:ED_ana_realign}).
Additionally, we had to subtract the noise from the $o_1$-term, which was reduced due to the realignments.
As shown at the right-hand side of Fig.~\ref{fig:ED_ana_realign}, we then find the same relative accelerations as measured after the realignments (green curve).
This simulation also supports our conclusion that the \gls{TM} alignments applied during the \gls{LPF} mission were not ideally chosen.
Note that we see, as in Fig.~\ref{fig:ED_ana_subtract}, a noise increase at high frequencies added due to the measurement noise included in the analytical model. Again, this is only an artefact of the modelling and would not occur in case of an actual realignment.

\subsection{TM realignments in June 2016}
\label{sec:TMalignments_June}

Since the \gls{TM} realignments during the \gls{ED} did not fully mitigate the \gls{TTL} coupling, the realignment angles were corrected on 19th and 25th June 2016 to the angles shown in Tab.~\ref{tab:realignments}.
Like in the previous sub-section concerning the realignment during the engineering days, we investigate here effect of these new alignments using the fit results and our analytical predictions.
We first analyse the changes of the \gls{TTL} noise contributors due to the realignments  (Sec.~\ref{sec:TMalignments_June_contributors}).
Furthermore, we directly compare the angular realignments set in June 2016 with the analytically computed alignment angles (Sec.~\ref{sec:TMalignments_June_angles}) as well as the fitted and the analytically derived coupling coefficients (Sec.~\ref{sec:TMalignments_June_coefficients}).

\subsubsection{Investigated timespans}
\label{sec:TMalignments_June_timespans}

For the computation of the \gls{TTL} coefficients and the therefrom derived noise contributors, we considered the noise runs closest to the \gls{TM} realignment times (Tab.~\ref{tab:realignments}).
The chosen noise runs all either ended about a day before or started about a day after the investigated realignment.

We could not use the result from the post-engineering phase for the analysis of the first realignment in June 2016 since the coupling coefficients were found to drift for long timespans. This is further investigated in \cite{LPFstabi23}.

The noise run before the realignment on 19th June started on 15th June at 1:30\,PM and lasted until 18th June at 8\,AM UTC.
Although no \gls{TM} realignments were performed between this noise run and the post-engineering phase, we found that the noise contributors changed in the meantime (compare Fig.~\ref{fig:ED_fit_contributors} and Fig.~\ref{fig:Realignment_DOY168_contributors}, which is further discussed in the following sub-section).
These changes are most likely related to relaxations in the optical system \cite{Armano2022_OMS}.
A more detailed discussion can be found in \cite{LPFstabi23}.

Our analysis of the effect of the first \gls{TM} realignment relies further on the noise run starting on 20th June at 8\,AM, i.e.\ subsequent to the realignment. 
This noise run lasted until 8\,AM at the day before the second and last alignment in June 2016. %

Unfortunately, there was no dedicated noise run for the investigation of the effect of this last \gls{TM} realignment.
The earliest following noise run took place two weeks later during the \gls{DRS} operations, and the measurements were additionally affected by temporary temperature increases \cite{Armano2019_temperature}. As we show in \cite{LPFstabi23}, these also changed the coupling coefficients such that the data is not suited for testing how the realignment affected the TTL coupling noise.
Therefore, we show here instead the results derived for a time segment of one hour (7:00--08:00\,AM on 26th June 2016) shortly after the realignments, having low noise.

\subsubsection{Noise contributor changes due to the realignments}
\label{sec:TMalignments_June_contributors}

\begin{figure}
\centering
\includegraphics[scale=0.31]{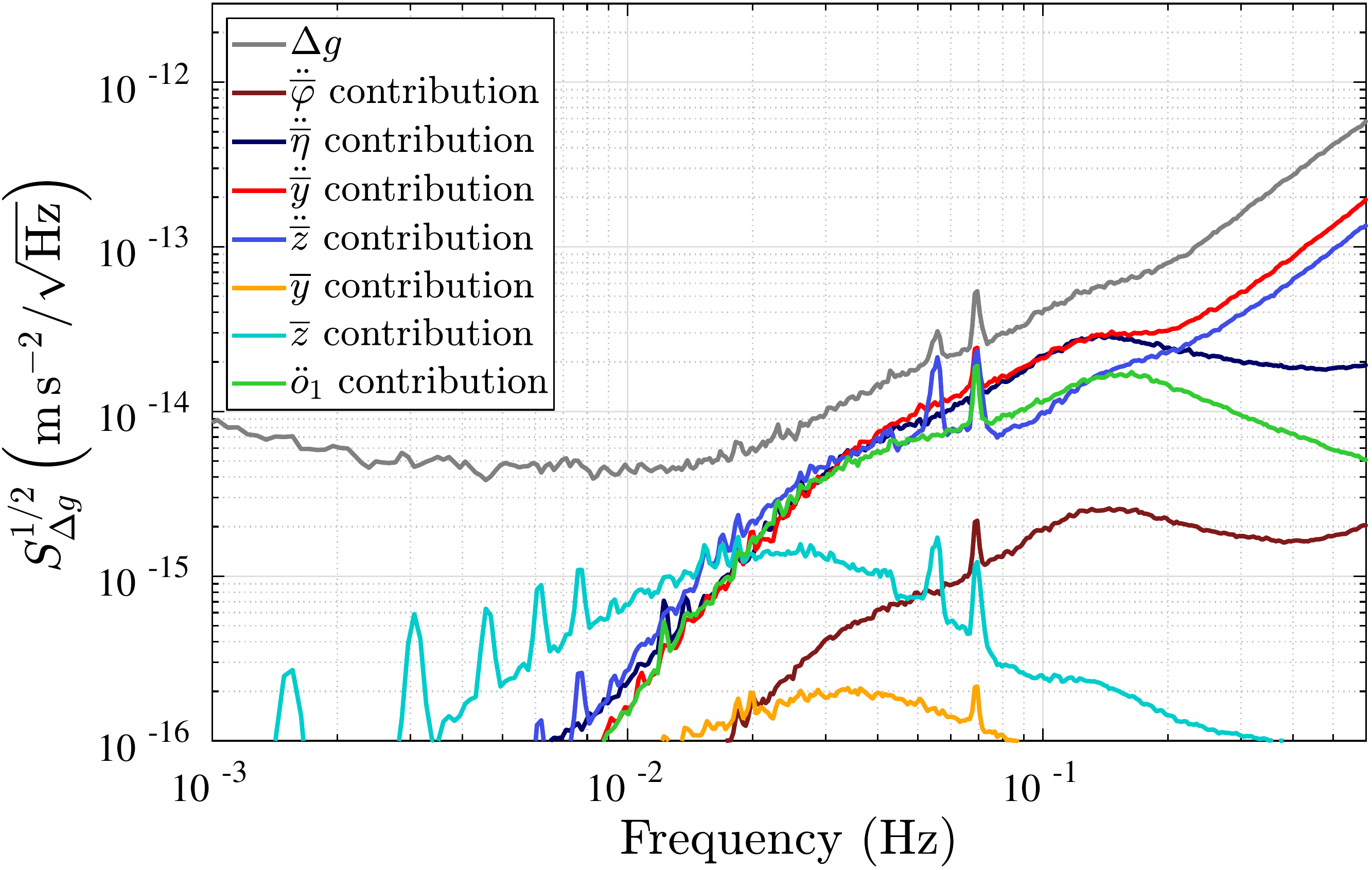}
\caption{Noise contributors during the noise run prior to the TM realignments in June 2016. The noise run started on 15th June and lasted almost three days until 18th June 2016.
  The coupling of the acceleration in $y$ and $\eta$ was the main cause of the TTL bump between 20 and 200\,mHz.}
\label{fig:Realignment_DOY168_contributors}
\end{figure}

\begin{figure}
\includegraphics[scale=0.31]{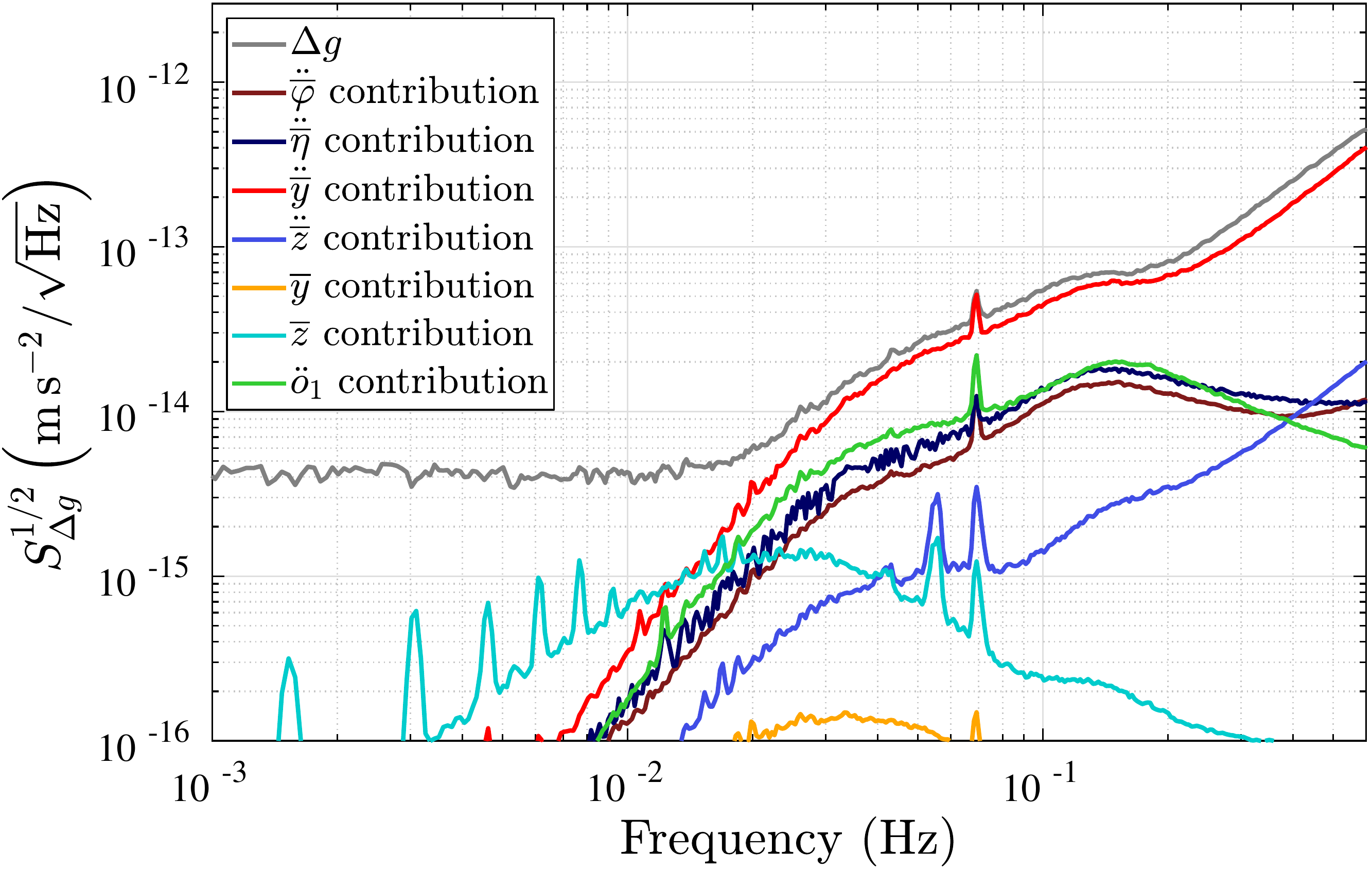}
\caption{Noise contributors during the noise run in between the two TM realignments in June 2016. It started a day after the realignment performed on 19th June 2016 and lasted for four days.
  In comparison to Fig.~\ref{fig:Realignment_DOY168_contributors}, we see a decrease of the $\eta$- and $z$-acceleration coupling but an increase of the $\varphi$- and $y$-acceleration noise. The overall TTL noise increased between 20 and 200\,mHz.}
\label{fig:Realignment_DOY172_contributors}
\end{figure}

In this sub-section, we analyse the effect of the \gls{TM} realignments performed in June 2016 on the \gls{TTL} noise contributors.
We start with comparing the noise contributions prior to the realignments (Fig.~\ref{fig:Realignment_DOY168_contributors}) with the noise contributors after the realignment on 19th June (Fig.~\ref{fig:Realignment_DOY172_contributors}).
As can be seen in these figures, the noise due to the $\varphi$- and $y$-accelerations significantly increased while the coupling of the $z$-accelerations decreased. 
This increase was due to a sign error in the application of the changes.
In total, it yielded larger \gls{TTL} coupling noise (bump of the grey curve between 20 and 200\,mHz). The \gls{ASD} increased by almost 50\,\% in this frequency regime.
 
\begin{figure}
\includegraphics[scale=0.31]{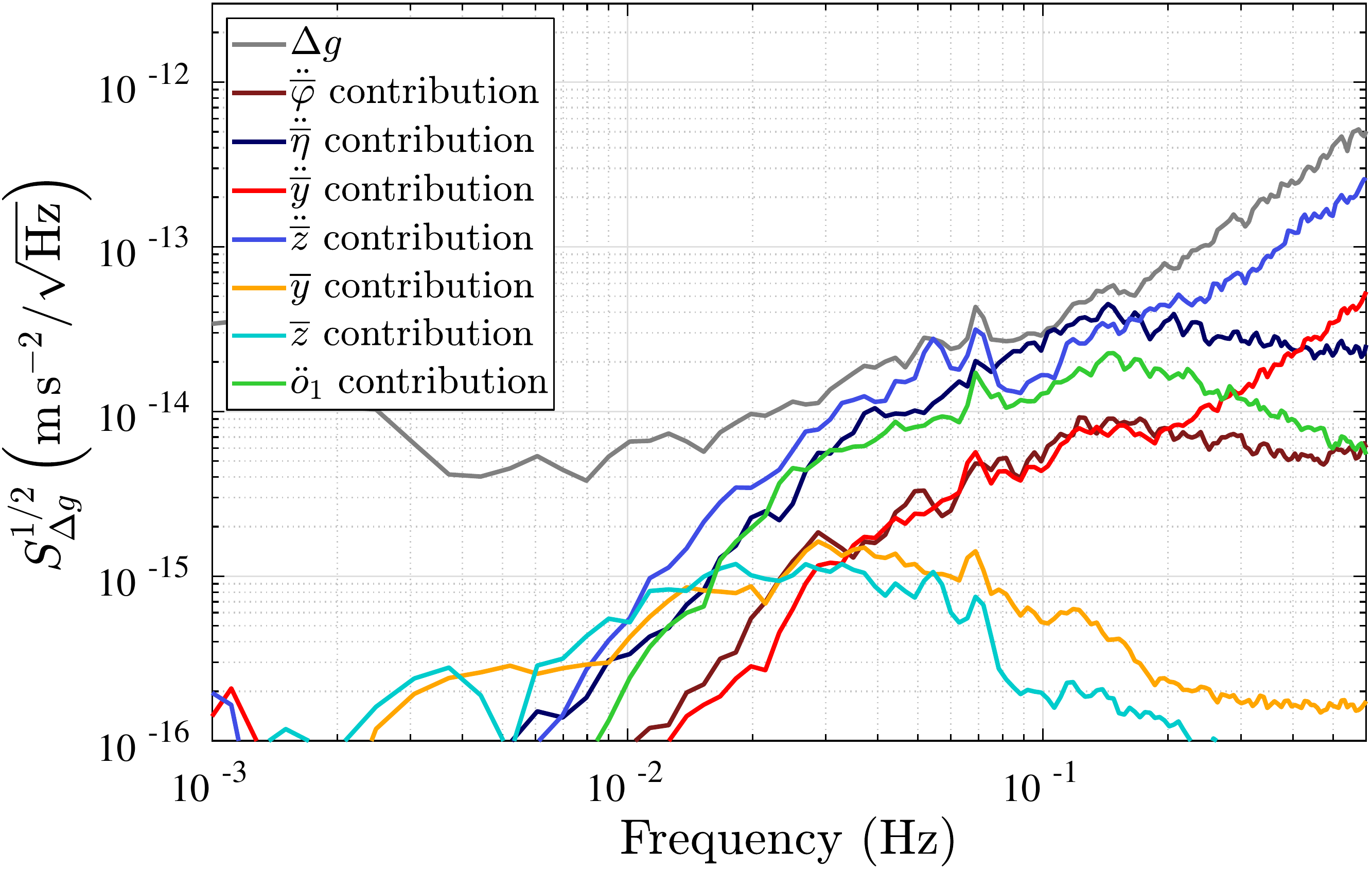}
\caption{Noise contributors after the realignment on 25th June 2016. Considered are data from 26th June 2016 from 7:00 to 8:00\,AM UTC.
  Due to this realignment, the $\varphi$- and $y$-acceleration noise decreased, but the $\eta$- and $z$-acceleration coupling increased again. In sum, the TTL bump is smaller than before this realignment.}
\label{fig:Realignment_DOY177_contributors}
\end{figure}

Consequently, the \glspl{TM} realignments were inverted on 25th June 2016.
This led to a mitigation of the $\varphi$- and $y$-acceleration coupling but unfortunately increased the noise in the orthogonal plane again, i.e.\ mostly the $z$-acceleration coupling, see Fig.~\ref{fig:Realignment_DOY177_contributors}.

Imagining a \gls{TM} alignment corresponding to the final $\varphi$-angles but the $\eta$-angle after the first June alignment, the full \gls{TTL} coupling noise would have been smaller than for each of the applied alignment settings
(combine the brown/red curves in Fig.~\ref{fig:Realignment_DOY177_contributors} and blue curves in Fig.~\ref{fig:Realignment_DOY172_contributors}).
From this observation, we conclude that an almost complete \gls{TTL} suppression via realignment could have been achieved during the mission if the \glspl{TM} would not have been realigned in $\eta$ on 25th June 2016.

\subsubsection{Comparison of the angular realignments set during the mission and derived analytically}
\label{sec:TMalignments_June_angles}

Since the cross-coupling of the $z$-accelerations into the $\Delta g$ readout increased in between the \gls{ED} and the realignments in June 2016 (see Figs.~\ref{fig:Realignment_DOY168_contributors} and~\ref{fig:Realignment_DOY172_contributors}), the differential $\eta$ angle must have increased, too.
Therefore, the analytically derived \gls{TM} alignment angles shown in Tab.~\ref{tab:realignments_ana} are outdated.
We rederived these \gls{TM} realignment angles. 
Analogously to before, we computed by Eq.~\eqref{eq:anamodel_C} the \gls{TM} alignments needed to counteract the fitted \gls{TTL} coefficients for the noise run prior to the realignments. This computation took the coupling coefficients into account, that were fitted to the prior noise run with \gls{TM} rotation angles as set during the \gls{ED}. Therefore, these angles are then added to these previous \gls{TM} alignments. The sum yields the total realignment angles, respectively. 
The result is shown in the right column of Tab.~\ref{tab:realignments_June_ana}.

The differential analytical $\eta$-angle ($\eta_1-\eta_2$) differs less than 1\,$\upmu$rad from the differential $\eta$-angle after the first realignment in June 2016 (cf.\ $\eta_1 - \eta_2$ in the first and third value columns of Tab.~\ref{tab:realignments_June_ana}). Due to this realignment, the $z$-acceleration noise was well suppressed (cf.\ Fig.~\ref{fig:Realignment_DOY172_contributors} and Eq.~\eqref{eq:anamodel_Cz}).
Likewise, the differential $\varphi$-angles ($\varphi_2-\varphi_1$) derived analytically and set on 25th June during the mission differ by less than 1\,$\upmu$rad. 
Correspondingly, the $y$-acceleration noise contribution was well suppressed (cf.\ Fig.~\ref{fig:Realignment_DOY177_contributors} and Eq.~\eqref{eq:anamodel_Cy}).

However, the individual analytical angles significantly differed from the ones set during the mission.
E.g., the analytically derived $\eta$-angles are both approximately 10\,$\upmu$rad larger than the angles set on 19th June. This difference increased even further after the last \gls{TM} realignment. 
Consequently, the \gls{S/C}-jitter coupling in $\eta$ was not well suppressed after both realignments (see Figs.~\ref{fig:Realignment_DOY172_contributors} and~\ref{fig:Realignment_DOY177_contributors}, compare with Eq.~\eqref{eq:anamodel_Ceta}), which supports our analytical result.
The $\varphi$-angles after the last \gls{TM} alignment differ by about 3\,$\upmu$rad from the corresponding analytical computations. 
By Eq.~\eqref{eq:anamodel_Cphi}, a small decrease of the $\varphi$-acceleration noise would have been possible by an adaption of these alignment angles.
However, the $\varphi$-acceleration noise was already low due to the chosen alignment angles.

\begin{table}
\begin{tabular}{ll|ccc}
\toprule
DoF & [unit] & 19.06.\ 8:20 & 25.06.\ 8:00 & analytical \\
\midrule
$\varphi_1$ & [$\upmu$rad] & -56.32 & -61.2 & -63.75 \\
$\varphi_2$ & [$\upmu$rad] & -33.01 &  -9.7 & -13.05 \\
$\eta_1$    & [$\upmu$rad] &  -2.14 &  -4.9 &   8.3  \\
$\eta_2$    & [$\upmu$rad] &  10.3  &  -3.3 &  21.6  \\
\bottomrule
\end{tabular}
\caption{Comparison of the TM realignment angles set in June 2016 for TTL suppression with the alignment angles derived analytically based on the TTL coupling during the noise run before both alignments (15.06.\ 13:30 until 18.06.\ 8:00).
Times are given in UTC.}
\label{tab:realignments_June_ana}
\end{table}

\subsubsection{Comparison of the fitted and the analytical coupling coefficients}
\label{sec:TMalignments_June_coefficients}

For further confirmation of the analytical \gls{TTL} model (Eq.~\eqref{eq:anamodel}), we compare the coupling coefficients fitted after the \gls{TM} realignments with their analytical prediction.
The latter was derived by replacing the constant coefficient offsets $C_{i,0}$ in Eq.~\eqref{eq:anamodel_C} by the fitted coupling coefficients computed for the time before the realignments.
Moreover, the \gls{TM} alignment angles in Eq.~\eqref{eq:anamodel_C} were substituted by the angles in Tab.~\ref{tab:realignments}. 

Instead of plotting the resulting noise reduction by the analytical model as shown in Fig.~\ref{fig:ED_ana_realign}, we compare here the coupling coefficients directly. 
If both are close to each other, the \gls{TTL} reduction during the mission and the analytical prediction would agree.

Both, the fitted and the analytically derived coupling coefficients, are summarised in Tab.~\ref{tab:realignments_June_coeffs}.
In the first column, we show the result for the noise run prior to the realignments. Here, only the fitted coefficients are given. The two following columns contain the computations for the noise run between the June realignments and the timespan of one hour after the last \gls{TM} realignment. %
The analytical coefficients show how the coefficients are expected to change due to the \gls{TM} rotations with respect to the fitted coefficients before the corresponding realignment.

The deviation between both sets of coefficients is in general small. The largest deviation occurs for the $C_y$ coefficients computed for the one-hour timespan after the last realignment. 
However, also this deviation would lie within 
the variations observed for the fitted $C_y$ coefficients during the noise run after the \gls{LXE} (Tab.~\ref{tab:LXE_errors}).
Although this noise run took place several months after the realignments, the $C_y$-coefficient has not changed much. 
Therefore, we assume the deviations to be applicable here.

In conclusion, the analytical model predicts within the 95\,\% confidence interval the correct \gls{TTL} noise coupling reduction due to the applied \gls{TM} realignments. 
Thus, we expect that a \gls{TTL} coupling suppression down to the fundamental noise limit would have been possible using the analytically computed alignment angles in Tab.~\ref{tab:realignments_June_ana}.

\begin{table}
\begin{tabular}{lrlrlrl}
\toprule
 & \multicolumn{6}{c}{TTL coefficients [fitted $\vert$ analytical]} \\
start time & \multicolumn{2}{c}{15.06.\ 13:30\ }
		   & \multicolumn{2}{c}{20.06.\ 08:00\ } 
		   & \multicolumn{2}{c}{26.06.\ 07:00\ } \\
end time   & \multicolumn{2}{c}{18.06.\ 08:00\ } 
		   & \multicolumn{2}{c}{24.06.\ 08:00\ } 
		   & \multicolumn{2}{c}{26.06.\ 08:00\ } \\
\midrule
$C_\varphi$ [$\upmu$m/rad] &\ -0.6 $\vert$& - & -3.3 $\vert$& -2.1 &  1.9 $\vert$&  1.2 \\ %
$C_\eta$ [$\upmu$m/rad]    &\ -5.7 $\vert$& - & -3.5 $\vert$& -4.2 & -7.5 $\vert$& -7.2 \\ %
$C_y$ [$10^{-6}$] &\ -12.8 $\vert$& - & -27.7 $\vert$& -27.5 &  3.3 $\vert$&  0.7 \\ %
$C_z$ [$10^{-6}$] &\   6.4 $\vert$& - &   0.9 $\vert$&   0.9 & 11.8 $\vert$& 11.8  \\
\bottomrule
\end{tabular}
\caption{Comparison of the fitted (left) and analytically computed (right) cross-coupling coefficients for timespans prior, in between and after the TM realignments in June 2016 (\ref{tab:realignments}). 
The analytical coefficients were computed via the Eqs.~\eqref{eq:anamodel_C}, where the constant offsets of the coefficients were substituted by the respective fitted coefficient prior to these realignments (first column) and the angular TM alignment changes due to the realignments were inserted for the TM angles.
Times are given in UTC.
}
\label{tab:realignments_June_coeffs}
\end{table}

\section{Summary}
\label{sec:summary}

In this work, we presented an extensive analysis of the \gls{TTL} coupling in \gls{LPF} with focus on the \gls{LXE} and the \gls{TTL} coupling reduction by realignment.
We described the coupling dependencies on the test mass alignment parameters.
This has been done using two different \gls{TTL} coupling models.

The analytical model we presented is the first \gls{TTL} model successfully explaining the dependency of the \gls{TTL} coupling into the $\Delta g$ measurement on the \gls{TM} alignment angles. 
Using this model, we have shown why the applied \gls{TM} realignments failed to fully mitigate \gls{TTL} coupling and derived alignment angles, which could have suppressed the \gls{TTL} coupling a priori. This would have made the subtraction of the coupling in post-processing redundant (as long as the coefficients do not change over long time scales). 

While the linear fit model used during the mission was sufficient for \gls{TTL} subtraction in noise runs, we find higher-order noise peaks in experiments with injected \gls{TTL} noise. 
The analytical analysis has shown that the models can be easily extended to second-order models, adding only two additional coupling terms. 
Also, the additional coupling terms are stable and therefore need only to be computed once and can then be applied during the entire mission duration.

The derivation of the analytical model for \gls{LPF} was complex and time-consuming. 
Within this work, we have introduced an alternative computation of an alignment-dependent model.
By evaluating the relation of the fitted coupling coefficients for the twelve sub-experiments on the corresponding \gls{TM} alignment, a model equivalent to the analytical model was found.
Such a procedure can also be applied in future missions like \gls{LISA}, yielding the physical dependency of the measured \gls{TTL} coupling on certain alignment parameters.
Mind that the larger number of jittering components in \gls{LISA} would also make a higher number of sub-experiments with different realignments each necessary. 
However, from the analysis of the \gls{TTL} coefficient changes due to the \gls{TM} realignments in March and June 2016 using noise run data, we deduce that the injections applied during the \gls{LXE} would not have been necessary for the fit of the changing coupling coefficients. 
Hence, a comparable calibration scheme in \gls{LISA} does not necessarily interrupt the scientific measurements.

In conclusion, we have now understood the \gls{TTL} coupling in \gls{LPF} to a large extend.
This drives our confidence that the \gls{TTL} suppression strategies planned for future missions will be successful.

\begin{acknowledgments}
    \label{S:acc}
    This work has been made possible by the LISA Pathfinder mission, which is part of the space-science programme of the European Space Agency.
    The Albert Einstein Institute acknowledges the support of the German Space Agency, DLR. 
    The work is supported by the Federal Ministry for Economic Affairs and Climate Action based on a resolution of the German Bundestag (FKZ 50OQ0501, FKZ 50OQ1601 and FKZ 50OQ1801). We also acknowledge the support by the Deutsche Forschungsgemeinschaft (DFG, German Research Foundation) under Germany’s Excellence Strategy (EXC-2123 QuantumFrontiers, project ID 390837967).
    The French contribution has been supported by the CNES (Accord Specific de projet CNES 1316634/CNRS 103747), the CNRS, the Observatoire de Paris and the University Paris-Diderot. E. Plagnol and H. Inchauspé would also like to acknowledge the financial support of the UnivEarthS Labex program at Sorbonne Paris Cité (ANR-10-LABX-0023 and ANR-11-IDEX-0005-02).
    The Italian contribution has been supported by ASI (grant n.2017-29-H.1-2020 ``Attività per la fase A della missione LISA'') and Istituto Nazionale di Fisica Nucleare.
    The Spanish contribution has been supported by contracts AYA2010-15709 (MICINN), ESP2013-47637-P, ESP2015-67234-P, ESP2017-90084-P (MINECO), and PID2019-106515GB-I00 (MICIN). 
    Support from AGAUR (Generalitat de Catalunya) contract 2017-SGR-1469 is also acknowledged.
    The Swiss contribution acknowledges the support of the ETH Research Grant ETH-05 16-2 and the Swiss Space Office (SSO) via the PRODEX Programme of ESA. L. Ferraioli is supported by the Swiss National Science Foundation.
    The UK groups wish to acknowledge support from the United Kingdom Space Agency (UKSA), the Scottish Universities Physics Alliance (SUPA), the University of Glasgow, the University of Birmingham and Imperial College London.
    J. I. Thorpe and J. Slutsky acknowledge the support of the US National Aeronautics and Space Administration (NASA).
    N. Korsakova would like to thank for the support from the CNES Fellowship.
    The LISA Pathfinder collaboration would like to acknowledge Prof. Pierre Binetruy (deceased 30 March 2017), Prof. José Alberto Lobo (deceased 30 September 2012) and Lluis Gesa Bote (deceased 29 May 2020) for their remarkable contribution to the LISA Pathfinder science. 
\end{acknowledgments}

\def\bibsection{\section*{References}} 
\bibliography{References.bib}
\end{document}